\pdfoutput=1
\documentclass[aps, prd, floats, floatfix, superscriptaddress, onecolumn,PRD,nofootinbib,amsmath,amssymb]{revtex4}

\usepackage[T1]{fontenc}
\usepackage[utf8]{inputenc}
\usepackage{lmodern}
\usepackage{lipsum}
\usepackage{enumerate}% http://ctan.org/pkg/enumerate

\usepackage{MnSymbol}
\usepackage{amssymb}
\usepackage{amsfonts}
\usepackage{eufrak}

\usepackage[dvipsnames, usenames]{xcolor}
\definecolor{linkcolor}{rgb}{0.0,0.3,0.5}
\usepackage[hypertexnames=false, unicode, colorlinks=true, linkcolor=linkcolor,
citecolor=linkcolor, filecolor=linkcolor,urlcolor=linkcolor,
pdfusetitle]{hyperref}

\usepackage[all]{hypcap}
\usepackage{graphicx}
\usepackage{xspace}
\usepackage{amssymb}
\usepackage[normalem]{ulem} %for \sout
\usepackage{bm} % boldmath
\usepackage{appendix}

\usepackage{microtype}
\usepackage{phoenician}

\usepackage[english]{babel}
\usepackage{blindtext}

\graphicspath{%
  {figs/}%
}

\usepackage{tikz}
\usetikzlibrary{shapes,snakes}
\usetikzlibrary{arrows,intersections,patterns,decorations.markings,shapes.geometric}
\usetikzlibrary{calc,fadings,decorations.pathreplacing,positioning}
\usetikzlibrary{decorations.pathmorphing}
\tikzset{snake it/.style={decorate, decoration=snake}}

\usepackage{pgfplots}
\usepgfplotslibrary{colormaps}
\usetikzlibrary{intersections,patterns,pgfplots.fillbetween}

\tikzset{->-/.style={decoration={
  markings,
  mark=at position .5 with {\arrow{>}}},postaction={decorate}}
}
\tikzset{-<-/.style={decoration={
  markings,
  mark=at position .5 with {\arrow{<}}},postaction={decorate}}
}

\tikzset{%
  >=latex, % option for nice arrows
  inner sep=0pt,%
  outer sep=2pt,%
  mark coordinate/.style={inner sep=0pt,outer sep=0pt,minimum size=3pt,
    fill=black,circle}%
}

\DeclareMathAlphabet{\mathpzc}{OT1}{pzc}{m}{it}

\renewcommand{\vec}[1] {\bm{#1}}
\newcommand{\vhat}[1]{\vec{\hat{#1}}}

\newcommand*{\df}  {\delta}

\newcommand*{\non} {\nonumber}
\newcommand*{\lb} {\left(}
\newcommand*{\rb} {\right)}

\newcommand*{\la} {\left\langle}
\newcommand*{\ra} {\right\rangle}

\newcommand{\bk}{{\bf k}}

\newcommand{\HS}{$\mathcal{HS} $ }

\newcommand{\OS}{$\mathcal{OS} $ }
\newcommand{\MS}{$\mathcal{MS} $ }

\newcommand{\eq}[1]{\begin{align}#1\end{align}}
\newcommand{\eeq}[1]{\begin{equation}#1\end{equation}}

\definecolor{darkred}{RGB}{175,0,0}
\definecolor{darkblue}{RGB}{14,0,185}

\begin{document}

\rightline{\scriptsize RBI-ThPhys-2023-13}

%\title{Correlators in the sky.\\ Power spectrum in the cave}
\title{Power spectrum in the cave}

\newcommand\alvisehome{
\affiliation{Dipartimento di Fisica Galileo Galilei, Universit\` a di Padova, I-35131 Padova, Italy}
\affiliation{INFN Sezione di Padova, I-35131 Padova, Italy}
\affiliation{INAF-Osservatorio Astronomico di Padova, Italy}
\affiliation{Theoretical Physics Department, CERN, 1 Esplanade des Particules, 1211 Geneva 23,
Switzerland}}
\newcommand\zvonehome{
\affiliation{Ru\dj er Bo\v{s}kovi\'c Institute, Bijeni\v{c}ka cesta 54, 10000 Zagreb, Croatia}
\affiliation{Kavli Institute for Cosmology, University of Cambridge, Cambridge CB3 0HA, UK}
\affiliation{Department of Applied Mathematics and Theoretical Physics, University of Cambridge, Cambridge CB3 0WA, UK }
}

\author{Alvise Raccanelli}
\email{alvise.raccanelli.1@unipd.it}
\alvisehome

\author{Zvonimir Vlah}
\email{zvlah@irb.hr}
\zvonehome

% Because hyperref only gets the *last* author, we need to be explicit.
%\hypersetup{pdfauthor={Raccanelli et al.}}

%\date{\today}

%==========================================================================
\begin{abstract}
Forthcoming galaxy surveys will provide measurements of galaxy clustering with an unprecedented level of precision, that will require comparably good accuracy.
%. However, precision is not enough if we don't also have accuracy.
Current models for galaxy correlations rely on approximations and idealizations that might be inadequate for ultra precise measurements. On the other hand, exact calculations have proven to be computationally too expensive to be efficiently implemented in real data analyses.
%Here we start a project to provide precise and accurate formalisms for galaxy correlations, including effects due to the geometry of the system and the fact that the radial direction in the sky is degenerate with cosmic time.
%Our goal is to understand the structure of the properly observable galaxy power spectrum and provide a prescription flexible enough to give the required level of accuracy for different cases and geometry.

We start a project to provide precise and accurate formalisms for galaxy correlations, and in this paper we investigate the 3D angular power spectrum including effects of unequal time correlations.
%, i.e.,~including modes along the line of sight.
We establish an explicit link between the full- and flat-sky spectra by performing an asymptotic expansion of the full-sky result around the equal time case.
%, and show the structure and magnitude of such corrections.
The limiting case coincides with the idealized spectrum that a meta-observer would measure if it had access to the entire 4D Universe.
%, where a single Fourier mode could be used, there are no projection effects, and clustering is rotationally and translationally invariant.

%We show that the most conventionally introduced flat-sky, beyond Limber, the result is a leading contribution of the asymptotic series. Moreover, 
The leading term in the obtained flat-sky expansion is the only translationally invariant term in the plane perpendicular to the line of sight, while the higher-order terms account for the deviation from this invariance.
%We argue that these higher-order asymptotic terms in the flat-sky expansion should be interpreted as the self-consistent error estimate when deviating from the full-sky result.
We study the behavior of such corrections for a simplified universe where we can analytically solve the power spectrum and have full control of the equations, therefore being able to understand the exact nature of all the terms and the origin of the corrections.
We highlight that the conclusions and the structure of the unequal time spectra are fully general and serve as lessons and guidance in understanding galaxy clustering in any cosmology.

Finally, we show that our flat-sky unequal time expression matches the exact full-sky calculation remarkably better than commonly adopted approximations, even at the largest scales and for both shallow and deep redshift bins. 
\end{abstract}

\maketitle

%==========================================================================
\section{Introduction}
\label{sec:intro}
Galaxy clustering will be one of the main focuses for astrophysics in the next decade, and a considerable effort is being made worldwide to obtain extremely precise measurements of it. We will need both precision and accuracy.

This decade and the next one will see new generations of galaxy survey experiments, which will bring an unprecedented amount of data over a vast range of wide areas of the sky and deep redshift ranges, and with that improve current tests on a variety of cosmological models.
For this reason, it is paramount to have a set of tools that allows us to model galaxy clustering with the desired precision, maintaining computational feasibility.
Clustering analyses have been so far performed mostly assuming specific approximations that speed up the calculations and still work reasonably well given the survey specifications.
For future surveys, therefore, the optimal formalism needs to be adaptable and include corrections that are relevant either for long radial or for wide transverse correlations.

Most of the effort in improving the theoretical modeling in the past two decades has been on small scale effects, using various approaches, from more phenomenological ones to perturbation theory and the effective field theory of Large-Scale Structure (see e.g.,~\cite{Peacock1996, Bernardeau2002, Scoccimarro2004, matsubara08a, matsubara08b, Taruya2010, Baumann2012, Carrasco2012, Porto2014, Carrasco2014, Senatore2014, Vlah2015, Senatore2015, Angulo2015, Vlah2016, Perko2016, Senatore2018, Simonovic2018, Ivanov2018, Vlah2019b, Gebhardt2020, Philcox2020, Elsner2020, Fasiello2022, Wang2022, Garny2022}).
The very large scales have received a bit less attention, as very widely separated galaxy pairs will be measured accurately and precisely only for forthcoming surveys. Initial attempts set the ground for a proper treatment of wide-angle surveys (see e.g.,~\cite{Zaroubi, Szalay:1997, Matsubara:1999, Szapudi:2004, Papai:2008}).
Those works provided a formalism that is correct at the largest scales, but are set in ways that make them computationally prohibitive to use.

Neglecting the information on the largest scales however would reduce the ability of future surveys to reach their maximum potential, in particular for some science cases, such as e.g.,~primordial non Gaussianity~\cite{Raccanelli:2018Doppler, Spezzati:23}.

The galaxy power spectrum has been used for decades to set limits on cosmological parameters and test models, see e.g.,~\cite{Yu1969, Peebles1973, Tegmark2006, Percival2002, Percival2004, Percival2007, Okumura2008, guzzo08, Abazajian2009, Percival2010, Kazin2010, Reid2010, Samushia2012, Raccanelli:2013growth, Ross2013, Beutler2014, Percival2014, GilMarin2015, Rodriguez-Torres2016, GilMarin2016, Alam_2017, Sanchez2017, Beutler2017a, Beutler2017b, GilMarin2017, Abbott2018, GilMarin2018, Kobayashi2020, dAmico2020, Ivanov2020, Alam_2021}.
Until now, galaxy surveys covered small fractions of the sky (or, in the case of full sky ones, the statistical power of large-scale modes have been very small and there was no redshift information available), and therefore a series of approximations have been employed to model the galaxy power spectrum in a simplified way that allowed very fast calculations. In most cases, even when largely separated pairs were available, they have not been included in the analyses.

The next generation of galaxy surveys, however, will focus on higher redshift observations (see e.g.,~the proposed ATLAS Explorer\footnote{https://atlas-probe.ipac.caltech.edu/}~\cite{Spezzati:23ATLAS} and Megamapper~\cite{Megamapper2022} surveys); in this case, the cosmological volume probed will allow to have extremely precise measurements on very large scales, and therefore require very precise modeling on large scales. Recently there has been a debate on the appropriate modeling of the power spectrum, and in particular on the use of its full shape or just focusing on BAO parameters (see e.g.,~\cite{Nishimichi_2020, Brieden2021, Brieden, Chen2022} for some discussions and examples).

Such approximations, on large scales, involve mainly assuming the flatness of the sky, i.e.,~taking the so-called plane-parallel approximation or the distant observer limit (the two being similar but not strictly the same).
Some of the forthcoming and planned future cosmological surveys, such as the Euclid satellite~\cite{Euclid}, DESI~\cite{DESI}, SKAO~\cite{SKA}, SPHEREx~\cite{SPHEREx}, PFS~\cite{PFS}, the Roman Space Telescope~\cite{WFIRST}, the Vera Rubin observatory~\cite{LSST}, will survey very large areas of the sky, with a high number density; therefore it is becoming necessary to model statistical observables on full sky, appropriately accounting for the curvature of the sky and all the geometrical effects.
This necessity has become more clear in the community in recent years, and some efforts have been dedicated to this issue. Following the pioneering works mentioned earlier, and later by~\cite{Raccanelli:2010wa, Montanari2012, Bertacca:2012}, there have been studies of large-scale correlations with preparations for future surveys in mind~\cite{Yoo2013, Raccanelli:2013growth, Raccanelli2014, Raccanelli:2015GR, Raccanelli2016, Reimberg2016, Raccanelli:2018Doppler, Borzyszkowski_2017, Castorina2018, Scaccabarozzi2018, Castorina2019, Taruya2020, Grimm:2020, Castorina2022, Elkhashab2022, Noorikuhani2022}.

The physics of galaxy clustering happens over a series of 3D snapshots and it can be described by the classically defined power spectrum, using one Fourier mode. However, the 3D power spectrum in Fourier space is not a proper observable once we drop the idealistic solution of particles moving in a box with no observer and where the radial coordinate is not degenerate with time. In a real situation, measurements happen on the lightcone, and the proper observables are redshifts and angular position on the (spherical) sky, so that the statistical quantity that needs to be used is the 3D angular power spectrum. While brute-force calculations for this observable are doable and have been studied in detail (see e.g.,~\cite{CMBFAST, CAMB, CLASS}), for practical purposes and real data analyses there are two possible approaches: {\it (i)} slicing the catalogs in many redshift bins, paying the price of having a prohibitively large number of spectra to compute (with non-diagonal covariances), or {\it (ii)} resort to just a few, deep, redshift bins, with the downside of losing precious radial information.
We aim to develop a formalism that allows the connection between the idealized, mathematically simple, case of the power spectrum ``in the box'' (see Figure~\ref{fig:HS}, that assumes negligible angular separation between galaxies and no redshift separation) and the fully 3D observed spectrum. This connection clarifies the relation between calculations of the gravitational dynamics and projection effects; we will see the mathematical details of such a relation in~\cite{RVIII}.

Therefore, the 3D Fourier power spectrum that is appropriate to compute dynamics of galaxies is not technically observable by us, but lives in what we call Hyperuranion Space (\HS). There is however the possibility to link what we observe on the lightcone (angular positions on a sphere and redshifts) in what we call Observed Space (\OS). We give more details and illustrations of the situation in Section~\ref{sec:spaces}, and in Table~\ref{tab:notation} we summarize the list of spaces and notation used in this work.

In this paper we start by analyzing the difference between the meta-space where gravitational effects drive the evolution of clustering and the space where observations happen.
The situation can be compared to Plato's ``allegory of the cave''\footnote{Republic, Plato, 375 BC}, where the prisoners see shadows (projections) of the ``real'' forms; in the same way, when we observe correlators of the Large-Scale Structure (LSS) of the Universe from our fixed point of view, we have to account for these projection effects, and link them to the real motion of galaxies.

%=======================================================================%
\begin{table*}
\centering
\begin{tabular}{l|l}
\hline
\hline
$\df^{\rm K}_{ij}$ & Kronecker symbol \\[3pt]
$\df^{\rm D} (\vec x)$ & Dirac delta function \\
$ W(\chi) $ & Window function; related to the specific observable and survey\\
\hline
$\df(\vec x)$ & 3D density field of matter or biased tracer \\
$\hat \df(\vec \theta)$ & 2D projected filed in the real space coordinates on the sky  \\
\hline
$\mathcal P(\vec k;\, z, z')$ & Unequal-time theoretical power spectrum of the 3D density field (unobservable) \\
$C_{\ell}$ & Projected angular power spectrum (with finite size window functions)  \\
$P(\vec k;\, z)$ & Equal-time observed power spectrum (constructed from observable fields) \\
\hline
\HS & Hyperuranion space (unaccessible) \\
\OS & Observational space (theoretical observables --- accessible in principle) \\
\MS & Measurement space (where real observations are performed) \\
\hline
\hline
\end{tabular}
\caption{List of notation and most important quantities used in this paper.}
\label{tab:notation}
\end{table*}

We study the inclusion of unequal-time correlations to the angular (3D) power spectrum. We derive an expansion around the equal time case and we show how including radial effects in flat-sky allows us to fit the full angular power spectrum extremely well in a computationally feasible way.

\section{Theory and observations}
\label{sec:spaces}
As we plan to develop a formalism to describe the true observable power spectrum, we need to properly define and clarify the framework for our derivations, the mathematical properties and the physics of our system.
We clarify what are the idealized and realistic situations, and what are the mathematical consequences of such differences.

\begin{figure*}[htbp]
\centering
\includegraphics[width=0.45\columnwidth]{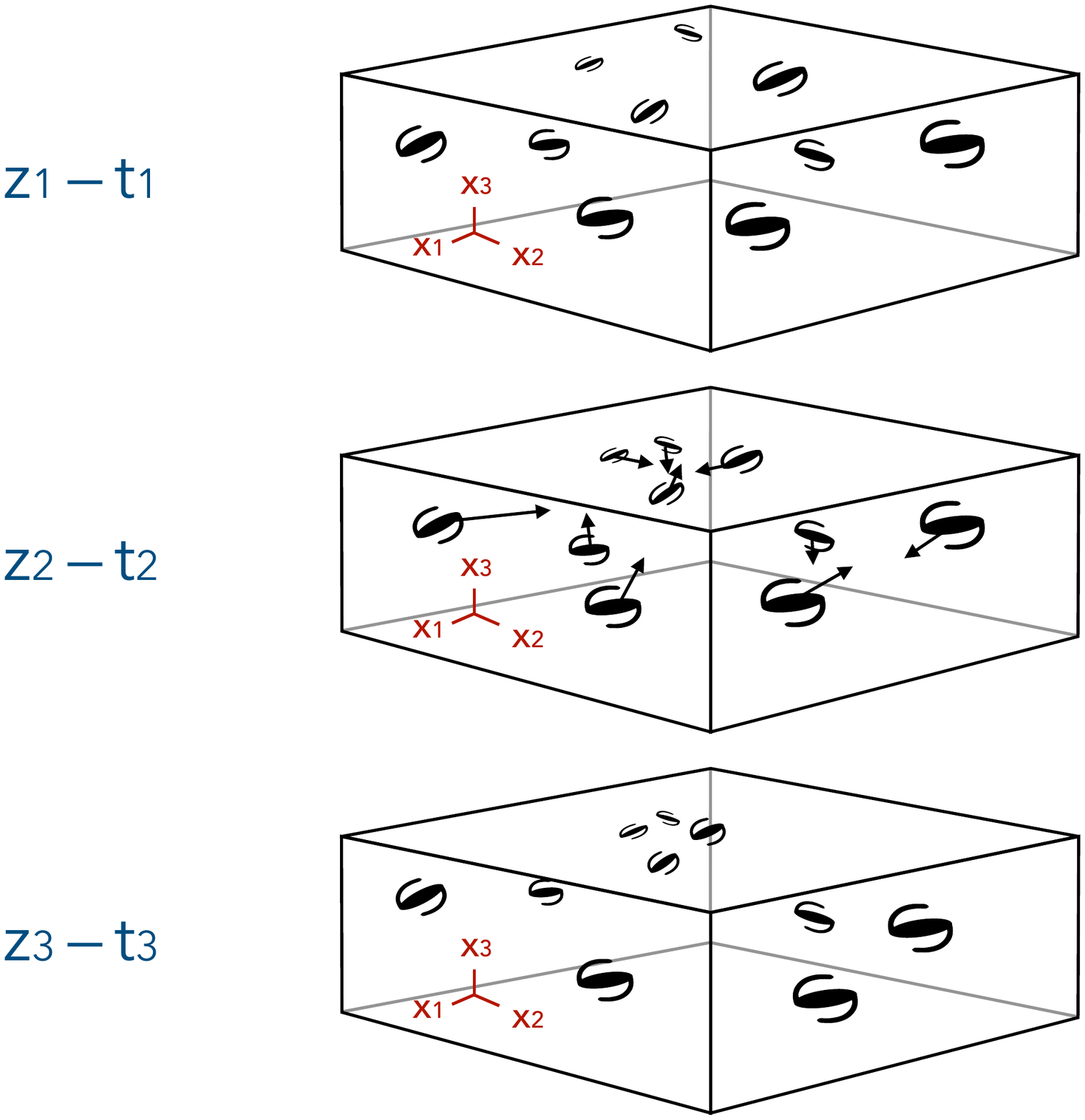}
\includegraphics[width=0.45\columnwidth]{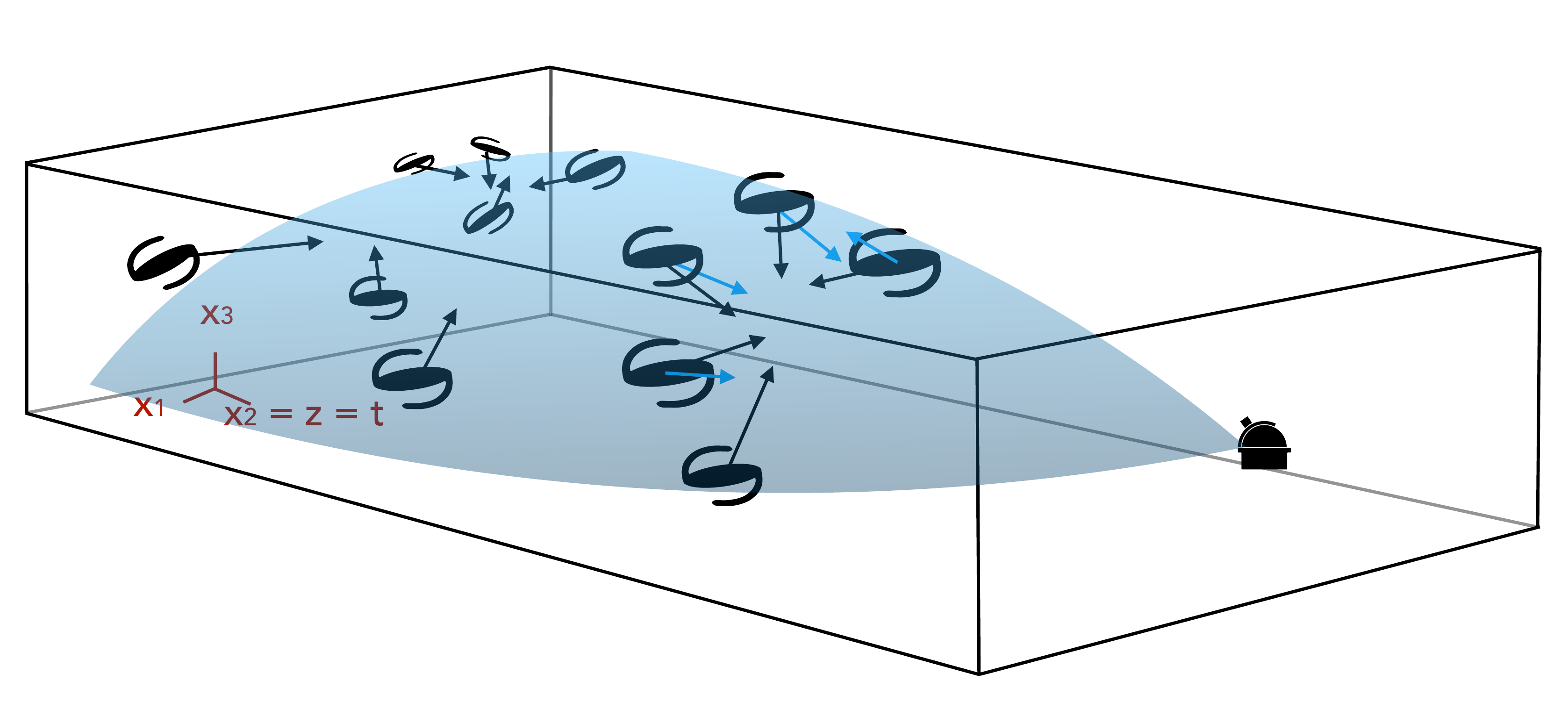}
\caption{{\it Left}: Hyperuranion-space (\HS). This is the Universe as it would be seen from an ``external'' (meta) observer. Inside the box, there are 3 spatial dimensions and there can be 3D snapshots in time. In this space one can define a proper 3D Fourier power spectrum, which would unambiguously describe the dynamics of the system. \\
{\it Right}: Observational-space (\OS). In this case, while we still ignore observational effects such as survey geometry and telescope specifications, we introduce an observer. This is the appropriate space to compute theoretical observables, and where redshift-space distortions can be correctly computed.
The introduction of an observer means that translational invariance is broken, there is a special direction in space, and the Fourier power spectrum cannot be defined in a mathematically unambiguous way.
Note that now one of the 3 spatial dimensions, the radial one, is degenerate with redshift and therefore time. To be rigorous, any radial correlation will also be an unequal time correlation.}
\label{fig:HS}
\end{figure*}

The most convenient (and used) way to calculate the dynamics of galaxy clustering is in the 3D Fourier space. In fact, this is what would be ideally used by an ``external'' {\it meta-observer} that could take a series of instantaneous snapshots of the whole Universe, without observing from one position within the box, and without the degeneracy between a spatial direction and the time coordinate (see Figure~\ref{fig:HS}).
In the rest of this series of papers we will distinguish between three different ``spaces'', that will help clarifying in which situation we will be for each calculation. We define:
\begin{enumerate}[i.]
\item Hyperuranion-space (\HS). This is the idealized theory space as it would be seen from the {\it meta} observer able to see every instant in time separately as in a series of screenshots. This is where the true dynamics should be calculated and therefore we compare it to the Platonic hyperuranion.
\item Observational-space (\OS). Here is where theoretical calculations leading to observations should be done, as this is where the {\it theoretically observed galaxy clustering} lives. This includes projection effects, as any realistic observations is performed from a point in space-time and therefore identifies a past lightcone.
\item Measurement-space (\MS). We identify this as the space where real data analyses should be performed. This includes not only projection effects but also experiment details and observational corrections due to geometry, masks, magnitude limits, local motions, etc.
\end{enumerate}
In this work we focus on the first two of these spaces.
We will leave to a future work the inclusion of  details needed to go into \MS.

In the \HS we can define a box that contains the objects we want to measure and define a proper power spectrum in Fourier space, with one Fourier mode ${\bf k}$. This is where the physical motions of galaxies happens, and therefore we should use all the symmetries and properties of the system. In particular, we can see that rotational and translational invariance are preserved, and so is statistical homogeneity. Here the best statistics to be used is the 3D power spectrum $P({\bf k})$,  defined as $P({\bf k_1}, {\bf k_2}) \propto \langle \delta({\bf k_1}) \delta({\bf k_2})\rangle$, that with statistical homogeneity and isotropy gives:
\begin{equation}
\label{eqPkhom}
P({\bf k_1}, {\bf k_2}) = \delta_D({\bf k_1}-{\bf k_2})P({|\bf k_1|}) \, .
\end{equation}
In Figure~\ref{fig:HS} we can see, on the left, the situation as it would be seen by the meta-observer, where at any moment in time there are 3 spatial directions and one can properly describe the gravitational dynamics in 4D.
The fact that Equation~\eqref{eqPkhom} is valid only for homogeneity and isotropy will be important for the rest of this work.

This is a result always valid for the meta-observer, but the situation is different when we include an observer in the box.
In \OS the presence of an observer means there is a lightcone, peculiar velocities affect the apparent radial position of sources, and translational invariance is broken (see the right panel of Figure~\ref{fig:HS}), while rotational invariance is still preserved. Moreover, the system can not be described anymore by using only one Fourier mode, but there would need to be one per galaxy.

This was already noted in e.g.,~\cite{Yoo2013} (but see also~\cite{zaroubi93} for an alternative formulation with two lines of sights), where the authors defined a spherical power spectrum $\mathcal{S}$ as:
\begin{equation}
\label{eq:spherpow}
\langle \delta_{\ell m} (k)\delta^*_{\ell' m'} (k') \rangle = \delta_{\ell \ell'}\delta_{m m'} \mathcal{S}_\ell(k,k') =
\frac{i^\ell(-i)^{\ell'} kk'}{(2\pi)^3} \int Y^*_{\ell m}(\hat{k})Y_{\ell' m'}(\hat{k}') \langle \delta(k)\delta^*(k') \rangle d^2\hat{k} \, d^2\hat{k}' \; .
\end{equation}

The connection between \HS and \OS strictly happens when the box containing the observed sources is reduced to a point as the observer is at infinite distance from the survey and the angular separation between galaxies reduces to zero. We will investigate the mathematics of this connections later on, and in detail in~\cite{RVIII}.
In terms of Equation~\eqref{eq:spherpow}, we reduce to the case:
\begin{equation}
\label{eq:spherpowsimm}
\mathcal{S}_\ell(k,k') = \delta^D(k-k')\mathcal{S}_\ell(k) = \delta^D(k-k') P(k) \, .
\end{equation}

This is the situation in which the standard ``Kaiser'' analysis is generally performed, and was a good enough approximation for past, (relatively) narrow and shallow surveys. In the last years, these approximations have been questioned for forthcoming surveys, focusing in particular on wide surveys, modeling the so-called ``wide-angle effects'' and studying their importance (see e.g.,~\cite{Szalay:1997, Szapudi:2004, Papai:2008, Raccanelli:2010wa, Desjacques:2021}). Following that, in consideration of the fact that future galaxy surveys will also be very deep in redshift, there have been studies on how to include the so-called projection effects which arise from cosmological perturbations due to gravitational potentials (gravitational lensing, integrated Sachs-Wolfe and time delay effects), see e.g.,~\cite{Yoo:2009, Yoo:2010ni, Bonvin:2011, Challinor:2011, Bertacca:2012, Jeong:2012GR, Maartens:2013, Raccanelli:2015GR, Tansella:2018, Yoo2019, Grimm:2020}.

%==========================================================================

\section{Galaxy clustering angular power spectrum}
In this work we argue that the most natural statistical analysis of the observed distribution of galaxies in the sky happens in spherical coordinates. In~\cite{RVII} we present a new observable that naturally includes transverse and radial mode effects in 3D angular space.
The idea of modeling the  galaxy power spectrum as distribution of points on a sphere dates back to the end of the 60s~\cite{Yu1969} and was then formalized in~\cite{Peebles1973}. The literature on the topic is vast and we will just mention the first application to the velocity field in~\cite{Regos1989}, and then to the IRAS survey in~\cite{Scharf1992}, followed by the addition of radial effects and RSD in~\cite{Scharf1993, Fisher1994}, where all the details of the original derivations can be found.

The idea is to define a function of angular position on the sphere and use spherical harmonics; the power spectrum for a distribution on a sphere is the ensemble average of the square of the sum of the spherical harmonics of all the objects in the catalog. The power spectrum we are looking for is then the covariance function of the harmonics transformation~\cite{peebles80}:
\begin{equation}
a_\ell^m = \int \chi^2 \ Y_\ell^m(\Omega) \rho(\chi-\chi_0) \, d \chi d \Omega \, .
\end{equation}
Note that in the 3D case including RSD, there is a radially varying weight to be included that breaks some symmetries, which will be important for our results in this series of papers.

Analyses in angular space have been performed in the 1990s and subsequently mostly fell out of fashion. For modern surveys, the focus moved to the Fourier-space $P(k)$ and the configuration space 2-point correlation function $\xi(s)$, with some notable exceptions such as e.g.,~\cite{Padmanabhan2007, ho08, Ross2011}, with angular correlations mostly used for photometric and radio surveys. More recently there has been some more push towards the use of $C_\ell$ (see e.g.~\cite{Matthewson2022, Tanidis2023}); some investigation of RSD effects on the angular power spectrum can be found in~\cite{Fisher:1993pz, Tadros1999, Padmanabhan2007, Tanidis2019, Tomlinson2020, Gebhardt2020, Assassi}.
The full expression of $C_\ell$s including all velocity and relativistic corrections has been calculated and coded up in~\cite{Challinor:2011, Bonvin:2011, DiDio:2013bqa, Bellomo2020} in the context of Boltzmann solvers.
Additional reasons to use the angular spectrum are the natural inclusion of wide-angle correlations, the relative easiness in including relativistic large-scale effects and lensing, and the fact that the covariance can be written in a simple way. Moreover, angular multipoles $\ell$ are (at a first approximation) uncorrelated.
The drawback of this approach is given by expensiveness of computations when analyzing deep surveys that have a good redshift resolution, therefore having to calculate a very large number of auto- and cross- bin correlations.
However, recently~\cite{Assassi, Tomlinson2020} developed some innovative approaches to make the calculation of $C_\ell$s more approachable, and so re-ignited the interest in the use of this observable.

At this point we want to mathematically connect the dynamics due to gravitational processes in \HS with the theoretically-observed angular power spectrum in \OS. We start by defining the coefficients for the spherical expansion of the overdensity field located at a radial distance from the observer:
\begin{equation}
a_{\ell m} (z) = i ^\ell \sqrt{\frac{2}{\pi}} \int \, dk \, \delta(k,z) j_\ell(kr) Y_{\ell m}^*(\hat{k}) \; .
\end{equation}
The overdensity field can be decomposed into (3D) spherical modes, and the 3D angular power spectrum is then the 2-point correlation of the coefficients $\delta_{\ell m}(k)$:
\begin{equation}
\label{eq:ensembleCls}
\left\langle \delta_{\ell m}(k_1 ; \bar{\chi}) \delta_{\ell^{\prime} m^{\prime}}^{*}\left(k_2 ; \bar{\chi}\right)\right\rangle=C_{\ell}(k ; \bar{\chi}) \delta_{D}\left(k_1-k_2\right) \delta_{\ell \ell^{\prime}} \delta_{m m^{\prime}} \, ,
\end{equation}
giving the well-known expression for the angular power spectrum:
\begin{equation}
\label{eq:}
C_{\ell}= \frac{2}{\pi} \int \, dk \, k^2 P(k) \int \, d\chi_1 W(\chi_1) j_\ell(k\chi_1) \int \, d\chi_2 W(\chi_2) j_\ell(k\chi_2) \, ,
\end{equation}
where $W(\chi_i)$ contain the physical effects kernels and the observational window functions.

However, it is important to note that the assumption of homogeneity and isotropy is broken in 3D redshift-space, and therefore Equation~\eqref{eq:ensembleCls} is not anymore valid.
Thus, the {\it observed} galaxy angular power spectrum is not diagonal, and the right hand side of Equation~\eqref{eq:ensembleCls} will not have a delta function.
One can connect the 3D $P(k)$ ``in the box'' to angular correlations in that way only for very large $\ell$ and infinitely thin radial windows.

It is important to notice a consequence of the presence of an observer, even without considering redsfhit-space distortions, that is not commonly appreciated.
In the full-sky treatment, isotropy guarantees the proportionality of the angular power spectrum to the Kronecker delta $\delta^{\rm K}_{\ell \ell'}$, while in the flat-sky approximation once can obtain this condition from the translational invariance in the plane. However, for two physical modes $\vec k_\perp$ laying on two different redshift planes to agree, we have to readjust the corresponding angles, as stated by the Dirac delta $\df^{\rm 2D} \big( \vec{\ell} - \vec{\ell}' \big)$. This generates off diagonal contributions, as a consequence of the fixed observer. We will investigate this in detail in~\cite{RVII}.

We can now calculate the expression of the (linear) angular power spectrum in redshift-space, including all peculiar velocity effects.
If, however, one plans to use, as we argue in this work, 3D angular correlations even for the case of narrow bins, velocity effects need to be included.
We start by writing the overdensity in redshift space as the one in real space corrected for peculiar velocities:
\begin{equation}
\delta^s({\bf n}, \chi)=\delta^r({\bf n}, \chi)-\frac{1}{\mathcal{H}(\chi)} \partial_{r}[{\bf V}({\bf n}, \chi) \cdot n] \; .
\end{equation}
To express the power spectrum in redshift-space, we need to first calculate the spherical coefficients. They can be written, for the real-space (intrinsic clustering), peculiar velocity (standard RSD-Kaiser term) terms, and the Doppler term, as:
\begin{align}
\label{eq:almrsd}
& a_{\ell m}^{\delta}(\chi)=i^{\ell} \sqrt{\frac{2}{\pi}} \int d {\bf k} \,D(\chi) j_{\ell}(k \chi) Y_{\ell m}^{*}(\hat{k}) \, , \\
& a_{\ell m}^{\partial V}(\chi)=i^{\ell} \sqrt{\frac{2}{\pi}} \int d {\bf k} \, k \, {\bf V} ({\bf k}, \chi) j_{\ell}^{\prime \prime}(k \chi) Y_{\ell m}^{*}(\hat{k}) \,  , \\
& a_{\ell m}^{\bf v}(\chi)=-i^{\ell} \sqrt{\frac{2}{\pi}} \int d {\bf k} \,  D(\chi) \frac{\alpha(\chi) \mathcal{H}(\chi) f(\chi)}{k \chi} j_{\ell}^{\prime}\left(k \chi\right) Y_{\ell m}^{*}(\hat{k}) \, .
\end{align}
The expression for the 3D, full sky angular 2-point correlation is then:
\begin{align}
\label{eq:Cls3D}
\left\langle a^{\chi_1}_{\ell m}a^{\chi_2 *}_{\ell' m'}\right\rangle &= \delta^K_{\ell\ell'}\delta^K_{mm'}C_\ell(\chi_1,\chi_2) =
\frac{2}{\pi} \int dk \, k^2 P(k) \times \nonumber \\
&\times \int d\chi_1 \left[b_1 D_1 j_\ell(k\chi_1) - f_1 \frac{\alpha_1}{k \chi_1} j_\ell^{'}(k\chi_1) -f_1 j_{\ell}^{''}(k\chi_1) \right] \times \nonumber \\ 
&\times \int d\chi_2 \left[b_2 D_2 j_\ell(k\chi_2) - f_2 \frac{\alpha_2}{k \chi_2} j_\ell^{'}(k\chi_2) -f_2 j_{\ell}^{''}(k\chi_2) \right] \; ,
\end{align}
where in the usual notation, $b$ is the galaxy bias, $f$ is the growth rate, and for an easier notation we indicate quantities at a specific radial distance $x(\chi_i) = x_i$. Note that we are here including only velocity effects for simplicity; a full expression including relativistic effects can be found in e.g.,~\cite{Bonvin:2011, Challinor:2011}.
In Appendix~\ref{app:cls} we write the full expression in a way that shows how projection effects are separated from the gravitational physics evolution, and can be moved to the window functions.

%%%%%%%%%%%%%%%%%%%%%%%%%%%%%%%%%%%%%%%%%%%%%%%%%%%%%%%%%%%%%%%%%%%%%%%%%%%%%%%%%%%%%%%%%%%%%%%%%%%%%%%%%%%%%%%%%%%%%%%%%%%%%%%%%%%%%%%%%%%%%%%%%%%%%%%%%%%%%%%%%%%%%%%%%%%%%%

\section{Connection between curved and flat sky}
\label{sec:}
\begin{figure}[!htbp]
\centering
\includegraphics[width=0.75\columnwidth]{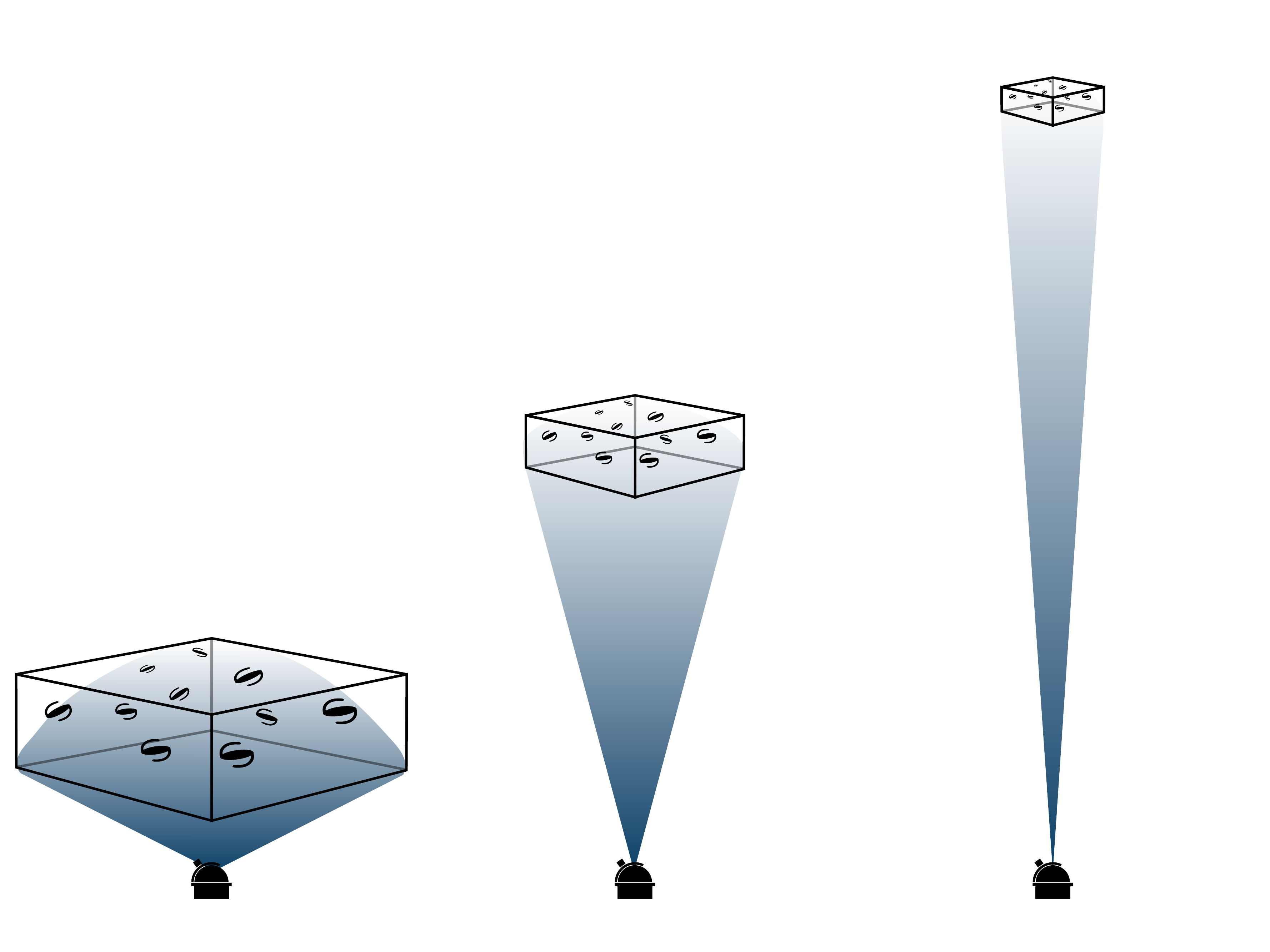}
\caption{Transition from lightcone case to Kaiser, equal time flat sky limit. When the survey is narrow and shallow, we recover the equal time and plane parallel limit.}
\label{fig:kaiserlimit}
\end{figure}

The connection between curved and flat sky happens when the observed volume is small and distant from the observer, to the point that {\it (i)} the curvature of the sky can be neglected, {\it (ii)} the aperture angle between sources is small, to the point that line of sights can be considered parallel, and {\it (iii)} the radial distance between sources is small compared to the distance to the observed volume. This is illustrated in Figure~\ref{fig:kaiserlimit}, and it also corresponds to observations made by the meta-observer.
Mathematically, this translates into assuming $\ell\rightarrow\infty$, $\Delta\theta\rightarrow0$ and $\Delta\chi/\chi\rightarrow0$. In the $C_\ell$ formulation, the first two conditions are simultaneously met by taking the limit of large-$\ell$, as normally assumed in flat-sky approximations, but we also have to take the limit of $\chi\rightarrow\infty$. This will recover the translational invariance and the Kaiser formula.

In this Section we derive formulas that show such connection mathematically, taking inspiration from some results initially obtained for the CMB (see e.g.,~\cite{Bond1987, Weinberg:CMB}).
The starting point here is to take the asymptotics for the limits above, which are in the limit of validity of the Nicholson approximation~\cite{Nicholson}, which is when both the order and the argument of the spherical Bessel functions are large~\cite{Gradshteyn, NIST:DLMF}; we can then write the spherical Bessel function as:

\begin{equation}
j_{\ell}(\tilde{\ell} x) \sim \sqrt{\pi}\left[\frac{
-\left(\frac{3}{2}\right)^{2/3} \left(\sqrt{x^2-1} - \operatorname{Arctan} \sqrt{x^2-1} \right)^{2/3}}
{1-x^{2}}\right]^{1 / 4} \frac{\operatorname{Ai}\left[ -\left(\frac{3 \tilde{\ell}}{2}\right)^{2/3}\left(\sqrt{x^2-1} - \operatorname{Arctan} \sqrt{x^2-1} \right)\right]}{\tilde{\ell}^{5 / 6} \sqrt{x}} \, ,
\end{equation}

where $\tilde{\ell}=\ell+1/2$ and $x$ is a positive number.
The crucial advantage of this approach is that such a limit can be rewritten in a much simpler way when we have squared spherical Bessel functions, where we have:

\begin{equation}
\label{eq:jl2}
j_{\ell}(\tilde{\ell} x)j_{\ell}(\tilde{\ell} x) \sim \pi \frac{\left\{\operatorname{Ai}\left[ -\tilde{\ell}^{\,2/3}\left(\frac{3}{2} \sqrt{x^2-1}-\operatorname{Arctan}\sqrt{x^2-1} \right)^{\frac{2}{3}}\right]\right\}^2}
{x \tilde{\ell}^{\,5/3}} 
\sqrt{\frac{-\left(\frac{3}{2} \sqrt{x^2-1}-\operatorname{Arctan}\sqrt{x^2-1} \right)^{\frac{2}{3}}}
{1-x^2}} \; .
\end{equation}

The next step in reducing this approximation to a simpler form uses the fact that for squared Airy functions there is an integral representation in the form:

\begin{equation}
\label{eq:Ai}
\rm{Ai}^2(z) = \frac{1}{2\pi^{3/2}} \int_0^\infty \cos{\left(\frac{1}{12}\varphi^3+z\varphi+\frac{\pi}{4}\right)} \frac{d\varphi}{\sqrt{\varphi}} \, \approx \frac{1}{2\pi(-z)^{1/2}}\Theta(-z) \, ;
\end{equation}

where we used the Riemann-Lebesgue lemma to perform the second step~\cite{Hamdan}, and $\Theta$ is the Heaviside step function.
Substituting Equation~\eqref{eq:Ai} into Equation~\eqref{eq:jl2}, we end up with the expression:

\begin{equation}
j_{\ell}(\tilde{\ell} x)j_{\ell}(\tilde{\ell} x) \sim
\frac{1}{2}
\frac{1}{x}
\frac{1}{\sqrt{1-x^2}}
\tilde{\ell}^{-2} \; .
\end{equation}

\begin{figure}[htb!]
\includegraphics[width=0.75\textwidth]{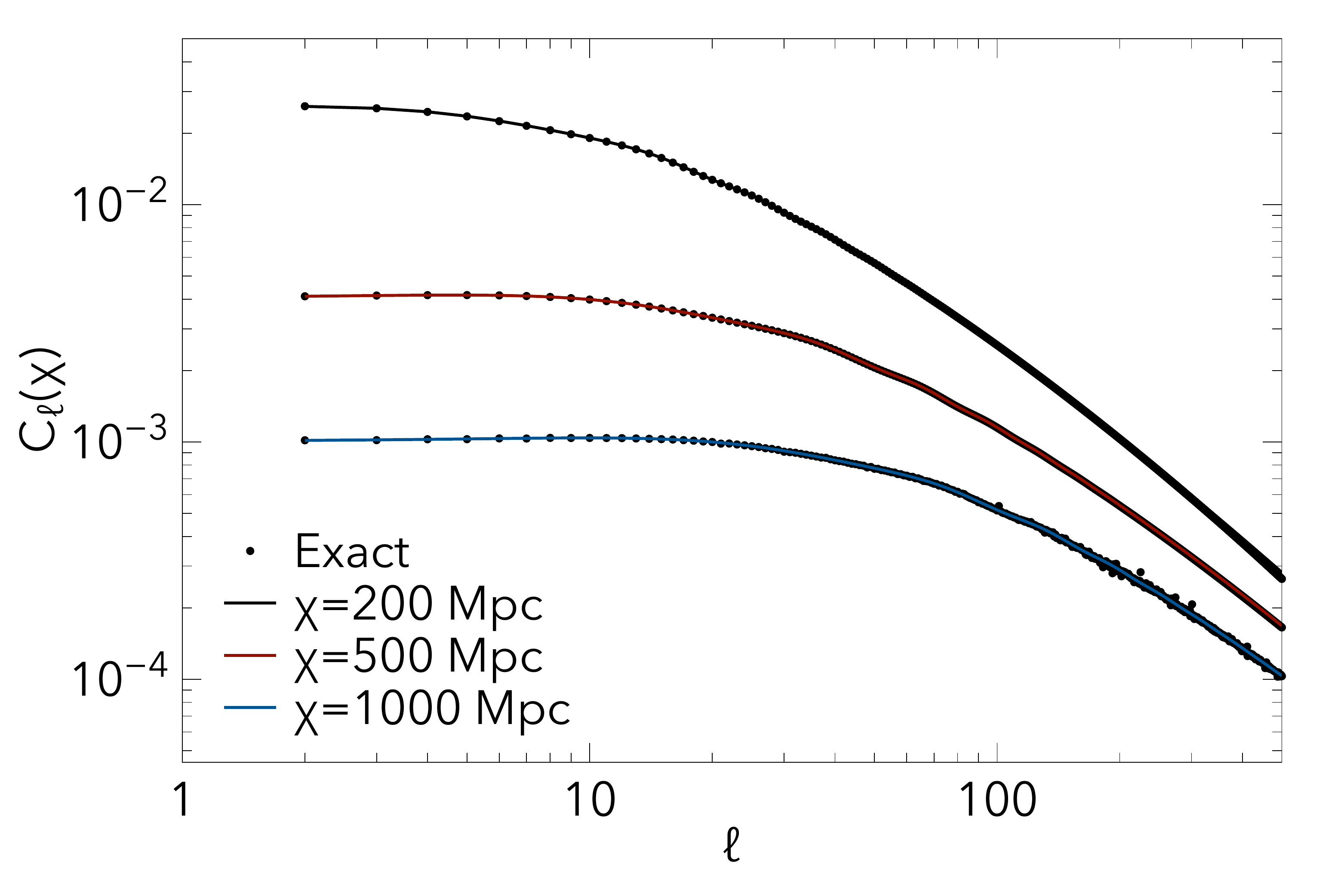}
\caption{Fit for our formula connecting curved to flat sky, for different comoving distances. Points are the exact calculation (the dispersion being due to numerical errors), lines from Equation~\eqref{eq:ETC_IR}.}
\label{fig:ETC_IR}
\end{figure}

Inserting the Nicholoson approximation for the squared spherical Bessel in this form into Equation~\eqref{eq:Cls3D}, we obtain our new expression for angular power spectra in the asymptotic, connecting case:
\begin{equation}
\label{eq:ETC_IR}
E_\ell = \frac{\tilde{\ell}}{4\chi^2} \int_0^{\infty} d(x^2-1) P\left(\frac{\tilde{\ell}}{\chi} \, x\right) \frac{1}{\sqrt{x^2-1}} \, .
\end{equation}
In Figure~\ref{fig:ETC_IR} we show the comparison between the exact calculation and Equation~\eqref{eq:ETC_IR}, using the $\Lambda$CDM power spectrum.

%%%%%%%%%%%%%%%%%%%%%%%%%%%%%%%%%%%%%%%%%%%%%%%%%%%%%%%%%%
%%%%%%%%%%%%%%%%%%%%%%%%%%%%%%%%%%%%%%%%%%%%%%%%%%%%%%%%%%
%%%%%%%%%%%%%%%%%%%%%%%%%%%%%%%%%%%%%%%%%%%%%%%%%%%%%%%%%%
%%%%%%%%%%%%%%%%%%%%%%%%%%%%%%%%%%%%%%%%%%%%%%%%%%%%%%%%%%
%%%%%%%%%%%%%%%%%%%%%%%%%%%%%%%%%%%%%%%%%%%%%%%%%%%%%%%%%%

\section{Expansion series for the angular power spectrum including radial modes}
We want to provide a formulation of the power spectra including the effects of radial modes. As we expect them to be small, and generally distances between correlated sources are much smaller than our average distance from them, in order to do that, we will derive an expression with terms of a series expansion around the case of equal time correlations, i.e.,~when the radial separation between the sources is zero and it's the limit where translational invariance is recovered.
We set ourselves in flat-sky, so that we can use projected fields and write:
\eeq{
\hat \df(\vec \ell)   
= \int \frac{d\chi}{\chi^2} ~ W(\chi) \int \frac{d k_{\hat n}}{2\pi}~ e^{- i \chi k_{\hat n}} \df \big( k_{\hat n} \vhat n, \vec{\tilde \ell}, z[\chi] \big) \, .
}
In a general way that keeps the radial modes contribution, and keeping in mind the different options for the choice of the average comoving distance $\bar \chi$ (see Appendix~\ref{app:mean}), the angular spectrum can then be written as:
\begin{equation}
\la \hat \df(\vec \ell) \hat \df^*(\vec \ell') \ra
= (2\pi)^2 \int d \bar \chi d \delta \chi \frac{|J(\bar \chi, \delta \chi)|}{ \bar \chi^4 f(\daleth)^2 } ~ W\lb \bar \chi, \delta \chi \rb W'\lb \bar \chi , \delta \chi \rb
\df^D \lb \vec{\tilde \ell} - \vec{\tilde \ell'} \rb
\int \frac{d k_{\hat n}}{2\pi} ~ e^{ i \delta \chi k_{\hat n}}
 \mathcal P \big( k_{\hat n} \vhat n, \vec{k}_\perp, \bar \chi, \delta \chi \big) \, , 
\end{equation}
where $f(\daleth)$ depends on the coordinate and the choice of the mean, $J(\bar \chi, \delta \chi)$ is the Jacobian of the transformation and we introduced a 
dimensionless variable $\daleth = \frac{1}{2}\delta \chi/\chi$.
Using the delta function representation in the new variables, we can write:
\begin{equation}
\label{eq:2Ddelta}
(2\pi)^2 \df^{\rm 2D} \lb \vec{ \tilde \ell} - \vec{\tilde \ell}' \rb 
= (2\pi)^2 \df^{\rm 2D} \lb \frac{\chi' \vec{\ell} - \chi \vec{\ell}'}{\chi\chi'} \rb
= \bar \chi^2 f(\daleth) (2\pi)^2  \df^{\rm 2D} \big[ \vec{\ell} - \vec{\ell}' + \tilde \delta (\vec{\ell}' + \vec{\ell}) \big] \, ,
\end{equation}
where $\tilde\delta$ is the off-diagonal phase of the Dirac delta function. The different expressions for the 2D delta function are reported in Appendix~\ref{app:2Ddelta}.
We have then:
\begin{align}
\label{eq:ellbra}
\la \hat \df(\vec \ell) \hat \df(\vec \ell') \ra
= \hspace{-0.1cm} (2\pi)^2 \hspace{-0.2cm}
\int d \bar \chi d \delta \chi\; 
W\hspace{-0.1cm}\lb \bar \chi, \delta \chi \rb W'\hspace{-0.1cm}\lb \bar \chi , \delta \chi \rb
\df^{\rm 2D} \big[ \vec{\ell} + \vec{\ell}' + \tilde\delta (\vec{\ell}' + \vec{\ell})  \big]  \frac{|J(\bar \chi, \delta \chi)|}{ \bar \chi^2} \hspace{-0.2cm}\int \hspace{-0.1cm}\frac{d k_{\hat n}}{2\pi} ~ e^{ i \delta \chi k_{\hat n}}
 \mathcal P \big( k_{\hat n} \vhat n, \vec{k}_\perp, \bar \chi, \delta \chi \big) \, .
\end{align}
We can rewrite the 2D delta function in Equation~\eqref{eq:ellbra} as $e^{\daleth  \bm{\mathcal{L}} \cdot \partial_{\vec \ell'}} \df^{\rm 2D} \big( \vec{\ell} + \vec{\ell}' \big)
= \df^{\rm 2D} \big( \vec{\ell} + \vec{\ell}' \big) + 
\lb e^{\daleth \bm{\mathcal{L}} \cdot  \overset{\rightarrow}{\partial}_{\vec \ell'}} - 1 \rb \df^{\rm 2D} \big( \vec{\ell} + \vec{\ell}' \big)$, with $\bm{\mathcal{L}} = (\vec{\ell}' + \vec{\ell})$. At this point we can start defining the expansion series for the angular spectrum. The expressions above show how the presence of an observer breaks translational invariance, as discussed in Section~\ref{sec:}. In the limit where the observer can be considered a meta-observer, the 3D power spectrum in \HS is recovered; mathematically, in Equation~\eqref{eq:2Ddelta} that means that the $\delta$ function does not depend on the comoving distances. In this limit, the distance between the observer and the observed system is much larger than the radial distance between the correlated sources, and we therefore expand over this, around the equal time case -- which preserves translational invariance, and would recover the standard flat-sky Kaiser formula. We can therefore write the correlator as:
\begin{align}
\label{eq:correxp}
\la \hat \df(\vec \ell) \hat \df(\vec \ell') \ra
&= (2\pi)^2
\df^{\rm 2D} \big( \vec{\ell} + \vec{\ell}' \big)
\int \frac{d\chi d \delta \chi}{  \chi^2} ~ 
W\lb \chi + \tfrac{1}{2} \delta \chi \rb W'\lb \chi - \tfrac{1}{2} \delta \chi \rb
\int \frac{d k_{\hat n}}{2\pi} ~ e^{ i \delta \chi k_{\hat n}}
\mathcal P \big( k_{\hat n} {\bf \hat n}, \vec{\tilde \ell}, \chi, \delta \chi \big) \, + \nonumber \\
&+ 
(2\pi)^2
\df^{\rm 2D} \big( \vec{\ell} + \vec{\ell}' \big)
\int \frac{d\chi d \delta \chi}{  \chi^2} ~ 
W\lb \chi + \tfrac{1}{2} \delta \chi \rb W'\lb \chi - \tfrac{1}{2} \delta \chi \rb
\lb e^{\delta \bm{\mathcal{L}} \cdot \overset{\leftarrow}{\partial}_{\vec \ell'}} - 1 \rb
\int \frac{d k_{\hat n}}{2\pi} ~ e^{ i \delta \chi k_{\hat n}}
\mathcal P \big( k_{\hat n} {\bf \hat n}, \vec{\tilde \ell}, \chi, \delta \chi \big) \non\\
&= (2\pi)^2
\df^{\rm 2D} \big( \vec{\ell} + \vec{\ell}' \big) \sum_{n=0}^\infty 
\frac{ ( \overset{\leftarrow}{\partial}_{\vec \ell'} \cdot \bm{\mathcal{L}} )^n }{2^n n!} C^{(n)}(\ell) \; ,
\end{align}
where we introduced:
\begin{equation}
C^{(n)}(\ell) = \int \frac{d\chi d \delta \chi}{  \chi^2} \lb \frac{\delta \chi}{\chi} \rb^n
W\lb \chi + \tfrac{1}{2} \delta \chi \rb W'\lb \chi - \tfrac{1}{2} \delta \chi \rb
\int \frac{d k_{\hat n}}{2\pi} ~ e^{ i \delta \chi k_{\hat n}}
\mathcal P \big( k_{\hat n} {\bf \hat n}, \vec{\tilde \ell}, \chi, \delta \chi \big) \, .
\end{equation}
Here we can see the familiar structure of flat-sky angular correlations, but with the addition of radial modes contributions in a series expansion form. The last line of Equation~\eqref{eq:correxp} is producing derivatives of the delta function, giving rise to off-diagonal contribution to the correlations. We will see more details and consequences of this in~\cite{RVII}.

In order to understand the relevance of unequal time contributions and the higher order terms in the series, we can investigate some simplified cases.
Assuming gaussian window functions, we can write:
\eeq{
W(\chi) = \frac{1}{\sqrt{2\pi} \sigma} e^{- \frac{(\chi - \chi_*)^2}{2 \sigma^2}} \, ,
}
so that we have:
\begin{equation}
\label{eq:Cnell}
C^{(n)_m}(\ell) = \frac{1}{2\pi \sigma^2}
\int \frac{d\chi }{  \chi^{n+2}} ~ D(\chi)^2
 e^{- \frac{(\chi - \chi_*)^2}{\sigma^2}}
\int \frac{d k_{\hat n}}{2\pi} ~ G^{(n,m)} \lb  k_{\hat n},\chi \rb
\mathcal P \lb k_{\hat n} \vhat n, \vec{\ell}/\chi \rb \, ,
\end{equation}
where:
\begin{equation}
\label{eq:Gn}
G^{(n,m)} \lb  k_{\hat n},\chi \rb
= \int d \delta \chi~ \delta \chi^n \left[ 1 + \sum_{m=1}^\infty c_m (\delta \chi)^m \right]
 e^{ i \delta \chi k_{\hat n} - \frac{ \delta \chi^2}{4 \sigma^2} } \; .
\end{equation}
We want to stress here the difference between the $n$ and $m$ indices. While the $n$ indices come from the off-diagonal part of the delta function, the $m$ indices come from the unequalness in time. These corrections ultimately come (the $n$ ones) from the fact that we are taking the ensemble average power spectrum in \HS and forcing it into the \OS, and (the $m$ ones) from the unequalness in time intrinsically inside radial modes.
While the expression in Equation~\eqref{eq:Gn} is not fully general, but specific of the choice of the window functions and model, the structure of the expansion is fully general.

The $c_m$ are coefficients for the expansion around the equal time case; in order to calculate them, we start by expanding the growth factor term around $\delta\chi=0$. We can write it as:
\begin{equation}
D(z[\chi_i(\chi,\df \chi)]) = D(z[\chi_i(\chi,0)]) 
+\frac{d}{d \df \chi} D(z[\chi_i(\chi,\df \chi)]) \bigg|_{\df \chi=0} \df \chi 
+\frac{1}{2} \frac{d^2}{d^2 \df \chi} D(z[\chi_i(\chi,\df \chi)]) \bigg|_{\df \chi=0} (\df \chi)^2 + \dots, 
\end{equation}
where $i\in\{1,2\}$ are indices of the sources, and we write:
\begin{equation}
\frac{d}{d \df \chi} D(z[\chi_i(\chi,\df \chi)])
= \frac{d \chi_i}{d \delta \chi}  \frac{dz}{d \chi_i} \frac{d}{dz} D(z[\chi_i(\chi,\df \chi)]) 
= F_a D H \frac{d \chi_i}{d \delta \chi} \; ,
\end{equation}
having introduced the factor

$F_a = - f /(1+z)$. 
Here we calculate this expansion up to second order; for the second derivative we have:
\eq{
\frac{d^2}{d^2 \df \chi} D(z[\chi_i(\chi,\df \chi)])
&= \frac{d}{d \df \chi} \lb F_a D H \frac{d \chi_i}{d \delta \chi} \rb 
= \bigg[ \frac{d \chi_i}{d \delta \chi}  \frac{dz}{d \chi_i} \frac{d}{dz}  \lb F_a D H \rb \bigg] \frac{d \chi_i}{d \delta \chi} + F_a D H \frac{d^2 \chi_i}{d \delta \chi^2} \non\\
&= H \lb  \frac{d}{dz} F_a D H \rb \lb \frac{d \chi_i}{d \delta \chi} \rb^2 + F_a D H \frac{d^2 \chi_i}{d \delta \chi^2}
= F_a F_b D H^2 \lb \frac{d \chi_i}{d \delta \chi} \rb^2 + F_a D H \frac{d^2 \chi_i}{d \delta \chi^2} \; ,
}
introducing the factor $F_b$:
\eq{
F_b = \frac{1}{F_a D H} \lb  \frac{d}{dz} F_a D H  \rb
= \frac{1}{F_a} \frac{d}{dz} F_a + \frac{d}{dz} \ln H + \frac{d}{dz} \ln D = - \frac{1}{1+z} + \frac{d}{dz} \ln (f H D) .
}

The Taylor expansion of the growth rate thus gives us:
\eeq{
\frac{D(z[\chi_i(\chi,\df \chi)])}{D(z[\chi_i(\chi,0)])}
=1 
+ F_a H \lb \frac{d \chi_i}{d \delta \chi} \rb_{\df \chi=0} \df \chi 
+\frac{1}{2} \left[ F_a F_b H^2 \lb \frac{d \chi_i}{d \delta \chi} \rb^2 
+ F_a H \frac{d^2 \chi_i}{d \delta \chi^2} \right]_{\df \chi=0} (\df \chi)^2 + \dots, \quad \; .
}

Here we will use the arithmetic average distance definition, $\chi = \frac{1}{2} \lb \chi_1 + \chi_2 \rb$,
which means $\chi_1 = \chi + \frac{1}{2} \delta \chi$, $\chi_2 = \chi - \frac{1}{2} \delta \chi$
and $d\chi_1/d \delta \chi = 1/2$, $d\chi_2/d \delta \chi = -1/2$.
All of this gives us:
\eq{
\frac{D(z[\chi_i(\chi,\df \chi)])}{D(z[\chi])}
&= 1 \pm \frac{1}{2} F_a H \df \chi 
+\frac{1}{8} F_a F_b H^2 (\df \chi)^2 + \dots \; ,
}
and the product for the correlation of two sources becomes, up to second order:
\begin{equation}
\label{eq:Dexp2}
\frac{D(z[\chi_1(\chi,\df \chi)])D(z[\chi_2(\chi,\df \chi)])}{D(z[\chi])^2}
\approx 1 + \frac{1}{4} F_a (F_b - F_a) H^2 (\df \chi)^2  \; .
\end{equation}
It is worth noting here that the first order correction due to unequal time correlations vanishes and we are left with only contributions of the order $\delta\chi^2$.
Equation~\eqref{eq:Dexp2} is therefore the expression for the corrections due to unequal time correlations, up to second order, for the dark matter case in real space.

\subsection{Biased tracers}
We now turn to the more realistic case of biased tracer, considering the general case where the tracers are not necessarily identical.
Using the same procedure as above, we now have, for the first order term:
\begin{equation}
\frac{b_X(\chi_i)D(z[\chi_i(\chi,\df \chi)])}{b_X(\chi)D(z[\chi])}
= 1 \pm \frac{1}{2} \left[ F_{a} + \frac{b_X'(\chi)}{b_X} \right] H \df \chi 
+ \dots ,
\end{equation}
So that now we have, for the correlation of two galaxy populations:
\begin{equation}
\label{eq:expbias1}
\frac{b_A(\chi_1)b_B(\chi_2)D(z[\chi_1(\chi,\df \chi)])D(z[\chi_2(\chi,\df \chi)])}{b_A(\chi)b_B(\chi)D(z[\chi])^2}
= 1 + \frac{1}{2} H \df \chi \left(\frac{b_A'}{b_A}-\frac{b_B'}{b_B} \right) %+ \frac{1}{8} (\df \chi)^2 (D+B-2AC) 
+  \dots \; ,
\end{equation}
and therefore we see that in this case we have a first order contribution.
This is a very interesting result: when performing a multi-tracer analysis, the differences in the bias and its evolution for different tracers cause a first order unequal time correction. We investigate this more in~\cite{RVII} and, more specifically for future surveys, in a paper in preparation.

\subsubsection{Redshift space distortions}
When we consider correlations in redshift-space, we have to add the effects of peculiar velocities. Here we limit ourselves to using the Kaiser term only~\cite{Kaiser:1987}; We can write the power spectrum in redshift space $P^s$ as (see e.g.,~\cite{Hamilton:1997}):
\begin{align}
    P^s(k,z,\mu) &= P_{\rm lin}(k) D_A(z) D_B(z) \left[b_A(z)+f(z)\mu^2 \right]\left[b_B(z)+f(z)\mu^2 \right] \, .
\end{align}
The first term is the intrinsic clustering, for which we derived unequal time corrections just above. As for the mixed term, we have:
\begin{align}
\frac{f(z[\chi_i(\chi,\delta\chi)])b_X(\chi_i)}{f(z[\chi])b_X(\chi)}
&= 1 \pm \frac{1}{2} H \delta\chi\left( fb_X'+f'b_X \right)
+ ... \; ,
\end{align}
and the $\mu^4$ term has no first order contributions. This is because there is no bias involved, and the growth rate $f$ is the same when we evaluate things at a common $\bar{\chi}$.
All of this gives us a power spectrum in the form (up to first order):
\begin{align}
\label{eq:UTPkMT1}
    P^s = \Big[P_{\rm lin} D_A D_B \left(b_Ab_B+(b_A+b_B)f\mu^2 +f^2\mu^4 \right) \Big]_{\bar \chi} + \frac{1}{2} H \delta \chi \left\{ \left(\frac{b_A'}{b_A}-\frac{b_B'}{b_B} \right) + \mu^2 \big[\left( fb_A'+f'b_A \right)-\left( fb_B'+f'b_B \right)\big] \right\} \; .
\end{align}
The terms multiplying $(\delta\chi)^m$ are the $c_m$ coefficients of our expansion~\eqref{eq:Gn}.
Note that in obtaining Equation~\eqref{eq:UTPkMT1} we are not considering relativistic and projection effects, which could bring important corrections; a further analysis including these terms will be presented in the future.

\subsection{Expansion for toy power spectrum}
\label{sec:exptoy}
\begin{figure*}[t!]
\includegraphics[width=0.47\textwidth]{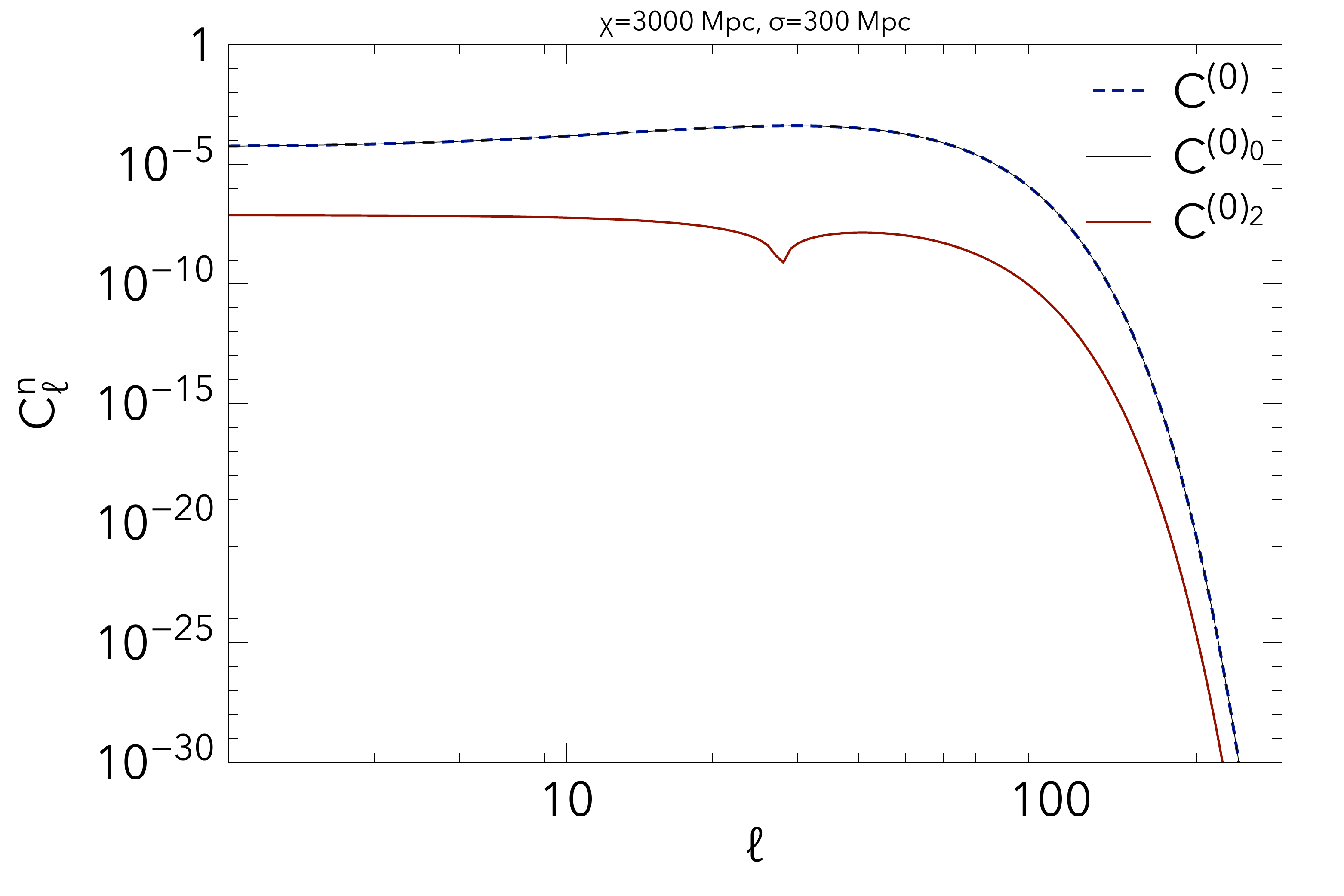}
\includegraphics[width=0.47\textwidth]{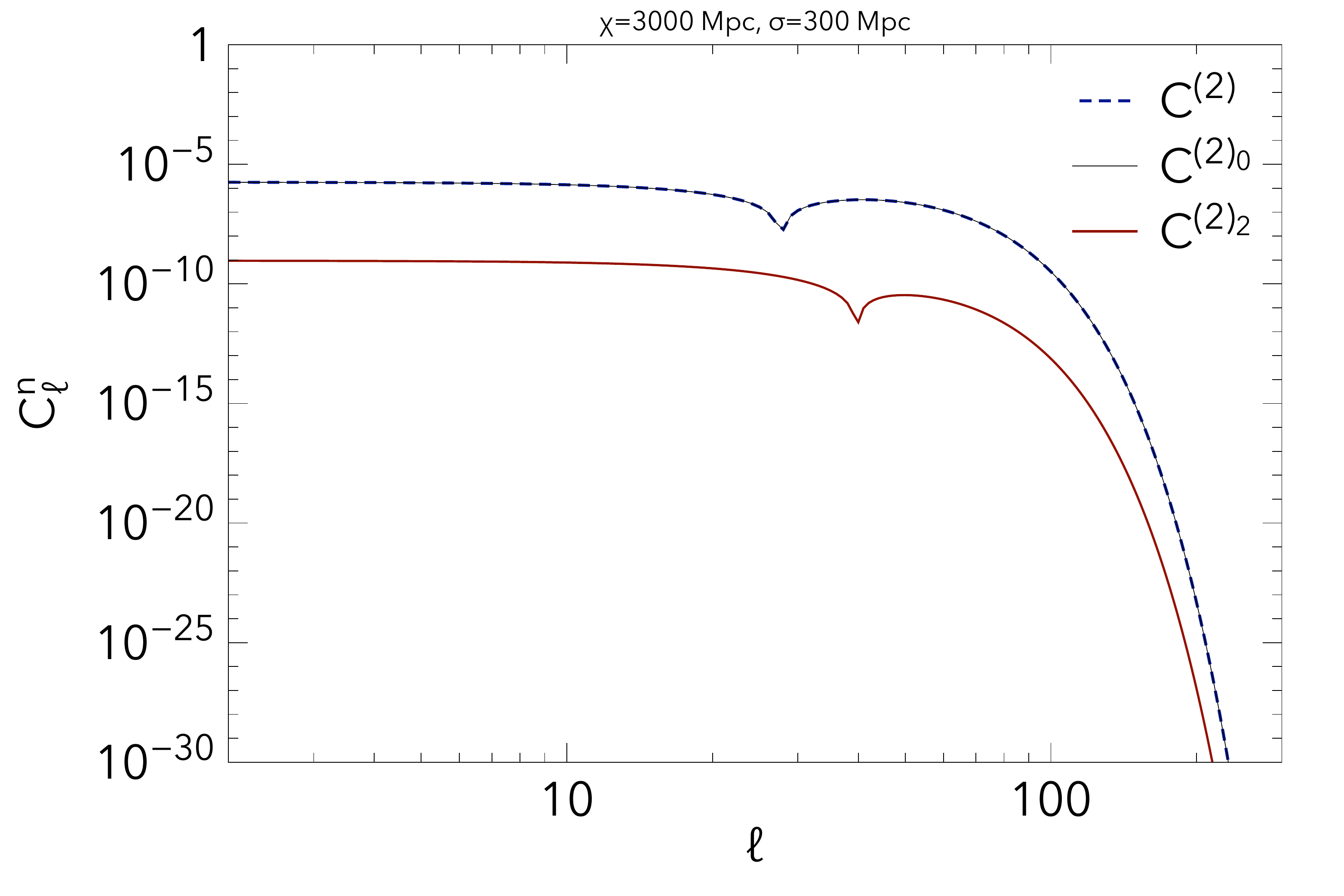}
\includegraphics[width=0.47\textwidth]{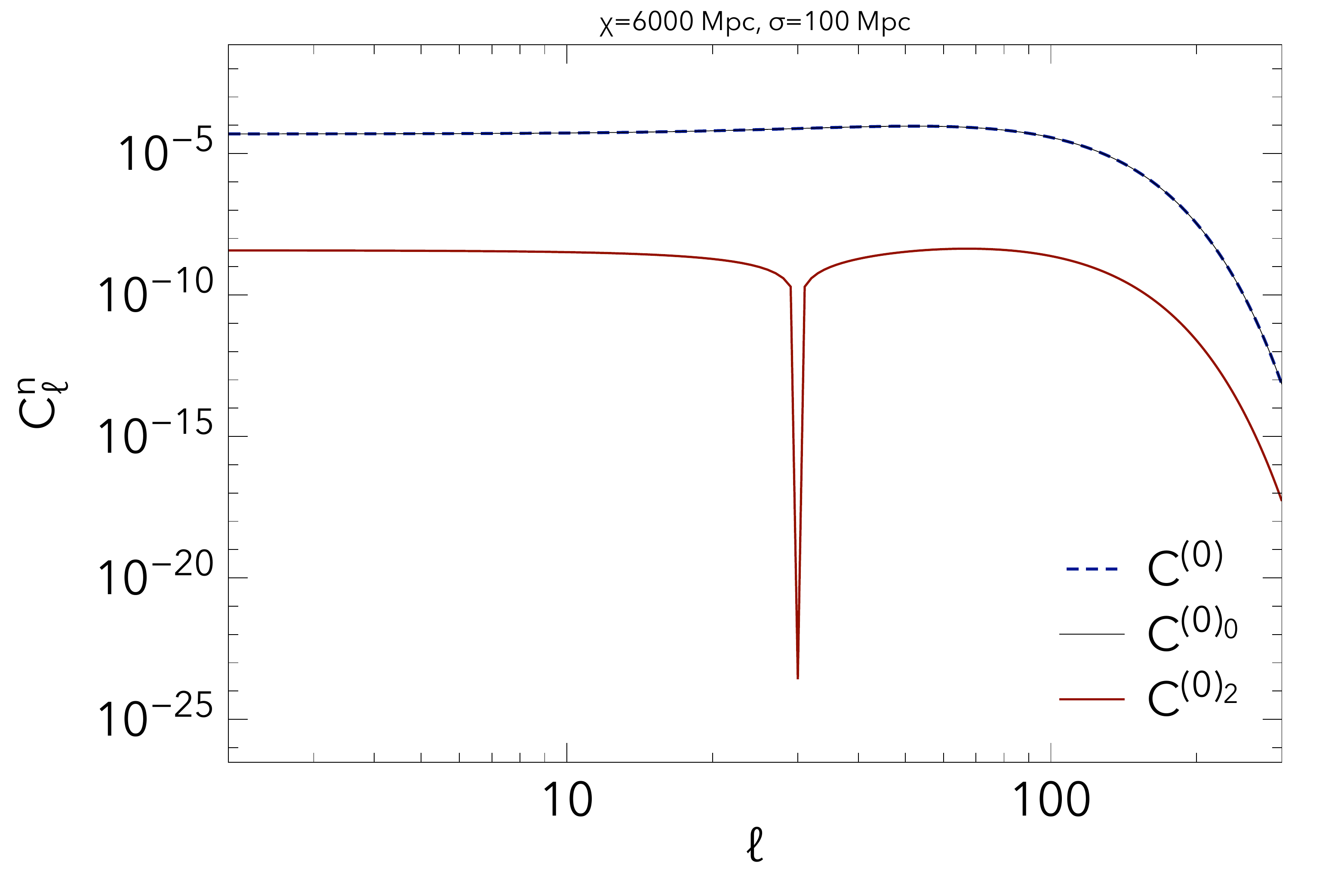}
\includegraphics[width=0.47\textwidth]{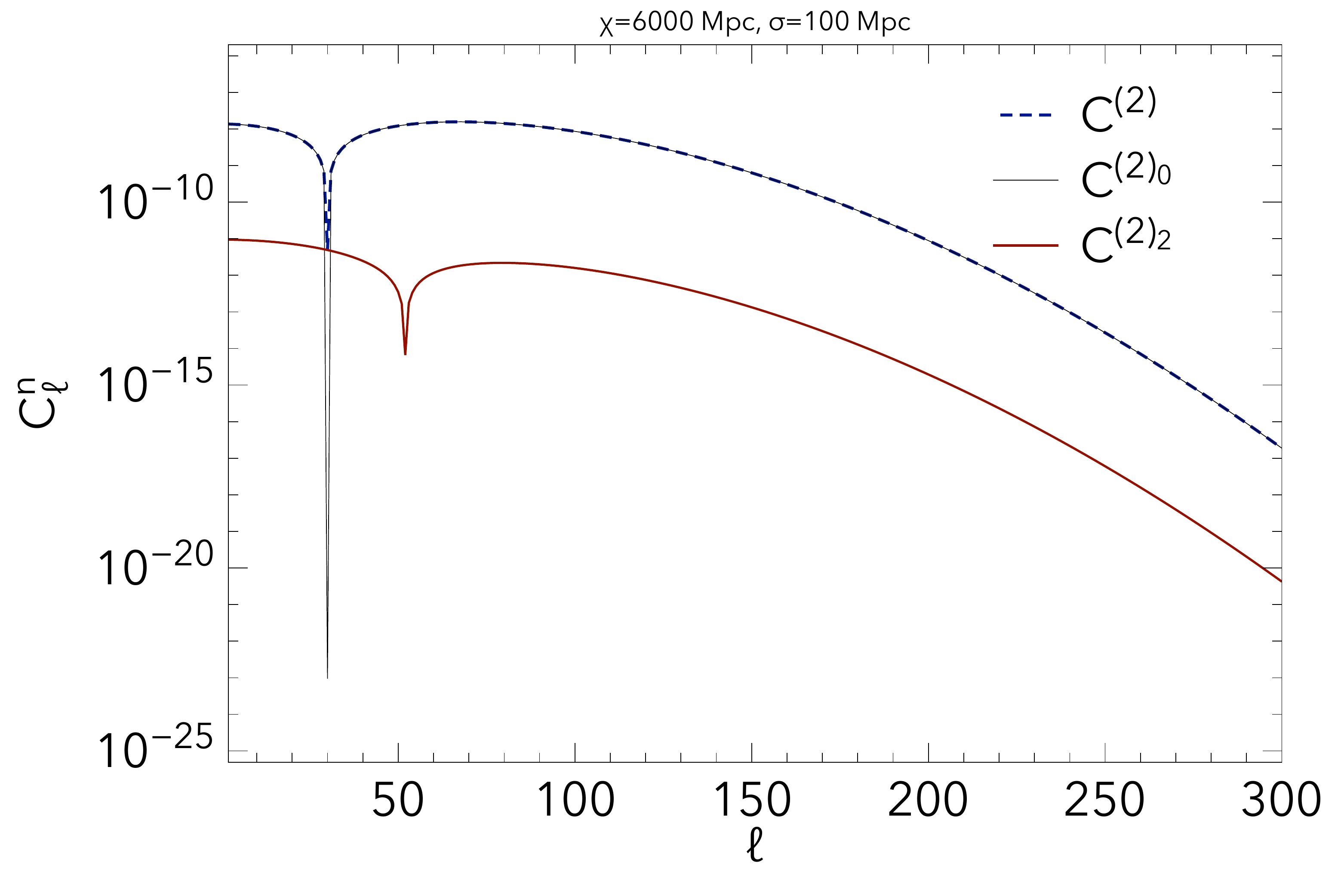}
\includegraphics[width=0.47\textwidth]{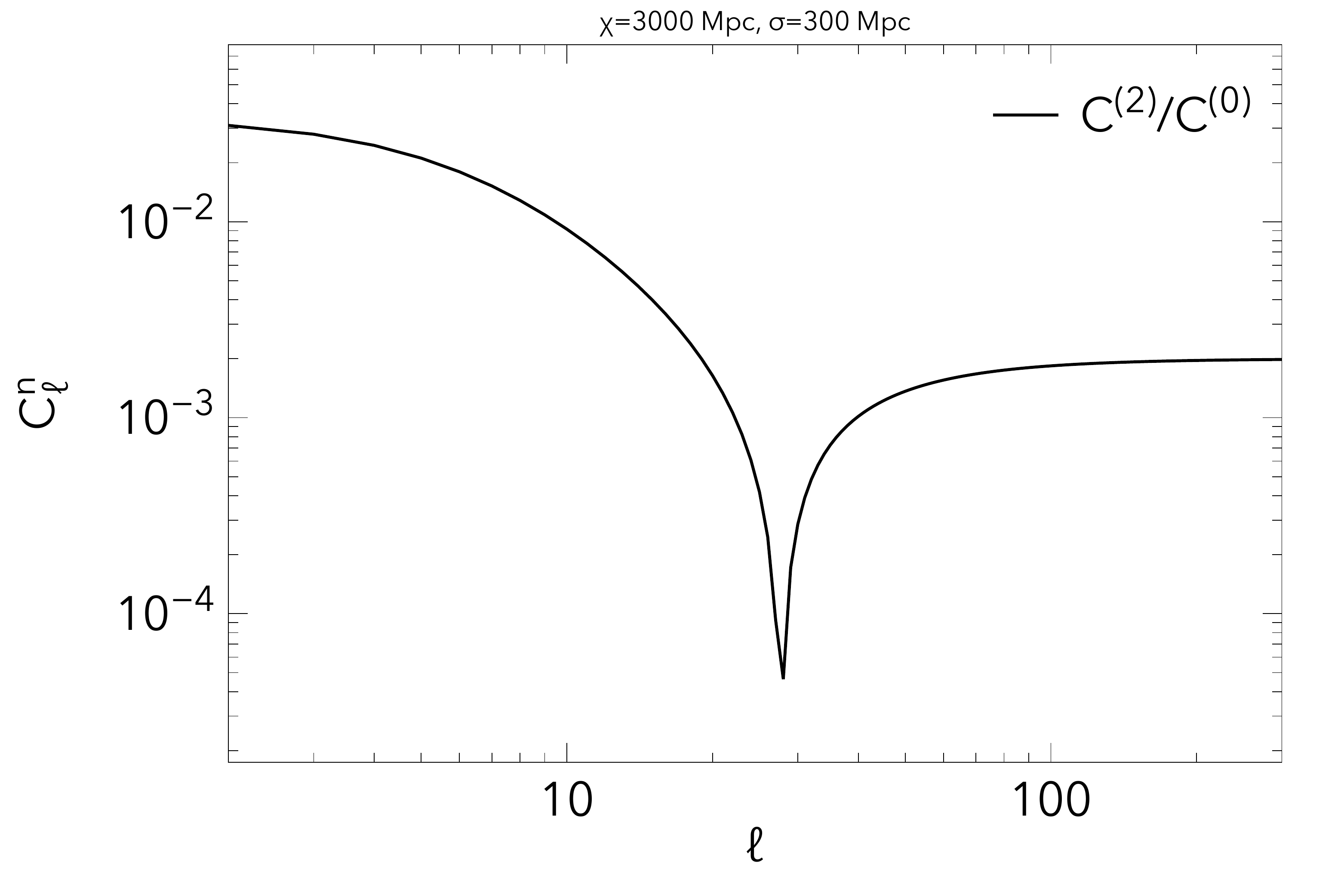}
\includegraphics[width=0.47\textwidth]{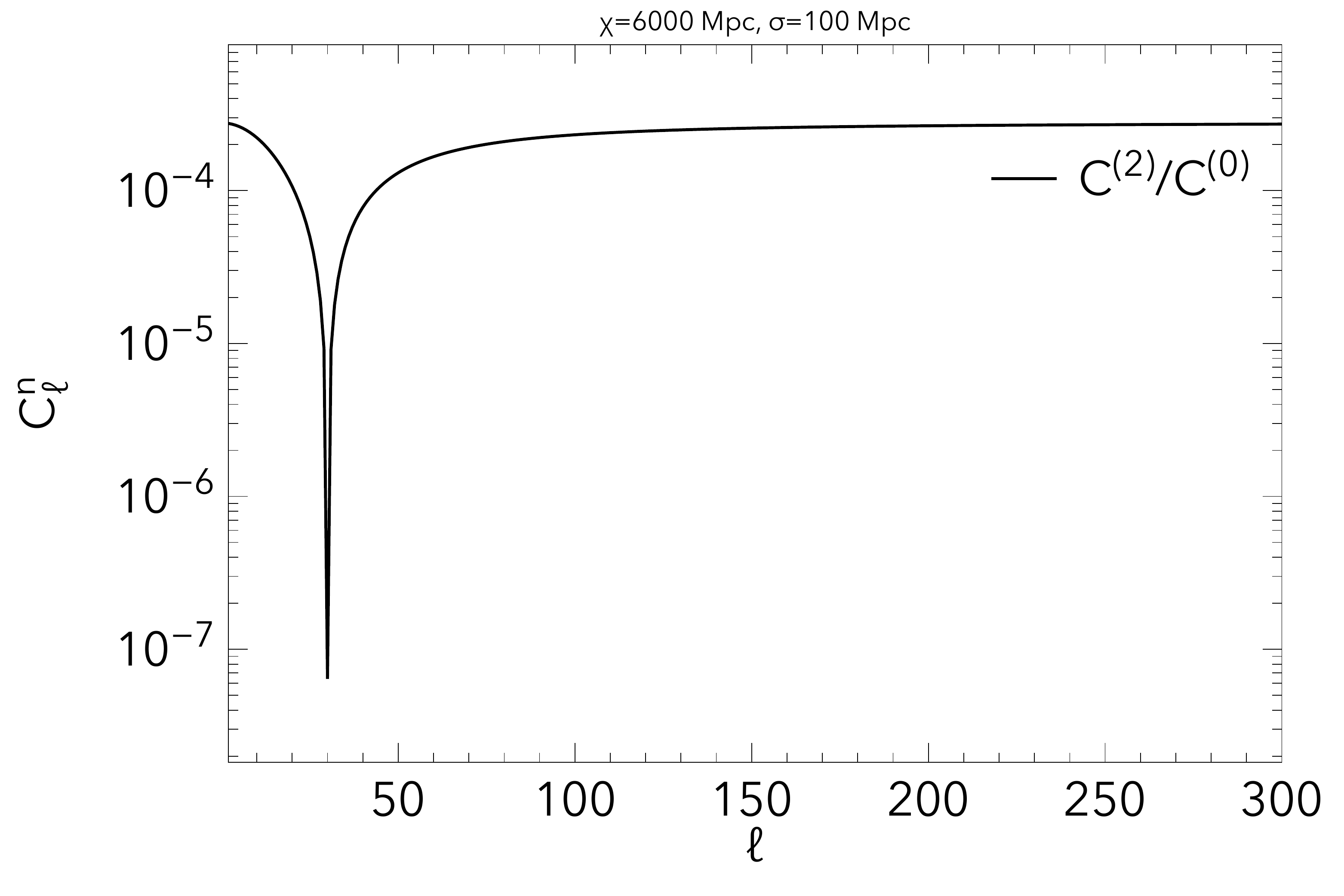}
\caption{Leading and second order contributions for the $C^{(n)_m}_\ell$ spectra, for two different cases of the average comoving distance and radial separation (plotted are absolute values). The top two rows show the impact of $m=2$ terms for the $n=0$ (top panel) and $n=2$ (middle panel) correlations. The bottom row shows the importance of the sky flattening ($n$ higher order terms). We can see how the $m=2$ terms are always negligible, while the $n=2$ contribution can be of the order of a few percent at very low-$\ell$ and lower redshift, and smaller otherwise.}
\label{fig:Cnmell}
\end{figure*}

At this point we can go back to Equation~\eqref{eq:Cnell} and calculate the behavior and importance of unequal time corrections. Here we start by introducing a toy power spectrum, that has a form chosen to allow us to perform analytical calculations.
We do so in order to be able to control all the steps, check that our procedure gives the expected results, and understand the various dependencies of new terms.
If we assume a power spectrum of the type:
\eeq{
P(k) = A \frac{k^2}{k^2_{\rm eq}} e^{-k^2/k^2_{\rm eq}} \;\; ,
}
we can explicitly calculate the expansion terms.
Here we report the results for the first few terms; stopping at $m=2$, we have:
\begin{align}
\label{eq:Gns}
%%% G0
G^{(0)} \lb  k_{\hat n},\chi \rb
&= 2 \sqrt{\pi} \sigma \Big[ 1 + c_1 i k_{\hat n} \sigma + c_2 \lb 1 - 2 k_{\hat n}^2 \sigma^2 \rb \sigma^2 \Big] e^{- k_{\hat n}^2 \sigma^2} \, ; \\
%%% G1
G^{(1)} \lb  k_{\hat n},\chi \rb
&= 4 \sqrt{\pi} \sigma^2 \left\{
i k_{\hat n} \sigma \Big[ 1 +c_2 \left(3-2 k_{\hat n}^2 \sigma^2 \right) 2 \sigma^2 \Big]
+\left[c_1\left( 1-2 k_{\hat n}^2 \sigma^2 \right) \sigma \right]
\right\}
e^{- k_{\hat n}^2 \sigma^2} \, ; \non\\
%%% G2
G^{(2)} \lb  k_{\hat n},\chi \rb
&= 4 \sqrt{\pi} \sigma^3 \Big[ 1 - 2 k_{\hat n}^2 \sigma^2 +2 c_1 i k_{\hat n} \sigma \left(3-2 k_{\hat n}^2 \sigma^2 \right) \sigma + 2 c_2 \left(3 - 12 k_{\hat n}^2 \sigma^2 + 4 k_{\hat n}^4 \sigma^4 \right) \sigma^2 \Big] e^{- k_{\hat n}^2 \sigma^2} \; .
\end{align}
Inserting these coefficients into the observed power spectrum, we have:
\begin{align}
\label{eq:Gint}
% G0
C^{(0)_m}_\ell(\chi,\sigma) = \int \frac{d k_{\hat n}}{2\pi} ~ G^{(0)} \lb  k_{\hat n},\chi \rb
\mathcal P \lb k_{\hat n} \vhat n, \frac{\vec{\ell}}{\chi} \rb
& = A
\frac{\sigma k_{\rm eq}}{2 \Xi^5} 
\Big\{
\Xi^2 \lb 1 + 2 \Xi^2 X^2 \rb
+2 c_2
\left[3 - 2\Xi^2 \lb 1 -X^2 \rb \right] \sigma^2 \Big\} \, e^{-X^2} \, ; \\ 
% G1
C^{(1)_m}_\ell(\chi,\sigma) = \int \frac{d k_{\hat n}}{2\pi} ~ G^{(1)} \lb  k_{\hat n},\chi \rb
\mathcal P \lb k_{\hat n} \vhat n, \frac{\vec{\ell}}{\chi} \rb
& = A \frac{\sigma k_{\rm eq}\sigma^2}{\Xi^5} c_1 \left[3+2\Xi^2(X^2-1)  \right] \, e^{-X^2} \, ;\\
% G2
C^{(2)_m}_\ell(\chi,\sigma) = \int \frac{d k_{\hat n}}{2\pi} ~ G^{(2)} \lb  k_{\hat n},\chi \rb
\mathcal P \lb k_{\hat n} \vhat n, \frac{\vec{\ell}}{\chi} \rb
& = A
\frac{ (\sigma k_{\rm eq}) \sigma^2}{2 \Xi^7} 
\Big\{
2\Xi^2 \left[ 3 - 2 \Xi^2 (1-X^2) \right]
+12 c_2
\left[ 5 - 2 \Xi^2 \lb 2 - X^2 \rb \right] \sigma^2 \Big\} \, e^{-X^2} \, .
\end{align}
Here $\Xi^2 = 1 + k^2_{\rm eq} \sigma^2$ and $X= \frac{\ell}{ k_{\rm eq} \chi}$.
In Figure~\ref{fig:Cnmell} we show the leading and second order contributions for two cases of average distance and width of the bin. We can see that the terms are well behaved and, in our formulation, the leading order in both $\{n,m\}$ should be very close to the full solution.
It is evident that the $m=2$ terms are always a few orders of magnitude smaller than the $m=0$ ones, and the $n=2$ contribution is at maximum of a few percent.
We will verify this when including window functions in the following Section, where we also compare with brute force calculations and the Limber approximation.

We can therefore now have our expression for the angular power spectrum including unequal time corrections.
In the next Section we will show the remarkable and in some aspect unexpected result that the inclusion of unequal time corrections allows us to use a flat-sky approximation that matches the exact results much better than currently employed approximations even at low multipoles $\ell$.

%%%%%%%%%%%%%%%%%%%%%%%%%%%%%%%%%%%%%%%%%%%%%%%%%
%%%%%%%%%%%%%%%%%%%%%%%%%%%%%%%%%%%%%%%%%%%%%%%%%
%%%%%%%%%%%%%%%%%%%%%%%%%%%%%%%%%%%%%%%%%%%%%%%%%
%%%%%%%%%%%%%%%%%%%%%%%%%%%%%%%%%%%%%%%%%%%%%%%%%
%%%%%%%%%%%%%%%%%%%%%%%%%%%%%%%%%%%%%%%%%%%%%%%%%

\section{Angular power spectrum from full- to flat- sky} 
\label{sec:angular_ps}
In this Section we illustrate a new expression for the calculation of $C_\ell$s in a way that captures all the correct behaviors of the full sky expression, but assuming a flat-sky approach.
In order to verify that our calculations are correct, we again start with an example of a toy power spectrum for which we can compute analytical results.
While we confirm here with analytical calculations that our procedure is correct (and independent on the details of the toy model), results for the $\Lambda$CDM power spectrum will be presented in a numerical work.
Let us assume that our power spectrum is in a form similar to the one of the previous Section: 
\begin{equation}
\label{eq:pktoy}
P(k, z_1, z_2) = A D(z_1)D(z_2) k^2 e^{- \aleph^2 k^2}
= - \frac{A}{\aleph^2} D(z_1)D(z_2) \partial_\epsilon e^{- \epsilon \aleph^2 k^2} \Big |_{\epsilon=1}  \, ,
\end{equation}
where A is an amplitude and $\aleph$ is a parameter that controls the slope of the power spectrum on small scales. Using a power spectrum in this form allows us to obtain analytical results, and therefore it is very instructive as a starting point and a way to keep under control everything that is happening. In Appendix~\ref{app:toyPK} we show an example of the behavior of this type of spectrum, and in Figure~\ref{fig:toypowerspectrum} we show a comparison with the $\Lambda$CDM power spectrum for a choice of the parameters in Equation~\eqref{eq:pktoy}. All our results are independent on the choice of such parameters, but we chose a toy model that had a similar behavior of the $\Lambda$CDM power spectrum.

From this power spectrum, we can calculate the angular spectrum as:
\begin{align}
\label{eq:cls38}
 \frac{\mathcal C_s( \ell, \chi, \df \chi)}{\Omega( \chi, \df \chi)} 
 &= A \, D(z[\chi_1(\chi,\df \chi)]) D(z[\chi_2(\chi,\df \chi)]) \lb - \frac{1}{\aleph^2} \partial_{\epsilon} \rb e^{ - \epsilon \aleph^2 (\ell/\chi)^2}
 \int \frac{d k_{\hat n}}{2\pi} ~e^{- i\delta \chi  k_{\hat n} - \epsilon \aleph^2 k^2_{\hat n} }  \Big |_{\epsilon=1} \non\\
 &= A \, D(z[\chi_1(\chi,\df \chi)]) D(z[\chi_2(\chi,\df \chi)]) \lb - \frac{1}{\aleph^2} \partial_{\epsilon} \rb e^{ - \epsilon \aleph^2 (\ell/\chi)^2}
 \frac{1}{2 \sqrt{\pi \epsilon }\aleph} e^{ - \frac{(\delta \chi)^2}{4 \epsilon \aleph^2}} \Big |_{\epsilon=1} \non\\
 &= D(z[\chi_1(\chi,\df \chi)]) D(z[\chi_2(\chi,\df \chi)]) \frac{A}{2 \sqrt{\pi} \aleph^3}
 \left[ \frac{1}{2} + \aleph^2 \left(\frac{\ell}{\chi}\right)^2 - \frac{\df \chi^2}{4 \aleph^2} \right] e^{ - \aleph^2 (\ell/\chi)^2 - \frac{\df \chi^2}{4\aleph^2}} \; ,
\end{align}
where $\Omega$ is a global window function, $\Omega(\chi, \df \chi) = \Omega(\chi_1, \chi_2) = W(\chi_1)W(\chi_2)$.
This expression allows us to identify what parameters control the behavior of the angular power spectrum, and what is the influence on radial modes.
We can see that the equivalence scale is entering in various places, controlling the amplitude but also the exponential cutoff.
The exponential cutoff reduces the amplitude of the correlation, given a distance $\chi$, for larger $\ell$ and (very rapidly) when the radial distance increases. For larger comoving distances, the cutoff moves to larger multipoles and radial separations, making the unequal time effects more important for future, high-redshift, surveys.
In Appendix~\ref{app:dep} we show some plots illustrative of these dependencies from of Equation~\eqref{eq:cls38} and its derivative with respect to $\delta\chi$ to better investigate such behaviors and dependencies.

\begin{figure*}[t!]
\includegraphics[width=0.47\textwidth]{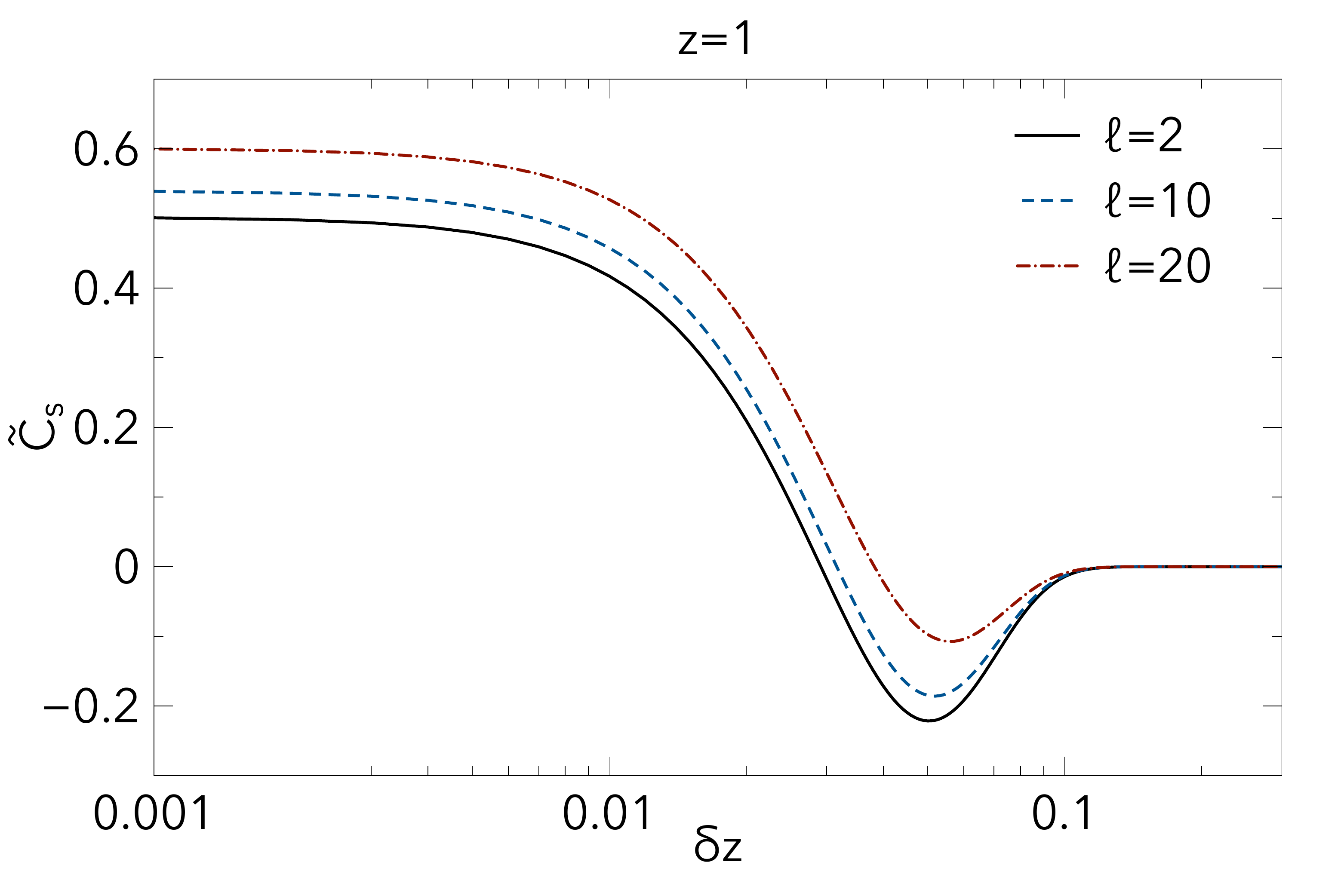}
\includegraphics[width=0.47\textwidth]{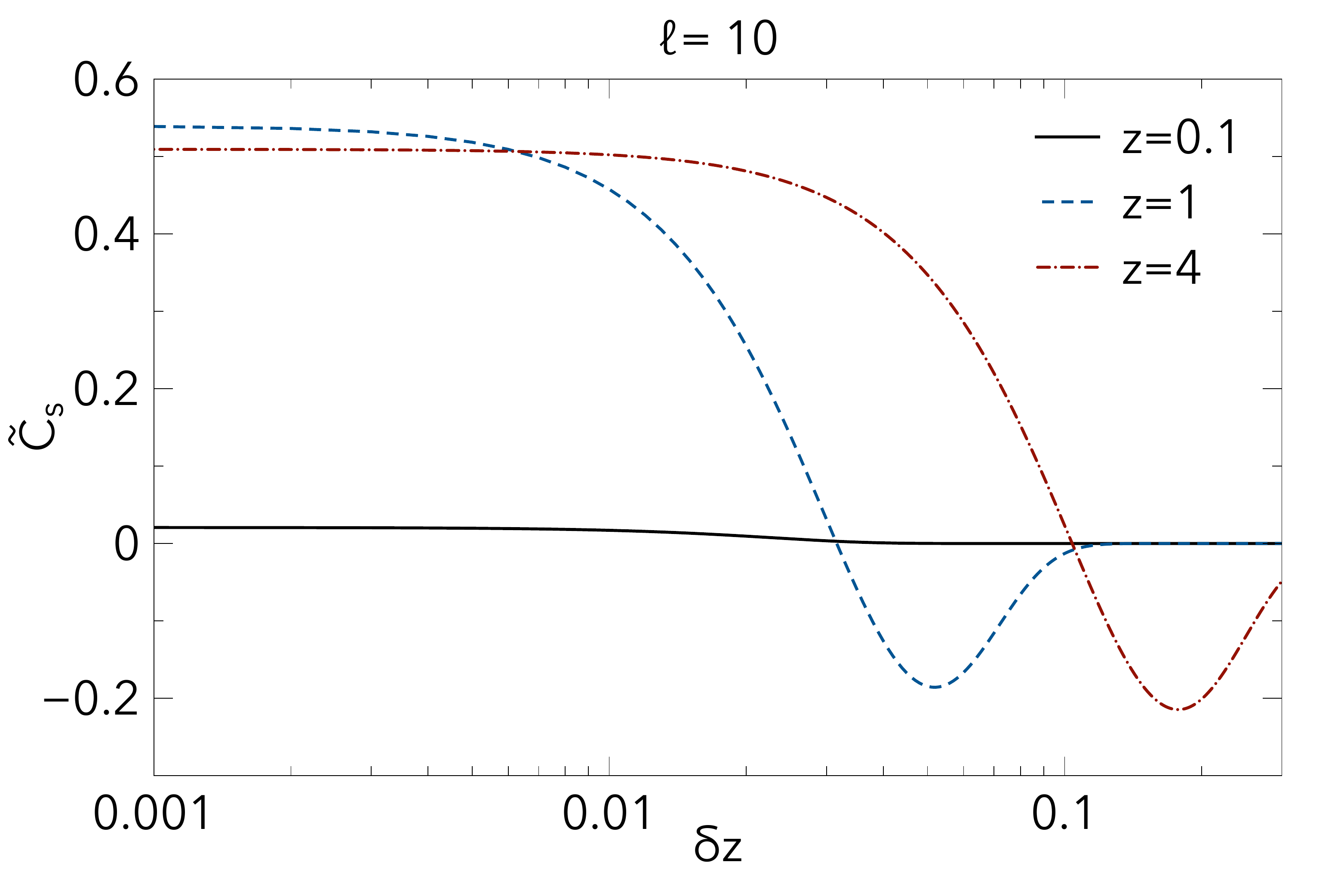}
\includegraphics[width=0.47\textwidth]{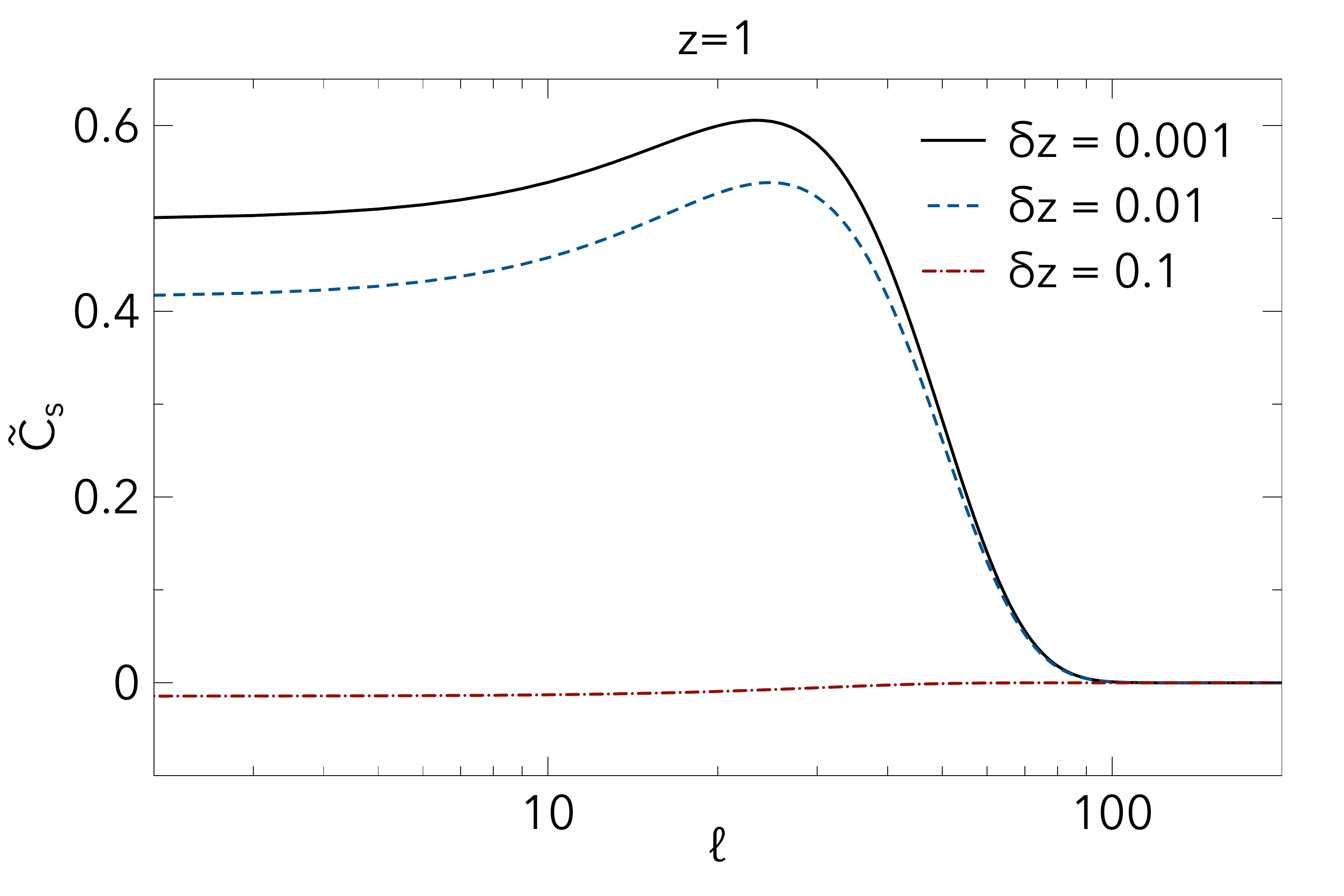}
\includegraphics[width=0.47\textwidth]{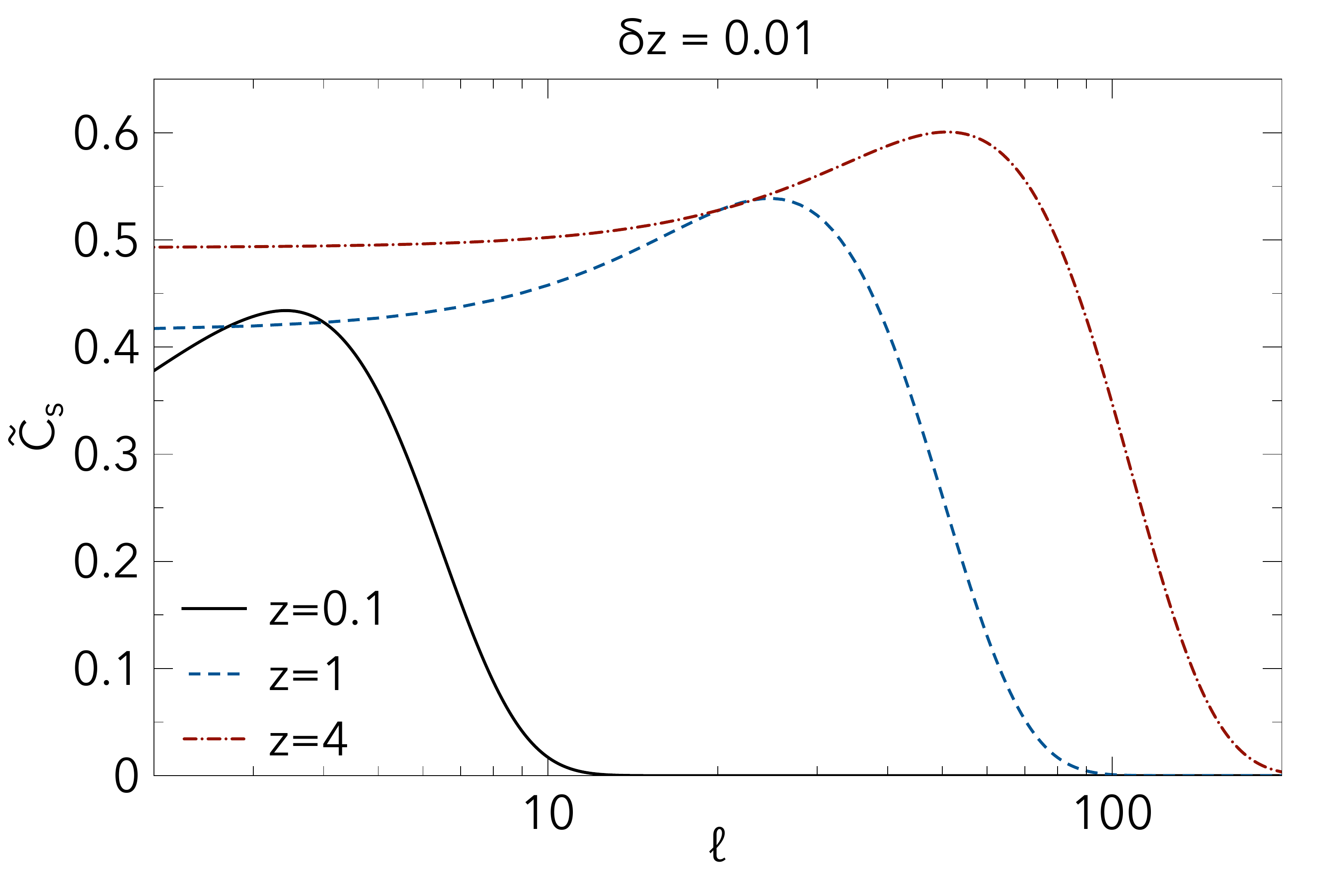}
\caption{$\tilde{C}_s$ as a function of the unequalness in time $\delta z$ (upper panels), for different combinations of average distances and angular multipoles.
Lower panels show $\tilde{C}_s$ as a function of $\ell$, for different combinations of average distances and unequalness in time.}
\label{fig:Cls39}
\end{figure*}

Introducing the variables $\Lambda = \aleph \ell/\chi$ and $\Delta = \frac{\df \chi}{\aleph}$, we can write:
\begin{equation}
\label{eq:CsLD}
 \tilde{C}_s = 2 \sqrt{\pi}  \frac{\aleph^3 \mathcal C_s( \ell, \chi, \df \chi)}{ A \, D_1 D_2 \Omega( \chi, \df \chi)} 
=
 \lb \frac{1}{2} + \Lambda^2 - \frac{1}{4}\Delta^2 \rb e^{ - \lb \Lambda^2 + \frac{1}{4} \Delta^2 \rb} \, ,
\end{equation}
where we indicate $D_i=D(z[\chi_i(\chi,\df \chi)])$ for compactness of notation.
If we consider Equation~\eqref{eq:CsLD}, we can easily separate where the unequal time effects enter in modifying the amplitude of the angular correlations. In Figure~\ref{fig:Cls39} we plot $\tilde{C}_s$ for different dependencies and values of the average comoving distance, the unequalness in time and angular multipoles.
We can directly see here how the unequal time corrections grow with redshift and are more important for thinner radial bins. We will see later how this will help us fit the full result by including radial modes.

Performing a Fourier transform, we can obtain the power spectrum including radial modes, which now reads:
\eq{
P_s \lb k_{\hat n}, \frac{\ell}{\chi}, \bar z \rb
&= \frac{A}{2 \sqrt{\pi} \aleph^3} e^{-\Lambda^2} 
\int_{-\infty}^{\infty} d( \delta \chi ) ~ D_1 D_2
\lb \frac{1}{2} + \Lambda^2 - \frac{\Delta^2}{4} \rb e^{i\delta \chi  k_{\hat n} - \frac{\Delta^2}{4}} = \frac{A}{2 \sqrt{\pi} \aleph^3} 
\lb \frac{1}{2} + \Lambda^2 + \frac{\partial^2_{k_{\hat n}}}{4 \aleph^2} \rb F(k_{\hat n}, \chi) ~ e^{ - \Lambda^2 },
}
where we have defined the quantity:
\eeq{
F(k_{\hat n}, \chi) = \int_{-\infty}^{\infty} d( \delta \chi ) ~ D_1 D_2 
e^{i\delta \chi  k_{\hat n} - \frac{\Delta^2}{4}} \, .
}
If $D(z[\chi_i(\chi,\df \chi)])$ would not depend on $\df \chi$, we would have:
\eeq{
F(k_{\hat n}, \chi) = 2 \sqrt{\pi} \aleph D(z[\chi])^2 e^{-\aleph^2 k^2_{\hat n}} \; ,
}
and consequently:
\eeq{
P_s \lb k_{\hat n}, \frac{\ell}{\chi}, \bar z \rb = A D(z[\chi])^2 \left[ k^2_{\hat n} + \left(\frac{\ell}{\chi}\right)^2 \right] e^{ - \alpha^2 \big[ k^2_{\hat n} + \left(\frac{\ell}{\chi}\right)^2 \big] } \; ,
}
which corresponds to our initial full power spectrum.

We want to compare our findings with the exact formula, so we write the full result using the toy power spectrum, which reads as:
\begin{align}
C_\ell(\chi_1,\chi_2) 
&= 4\pi \Omega( \chi, \df \chi)\chi_1\chi_2 \int d k \, \frac{k^2}{2\pi^2} ~ j_{\ell} (k\chi_1) j_{\ell} (k\chi_2) P \lb k, z[\chi_1], z[\chi_2] \rb \non\\
&= - \frac{2}{\pi} A \, D(\chi_1) D(\chi_2) \Omega( \chi, \df \chi) \chi_1\chi_2 \, \partial_\kappa \int k^2 d k ~ j_{\ell} (k\chi_1) j_{\ell} (k\chi_2) e^{- \kappa k^2} \Big |_{\kappa = \aleph^2} \, .
\end{align}
To derive an expression for the angular spectrum that includes radial modes and gets rid of the double spherical Bessel integration, we can use the result for the modulated product of spherical Bessel functions related to the modified Bessel function of the first kind:
\eeq{
\int k^2 d k ~ j_{\ell} (k\chi_1) j_{\ell} (k\chi_2) e^{- \kappa k^2} = \frac{\pi}{4\sqrt{\chi_1 \chi_2}} \frac{1}{\kappa} 
e^{- \frac{\chi_1^2+\chi_2^2}{4 \kappa}} I_{\ell+1/2} \lb \frac{\chi_1 \chi_2}{2 \kappa} \rb \, ;
}
substituting we then have:
\begin{equation}
\mathcal C_\ell(\chi_1,\chi_2) 
 =  - \frac{A}{2} D_1 D_2 \Omega( \chi, \df \chi)\sqrt{\chi_1 \chi_2}  \partial_\kappa \left[ \frac{1}{\kappa}
e^{- \frac{\chi_1^2+\chi_2^2}{4 \kappa}} I_{\ell+1/2} \lb \frac{\chi_1 \chi_2}{2 \kappa} \right) \right]_{\kappa = \aleph^2} \, .
\end{equation}
Using the dimensionless variables introduced above, this gives:
\begin{align}
\label{eq:C_full}
2 \sqrt{\pi}  \frac{\aleph^3  \mathcal C_s( \ell)}{ A D_1 D_2 \Omega( \chi, \df \chi)}
&=  - \aleph^3 \sqrt{\pi \chi_1 \chi_2}  \partial_\kappa \left[ \frac{1}{\kappa} 
e^{- \frac{\chi_1^2+\chi_2^2}{4 \kappa}} I_{\ell+1/2} \lb \frac{\chi_1 \chi_2}{2 \kappa} \rb \right]_{\kappa = \aleph^2} \non\\
&=  - \aleph^3 \sqrt{\pi \lb 4 \chi^2 - (\aleph x)^2 \rb} \partial_\kappa \left[ \frac{1}{2 \kappa} 
e^{- \frac{4 \chi^2 + \aleph^2 \Delta^2}{8 \kappa}} 
I_{\frac{1}{2} + \frac{\chi}{\aleph} \Lambda} \lb -\frac{\Delta^2 \aleph^2}{8\kappa} + \frac{\chi^2}{2\kappa} \rb \right]_{\kappa = \aleph^2}.
\end{align}
Equation~\eqref{eq:CsLD} is thus the flat-sky expression for the full-sky, unequal-time, angular spectrum given in Equation~\eqref{eq:C_full}. We derived it analytically for the toy power spectrum, in order to confirm the validity of our approach. One of the advantages of having a power spectrum of this type is given by the fact that we can use this formula to understand various dependencies on angular multipoles, average distance, and unequalness in time.
In a $\Lambda$CDM-like power spectrum, the functions $\{\Lambda, \Delta\}$ play the role of scale ($\Lambda$ is a rescaled version of the $\ell/\chi$ scale present in the standard flat-sky power spectrum) and unequalness in time ($\Delta$ is a rescaled radial distance between the two sources to be correlated). The rescaling factor plays a role similar to $k_{\rm eq}$ for the standard power spectrum.

We can already see that the unequalness in time $\delta\chi$ enters in the exponential cut-off, with the result that long radial modes are decorrelated. This is somewhat known, but it is interesting to see here the origin of this effect.
When changing the rescaling $\aleph$ we notice that for very small and very big values, the power spectrum goes to zero, while with intermediate values, it causes $\mathcal C_s(\ell, \chi, \delta\chi)$ to become negative or positive, depending on its value. It is very instructive to see how this parameter causes sources to become correlated or anti-correlated, the crossing point depending on the multipole and the average redshift.
In Appendix~\ref{app:dep} we show some of these behaviors.

In order to check how well our expression of the unequal time flat sky power spectrum fits the full calculation in a more realistic setup, the last step to be added is the inclusion of the window function; now the final observed power spectrum can be written as:
\begin{equation}
    C_\ell^{UT} = \int \frac{W(\chi-1/2\,\delta\chi)W(\chi+1/2\,\delta\chi)}{\chi^2} \mathcal C_s(\ell, \chi, \delta\chi) d\chi \; .
\end{equation}

\begin{figure*}[t!]
\label{fig:Clscomp}
\includegraphics[width=0.47\textwidth]{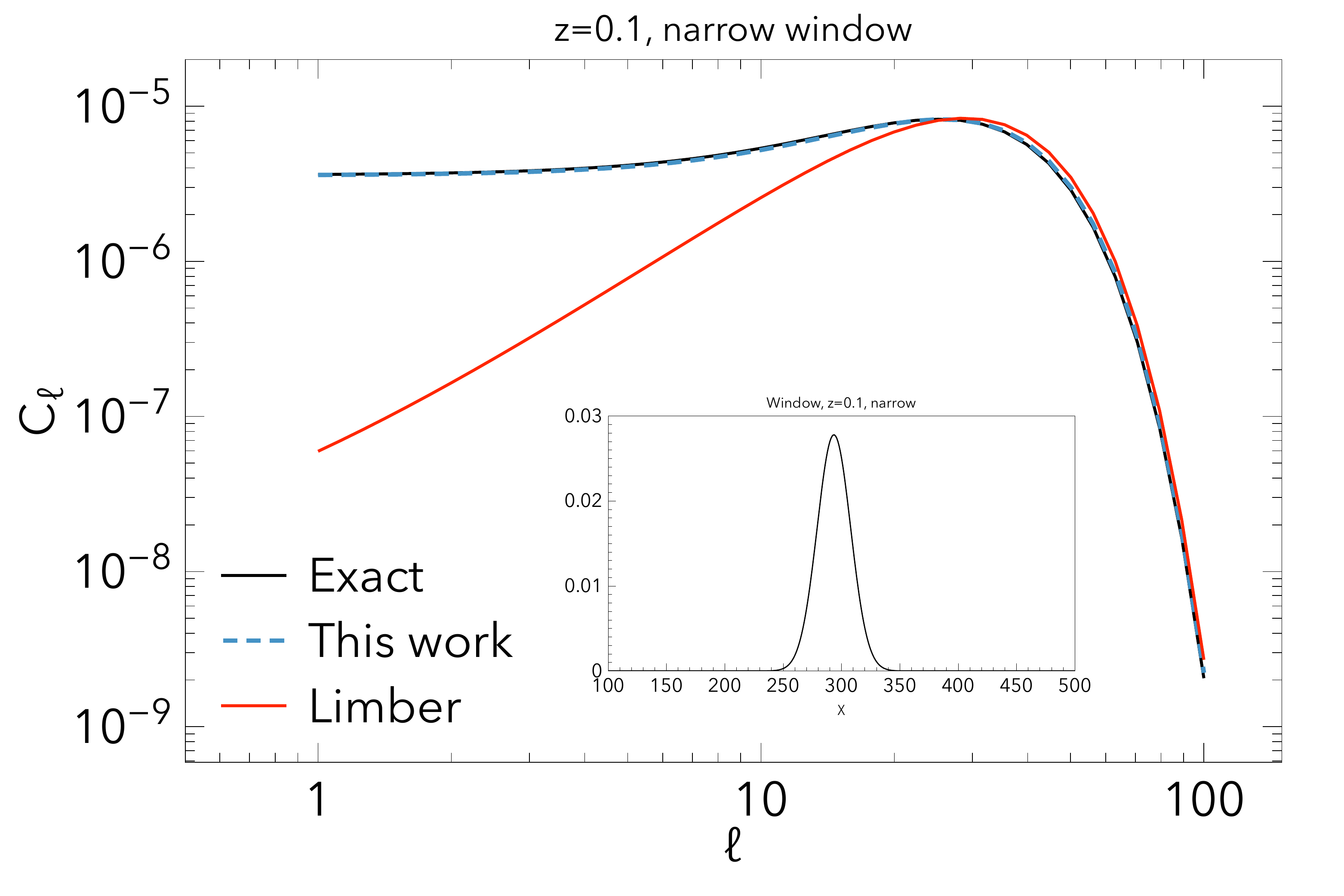}
\includegraphics[width=0.47\textwidth]{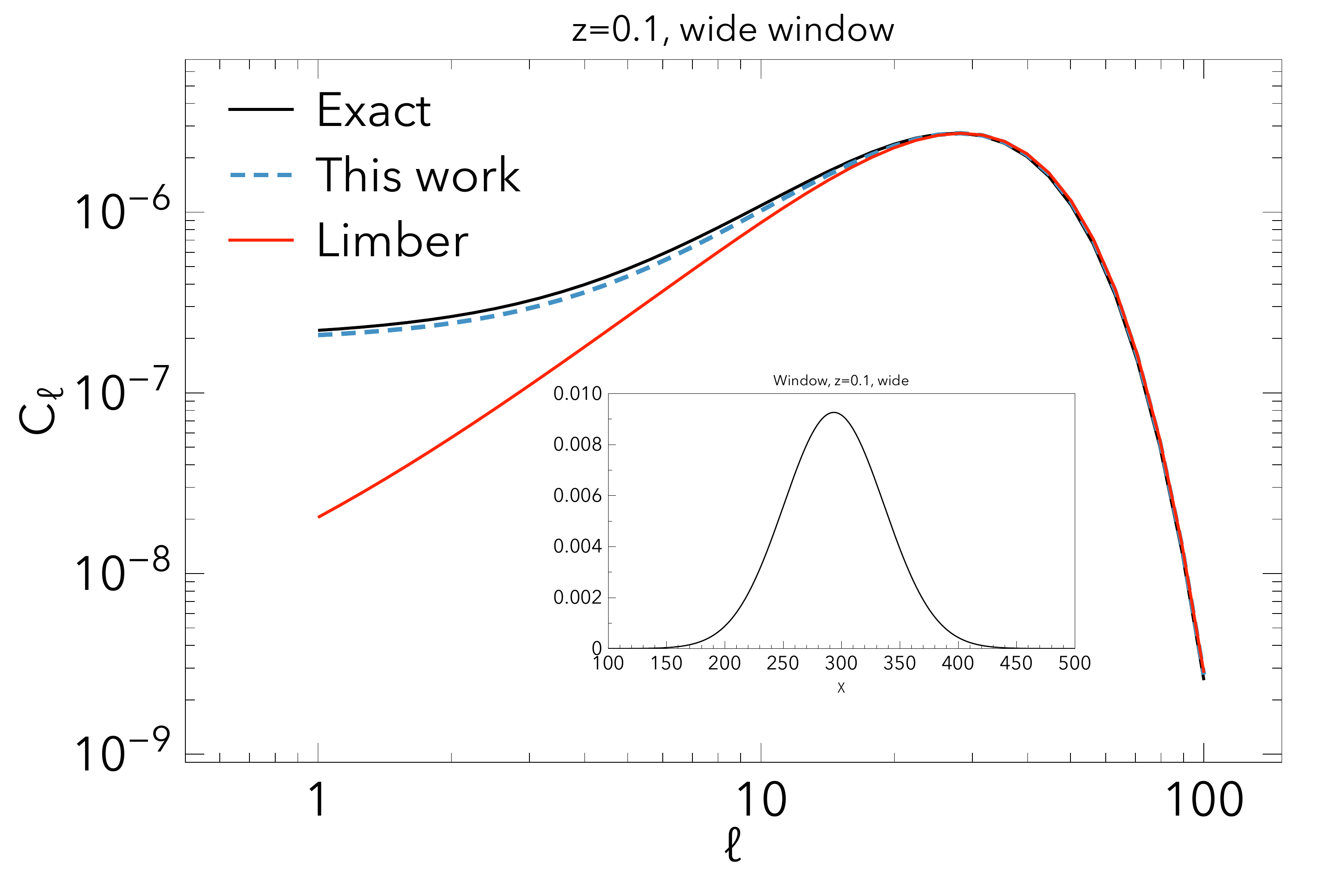}
\includegraphics[width=0.47\textwidth]{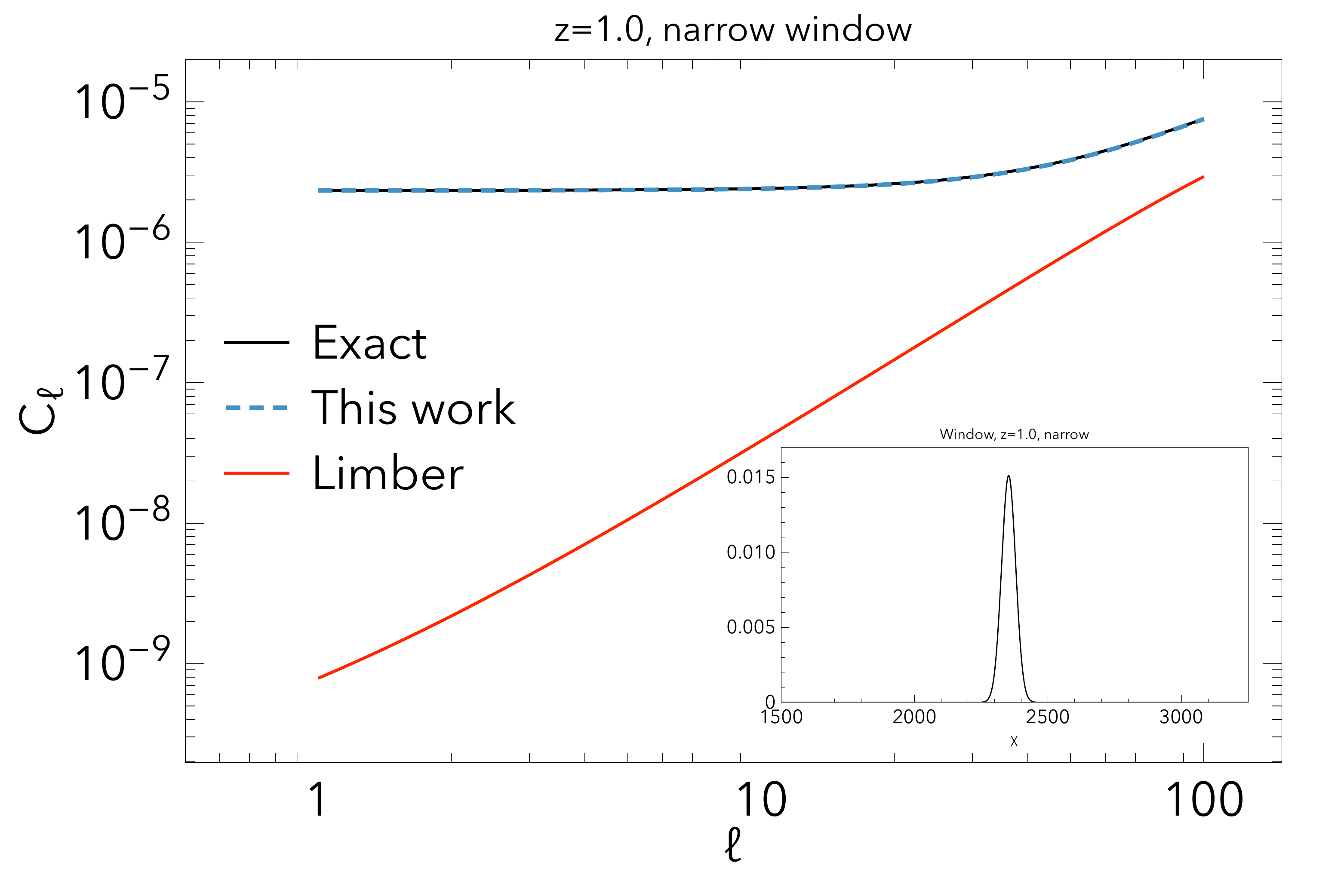}
\includegraphics[width=0.47\textwidth]{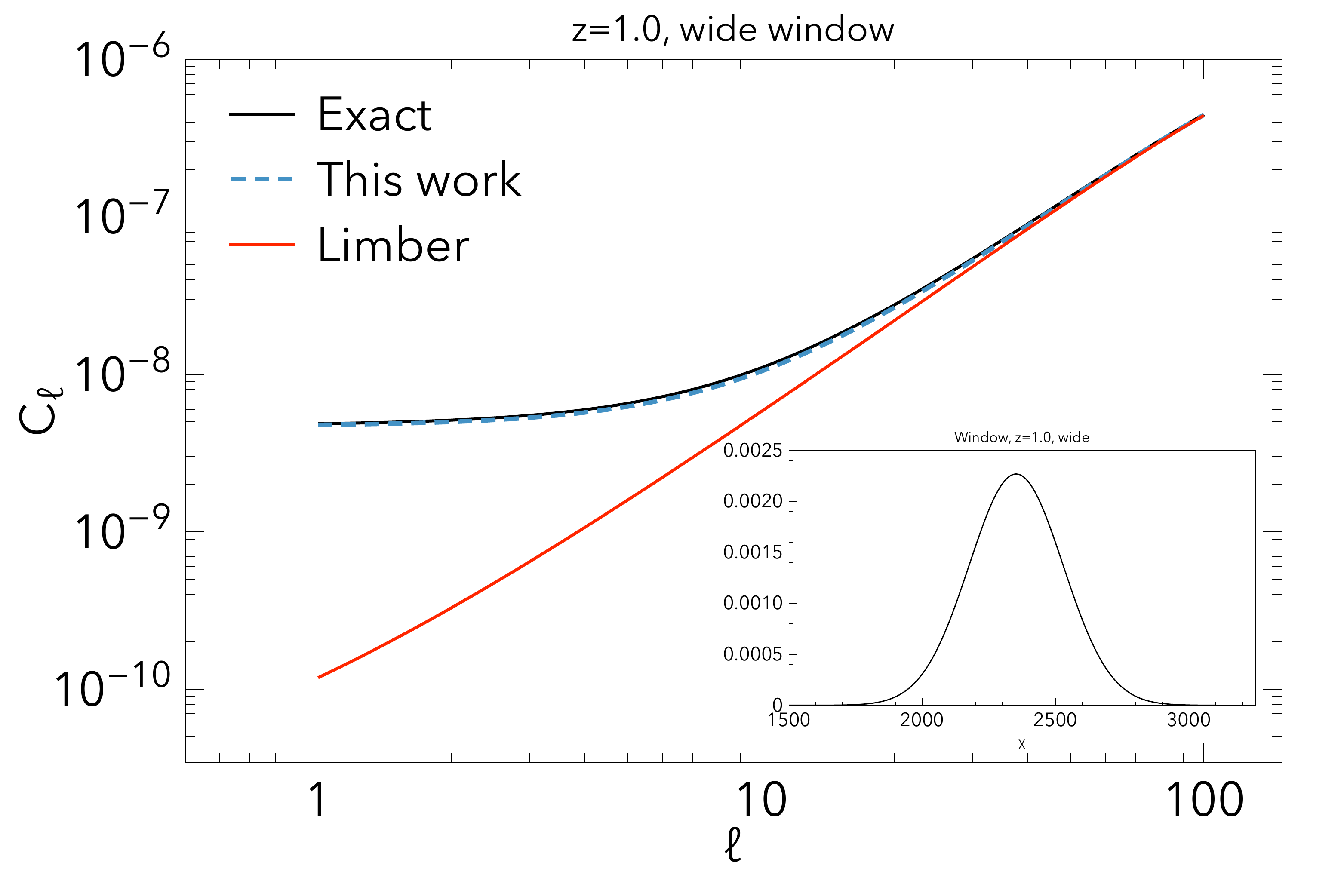}
\caption{Comparison of angular spectra $C_\ell$s: for different redshift bins and window function thickness, we show the exact full calculation (solid black lines), the results using the standard Limber approximation (solid red lines) and our results (dashed blue lines) for the unequal-time flat sky.}
\end{figure*}

We are now ready to present results for the angular power spectra including unequal time corrections; in Figure~\ref{fig:Clscomp} we compare the exact calculation (in black) with the Limber approximation (in red) and our result $\mathcal C_s(\ell, \chi, \delta\chi)$ (in dashed blue), for two different redshifts and narrow or wide window functions.
As expected, we notice that the Limber approximation works well for large $\ell$ and better for thicker than thinner redshift bins.
Our result instead works very well for any angular multipole, at both redshifts, and for thick as well as thin windows.

\label{sec:windows}
\begin{figure}[htb!]
\includegraphics[width=0.47\textwidth]{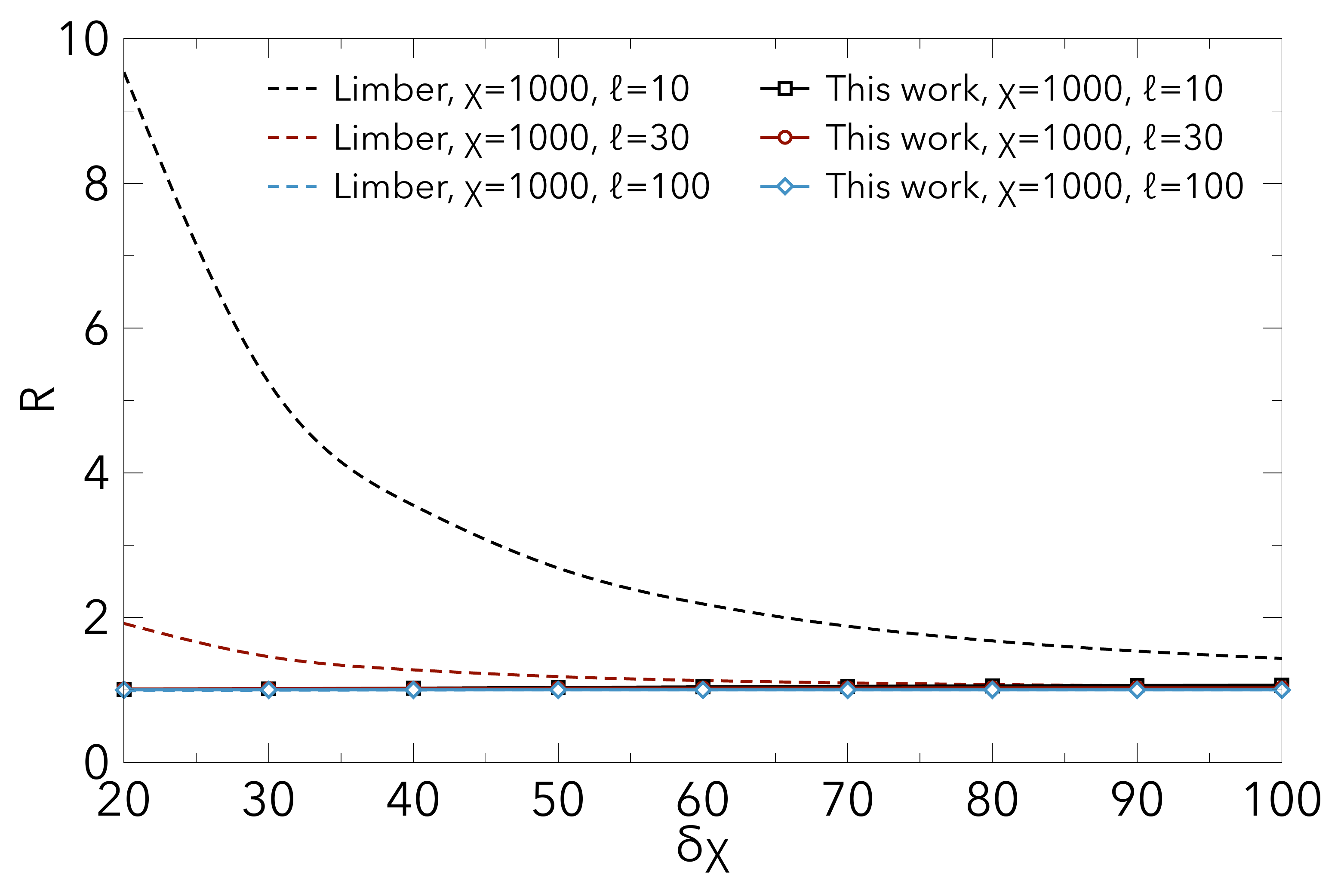}
\includegraphics[width=0.47\textwidth]{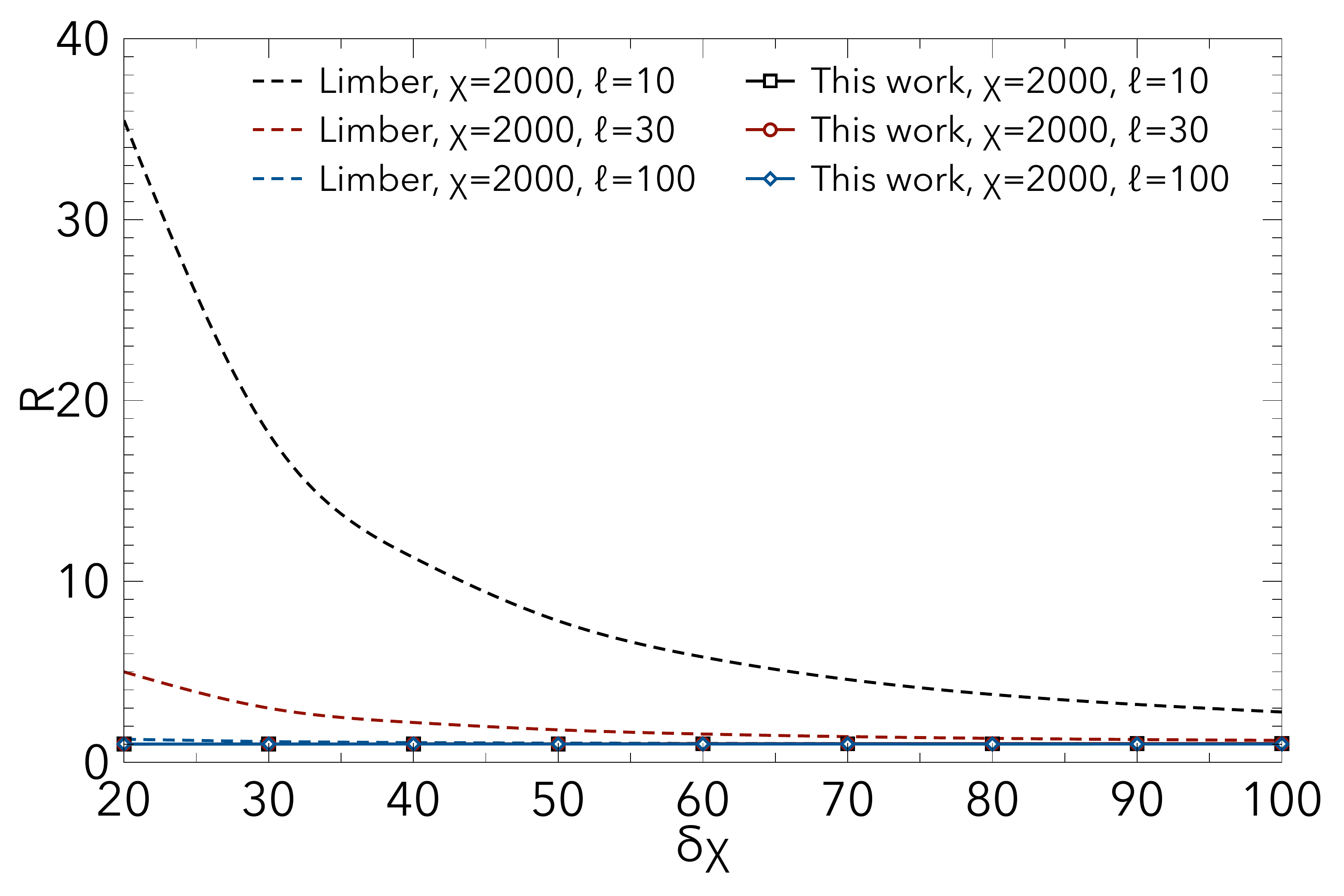}
\caption{Ratio of exact calculation 
 over approximations as a function of the radial bin thickness.}
\label{fig:ratio_windows}
\end{figure}
In Figure~\ref{fig:ratio_windows} we show the ratio $R=C_\ell^{\rm exact}/C_\ell^{\rm model}$ for the Limber approximation and the results of this work, for two values of the average bin comoving distance and at different multipoles $\ell$s, as a function of $\delta\chi$.
We can see how the Limber approximation works better for large $\ell$, as widely known in literature (see e.g.,~\cite{Assassi}), and the precision is degraded when the bin thins.
The reason for that is in the fact that the corrections in Equation~\eqref{eq:CsLD} are larger for smaller $\delta\chi$ (see Figure~\ref{fig:Cls39}), therefore indicating that neglecting unequal time corrections is less appropriate for thin bins.
Thanks to the results of this work we can therefore see the reason for the Limber approximation to be less and less correct for thin windows.

%==========================================================================
\section{Conclusions} 
\label{sec:conclusion}
In this paper we develop a formalism to describe 
the galaxy angular power spectrum including the effects of unequalness in time (equivalent to including the normally neglected radial modes).
After introducing our setup and formalism, setting up the basics for the proper understanding of the behavior of galaxy correlators in different spaces (the separation between dynamics and projection effects), we derive an expansion for the angular power spectrum around the equal time case.

We show that the connection between the theoretical, translational invariant power spectrum and the observed one happens in the limit of small separation angles and small relative radial separation between the sources.
Mathematically we set ourselves in the flat sky limit and Taylor expand around the radial separation $\delta\chi=0$. We find a mathematically consistent expression that includes radial modes and accounts for the transition from curved to flat sky, and we study its structure, verifying that higher order corrections are small.

Interestingly, two types of corrections arise: other than standard corrections due to the higher order terms in the sources' relative radial separation, there are corrections due to the off-diagonal terms of the delta function for the multipoles.

We explicitly add the contribution of radial modes, and using a toy power spectrum we are able to control the details of the structure of this expression and investigate the behavior of the resulting power spectrum in different regimes and cases, as a function of distance, scale and unequalness in time. 
While we show explicit examples using a toy power spectrum in order to follow and understand all the steps taken in the derivation, the structure of our expression is fully general.

We introduce the coefficients of the expansion and notice that corrections are of the order $\delta\chi^2$ or higher for the pure dark matter case and for single biased tracers, but when performing a multi-tracer analysis, there are first order unequal time corrections. A possibly very useful finding is that these corrections depend not only on the difference between the biases of the tracers, but also on their time derivative. Moreover, in redshift space, such corrections depend on the growth rate and its derivative.
This opens up the possibility of using such corrections to have additional handles to study the bias of galaxies and it evolutions, in addition to the temporal evolution of the growth rate.

Finally, we show that our expression that includes unequal time contributions fits the full exact calculation even at low multipoles $\ell$, and for both shallow and deep redshift bins.
This result showcases the importance of including modes along the line of sight, and provides a new angular power spectrum approximation that is computationally faster than the full calculation but much more precise than currently employed approximations.

In follow-up companion papers we investigate unequal time correlations in other spaces (in particular, the Fourier-space $P(k)$ in~\cite{RVII} and deepen the understanding of the mathematical structure of the limits we calculate in~\cite{RVIII}), observational strategies and consequences, and generalizations of this calculation.

%==========================================================================
\begin{acknowledgments}
We thank Jose Luis Bernal, Daniele Bertacca, Olivier Dore, Henry Grasshorn Gebhardt, Eric Huff, Dida Markovic, Federico Semenzato, Marko Simonovic, Francesco Spezzati, Eleonora Vanzan, Licia Verde, Martin White for useful discussions.
AR acknowledges funding from Italian Ministry of University and Research (MIUR) through the ``Dipartimenti di eccellenza'' project ``Science of the Universe''.
ZV is partially supported by the Kavli Foundation.
\end{acknowledgments}

%==========================================================================
\appendix

\section{Full angular power spectrum including velocity effects}
\label{app:cls}
It is important to note that in this work we are still implicitly making assumptions about some mathematical properties of the system; in particular, a proper treatment of the RSD spectrum in \OS involves accounting for the breaking of translational invariance.
This is the reason why the RSD operator becomes non-diagonal and there are mode-coupling effects~\cite{Szalay:1997, Hamilton:1997, Papai:2008, Raccanelli:2010wa}, because redshift distortions cause a phase shift to some modes when passing from real to redshift space.
In Fourier space, the full correlator deviates from the commonly defined power spectrum, and would be depending on two Fourier modes, $\{k_1, k_2\}$.
This happens because, without translation symmetry, Fourier modes are no longer eigenmodes of the redshift distortion operator, and thus the redshift space power spectrum
$\langle
\delta^{s}(\bk_1) \delta^{s}(\bk_2)
\rangle$
is no longer a diagonal matrix, while it is in the Kaiser approximation, which imposes a single Fourier mode.
Even the full RSD operator does however preserve angular symmetry about the observer.

In a similar way, the angular correlator deviates from being diagonal, and the $C_\ell$s should in principle depend on two angular wavenumbers $\{\ell_1, \ell_2\}$.
Mathematically, we can easily see this by noting that the derivatives of $j_\ell$ do no represent an orthonormal basis.

The derivatives of the spherical Bessel functions can be rewritten as $j^{(n)}(k\chi) = \partial^n/k^n\partial \chi^n j(k\chi)$.
In almost all realistic situations, we can consider the redshift-dependent quantities (such as bias, growth, etc.) and the observational windows (i.e.,~the integration extremes in the $\chi_i$ integrals) as smooth and move the derivatives to the radial integrals:

\begin{align}
\label{eq:Cls3Dder}
C_\ell(\chi_1,\chi_2) =& \frac{2}{\pi} \left\{
\left(\int d\chi_1 \left[b_1 D_1\right] \int d\chi_2 \left[b_2 D_2\right]\right) \int dk \, k^2 P(k) j_\ell(k\chi_1)j_\ell(k\chi_2) - \right. \nonumber \\
&-\left(\int d\chi_1 \left[b_1 D_1\right] \int d\chi_2 \frac{\partial}{\partial \chi_2}\left[f_2 \frac{\alpha_2}{\chi_2}\right] + \int d\chi_1 \left[b_1 D_1\right] \int d\chi_2 \frac{\partial^2}{\partial \chi_2^2}\left[f_2\right] \right.+\nonumber \\
&+ \left.\int d\chi_1 \frac{\partial}{\partial\chi_1}\left[f_1 \frac{\alpha_1}{\chi_1}\right] \int d\chi_2 \left[b_2 D_2\right] + \int d\chi_1 \frac{\partial^2}{\partial\chi_1^2}\left[f_1\right] \int d\chi_2 \left[b_2 D_2\right]\right)\times \nonumber \\
& \times \int dk \, P(k) j_\ell(k\chi_1)j_\ell(k\chi_2) + \nonumber \\
&+\left(\int d\chi_1 \frac{\partial}{\partial\chi_1}\left[f_1 \frac{\alpha_1}{\chi_1}\right] \int d\chi_2 \frac{\partial}{\partial \chi_2}\left[f_2 \frac{\alpha_2}{\chi_2}\right]
+ \int d\chi_1 \frac{\partial}{\partial\chi_1}\left[f_1 \frac{\alpha_1}{\chi_1}\right] \int d\chi_2 \frac{\partial^2}{\partial \chi_2^2}\left[f_2\right]\right. + \nonumber \\
&\left.+ \int d\chi_1 \frac{\partial^2}{\partial\chi_1^2}\left[f_1\right] \int d\chi_2 \frac{\partial}{\partial \chi_2}\left[f_2 \frac{\alpha_2}{\chi_2}\right]+ \int d\chi_1 \frac{\partial^2}{\partial\chi_1^2}\left[f_1\right] \int d\chi_2 \frac{\partial^2}{\partial \chi_2^2}\left[f_2 \right] \right) \times\nonumber\\
&\int dk \, \frac{1}{k^2}P(k) j_\ell(k\chi_1)j_\ell(k\chi_2) \; .
\end{align}
It can then be seen how peculiar velocity effects are in the $\chi$ integrals and really just affect the apparent radial position of galaxies; therefore their effect can be calculated with line-of-sight derivatives of the observational windows.
This reflects the fact that RSD are a projection effect, and we can separate them from the clustering dynamics happening in the \HS box.

\section{Choice of the mean distance }
\label{app:mean}

%=================%
\noindent\textbf{Arithmetic}

With the arithmetic mean definition of $\chi$,
$\chi = \frac{1}{2} \lb \chi_1 + \chi_2 \rb$,
we have:
\eeq{
\chi_1 = \chi + \frac{1}{2} \delta \chi, \qquad
\chi_2 = \chi - \frac{1}{2} \delta \chi
}
and $d\chi_1/d \delta \chi = 1/2$, $d\chi_2/d \delta \chi = -1/2$. \\

%=================%
\noindent\textbf{Geometric}

If we use the geometric mean definition of $\chi$,
$\chi = \sqrt{\chi_1 \chi_2}$,
\eeq{
\chi_1 = \sqrt{\chi^2 + \delta \chi^2/4} + \delta \chi/2, \qquad
\chi_2 = \sqrt{\chi^2 + \delta \chi^2/4} - \delta \chi/2.
}

%=================%
\noindent\textbf{Harmonic}

The harmonic mean definition of $\chi$,
$\chi = \frac{2 \chi_1 \chi_2}{\chi_1 + \chi_2}$ gives us:
\eeq{
\chi_1 = \frac{1}{2} \lb \chi + \delta \chi + \sqrt{\chi^2 + \delta \chi^2} \rb, \qquad
\chi_2 = \frac{1}{2} \lb \chi - \delta \chi + \sqrt{\chi^2 + \delta \chi^2} \rb.
}

\section{2D delta function}
\label{app:2Ddelta}
For arithmetic, geometric and harmonic coordinates, respectively, the relations to the 2D delta function are:
\eq{
\df^{\rm 2D} \lb \vec{ \tilde \ell} - \vec{\tilde \ell}' \rb 
&=  \df^{\rm 2D} \lb \frac{\vec{\ell} - \vec{\ell}' + \daleth \mathcal{L}}{\chi_{\rm a} (1-\daleth^2)} \rb
= \chi^2_{\rm a} (1-\daleth^2)^2 \df^{\rm 2D} \lb \vec{\ell} - \vec{\ell}' + \mathcal{L} \daleth \rb \, , \\
\df^{\rm 2D} \lb \vec{ \tilde \ell} - \vec{\tilde \ell}' \rb 
&= \df^{\rm 2D} \lb \frac{(\vec{\ell} - \vec{\ell}')\sqrt{1-\daleth^2} + \daleth \mathcal{L}}{\chi_{\rm g}} \rb
=  \frac{\chi^2_{\rm g}}{(1-\daleth^2)} \df^{\rm 2D} \lb \vec{\ell} - \vec{\ell}' + \mathcal{L} \daleth/\sqrt{1-\daleth^2}  \rb \, , \non\\
\df^{\rm 2D} \lb \vec{ \tilde \ell} - \vec{\tilde \ell}' \rb 
&= \df^{\rm 2D} \lb \frac{(\vec{\ell} - \vec{\ell}')\lb 1 + \sqrt{1+4\daleth^2} \rb + 2 \daleth \mathcal{L} }{\chi_{\rm h} \lb 1 + \sqrt{1+4\daleth^2} \rb } \rb
=  \chi^2_{\rm h} \df^{\rm 2D} \lb \vec{\ell} - \vec{\ell}' + 2 \mathcal{L} \daleth/\big(1 + \sqrt{1+4\daleth^2} \big) \rb \, . \non
}

\section{Toy power spectrum}
\label{app:toyPK}
\begin{figure}[htb!]
\includegraphics[width=0.75\textwidth]{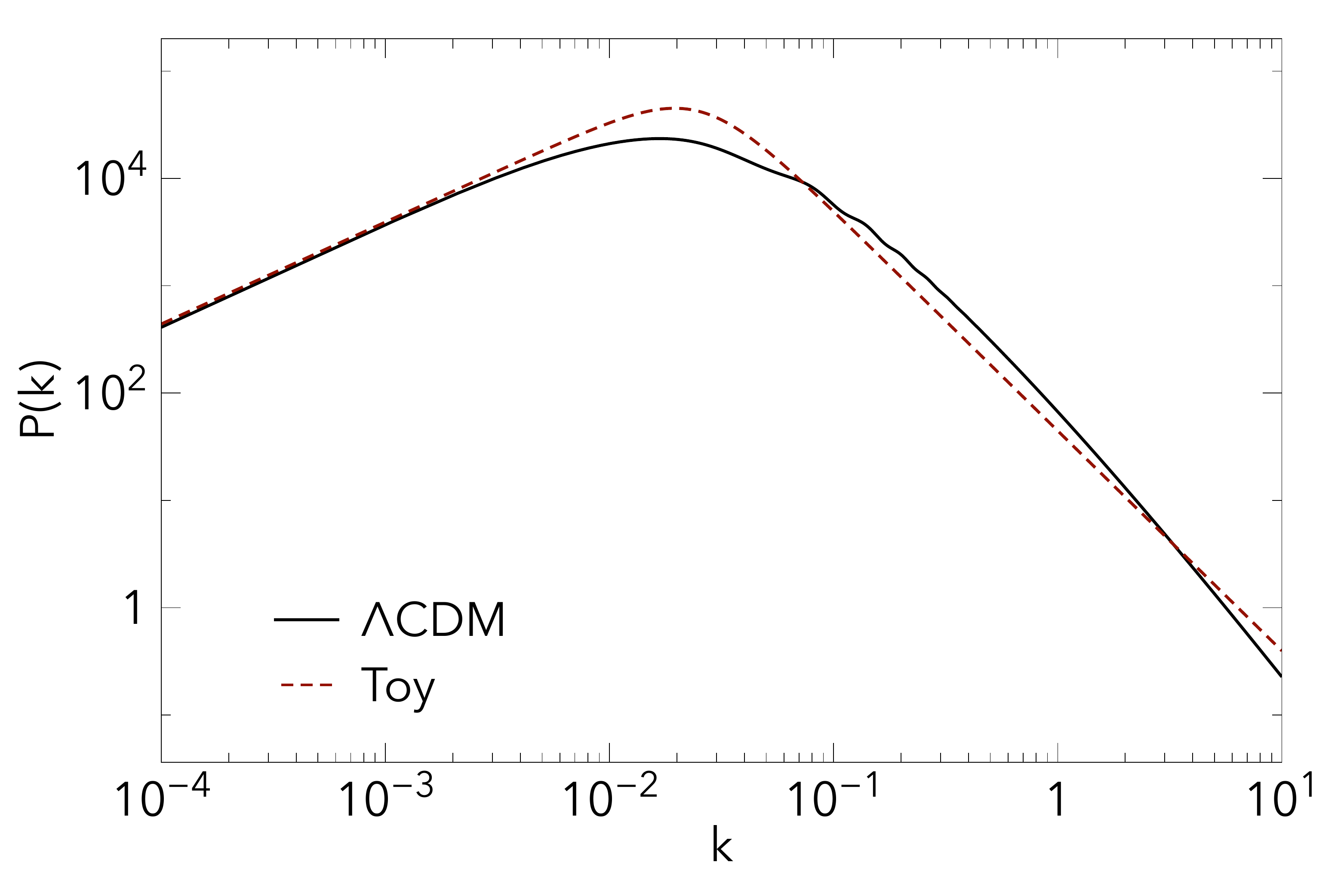}
\caption{$\Lambda$CDM power spectrum (solid black line) and ``power-law toy power spectrum'' (red dashed line).}
\label{fig:toypowerspectrum}
\end{figure}

The results obtained in Section~\ref{sec:angular_ps} are for a toy power spectrum that we used because it is in a form that is possible to solve analytically, and therefore maintain control of all the aspects of the derivation and check every step. We stress that our general results do not depend on the parameters and shape of the power spectrum, but in any case
we show here that such power spectrum (as in Equation~\eqref{eq:CsLD}) does not substantially deviate from the $\Lambda$CDM spectrum. But we note that all results are obtained comparing the same power spectra (be it the $\Lambda$CDM or the toy one) in a consistent way.
In Figure~\ref{fig:toypowerspectrum} we show the $\Lambda$CDM spectrum in solid black compared to the toy spectrum in dashed red.

\section{Power spectrum dependencies}
\label{app:dep}
\begin{figure*}[t!]
\includegraphics[width=0.47\textwidth]{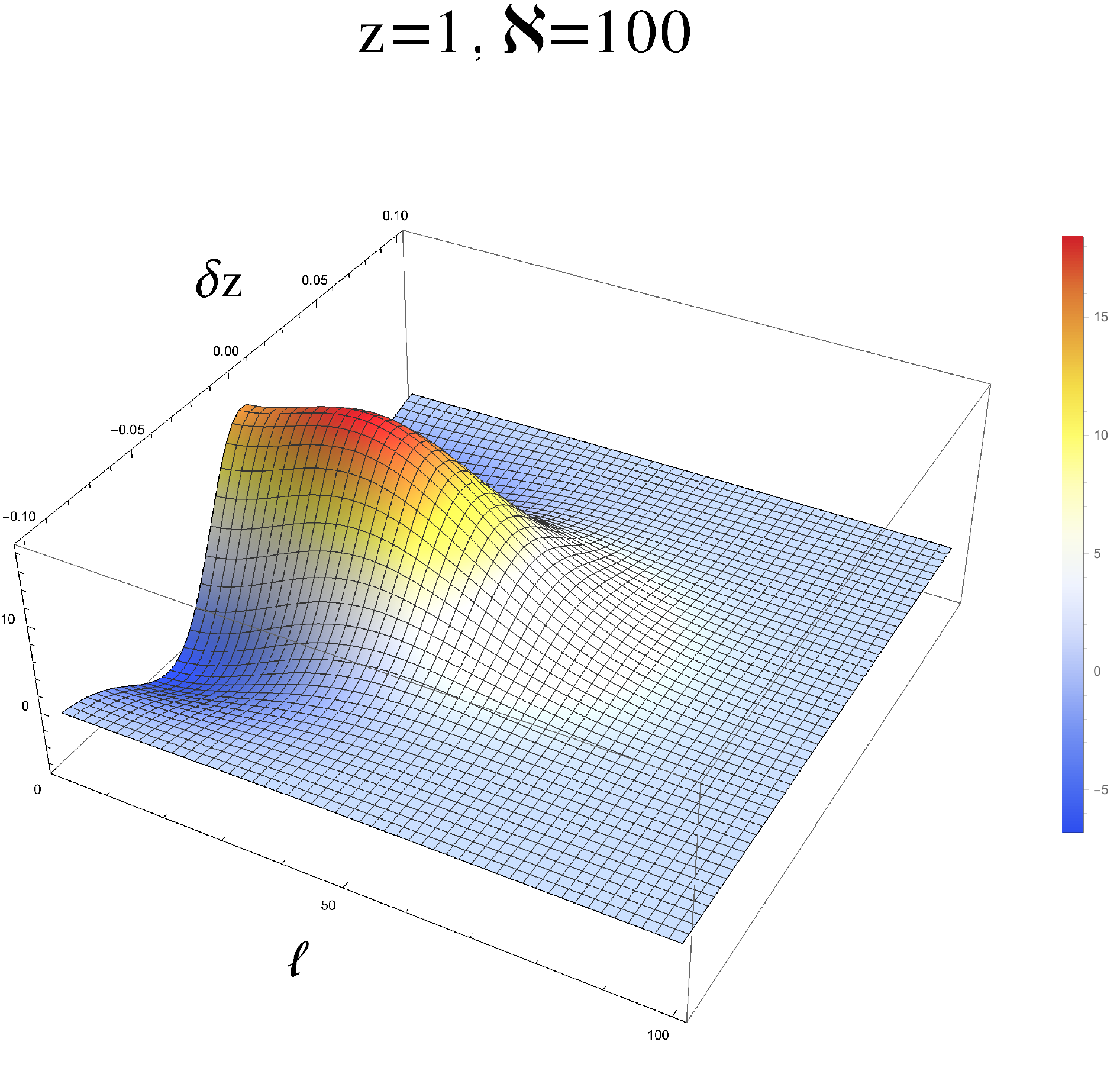}
\includegraphics[width=0.47\textwidth]{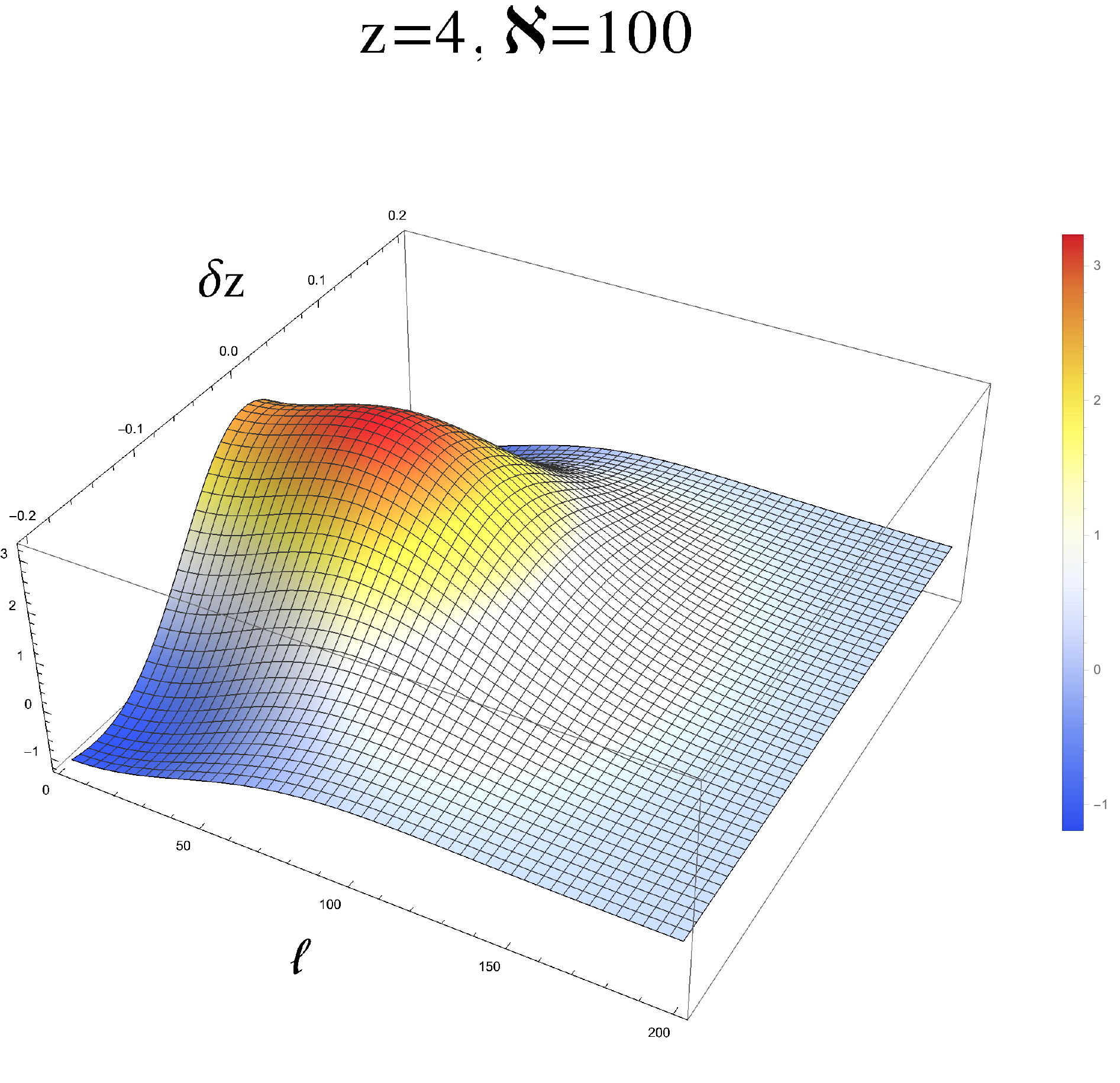}
\caption{Angular spectra from Equation~\ref{eq:cls38}, for two different values of $z$, as a function of the multipole $\ell$ and the unequal time thickness.}
\label{fig:Cls38}
\end{figure*}

\begin{figure*}[t!]
\includegraphics[width=0.47\textwidth]{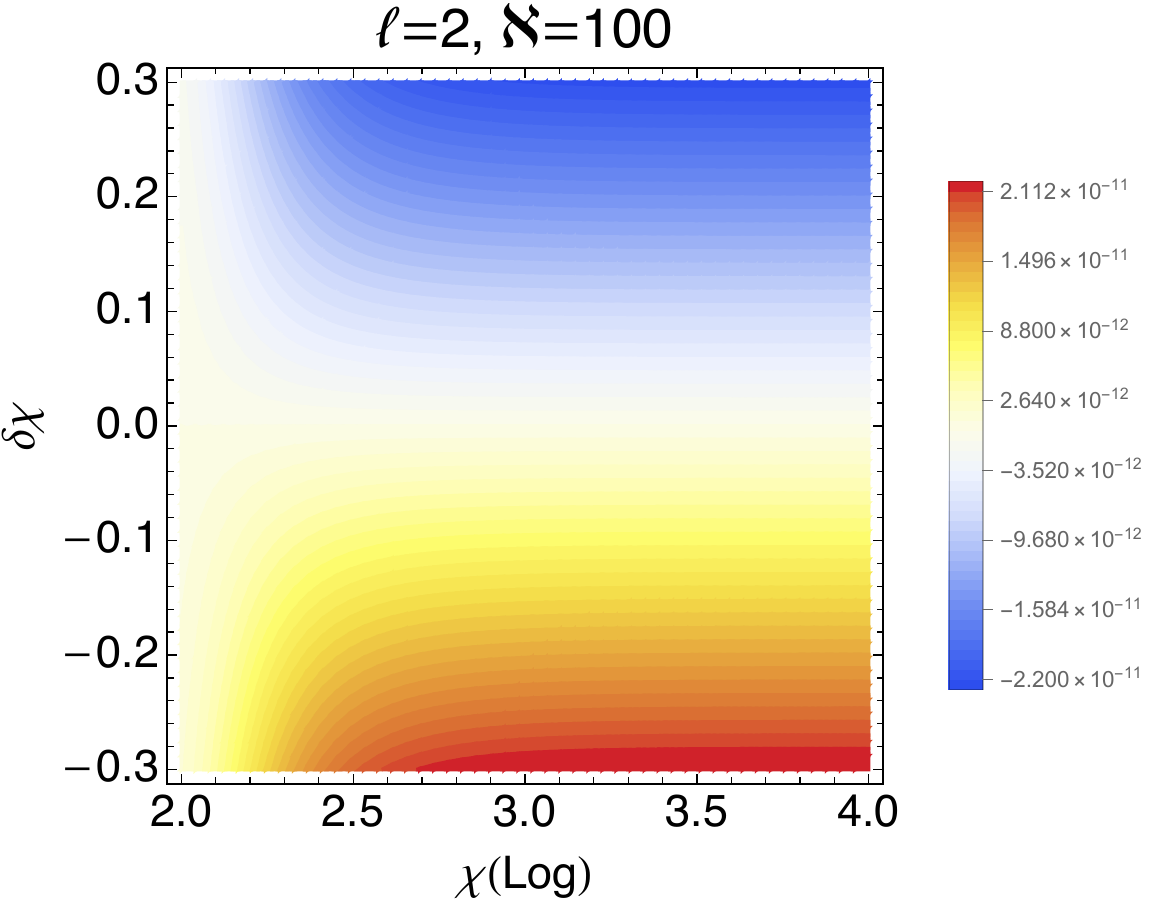}
\includegraphics[width=0.47\textwidth]{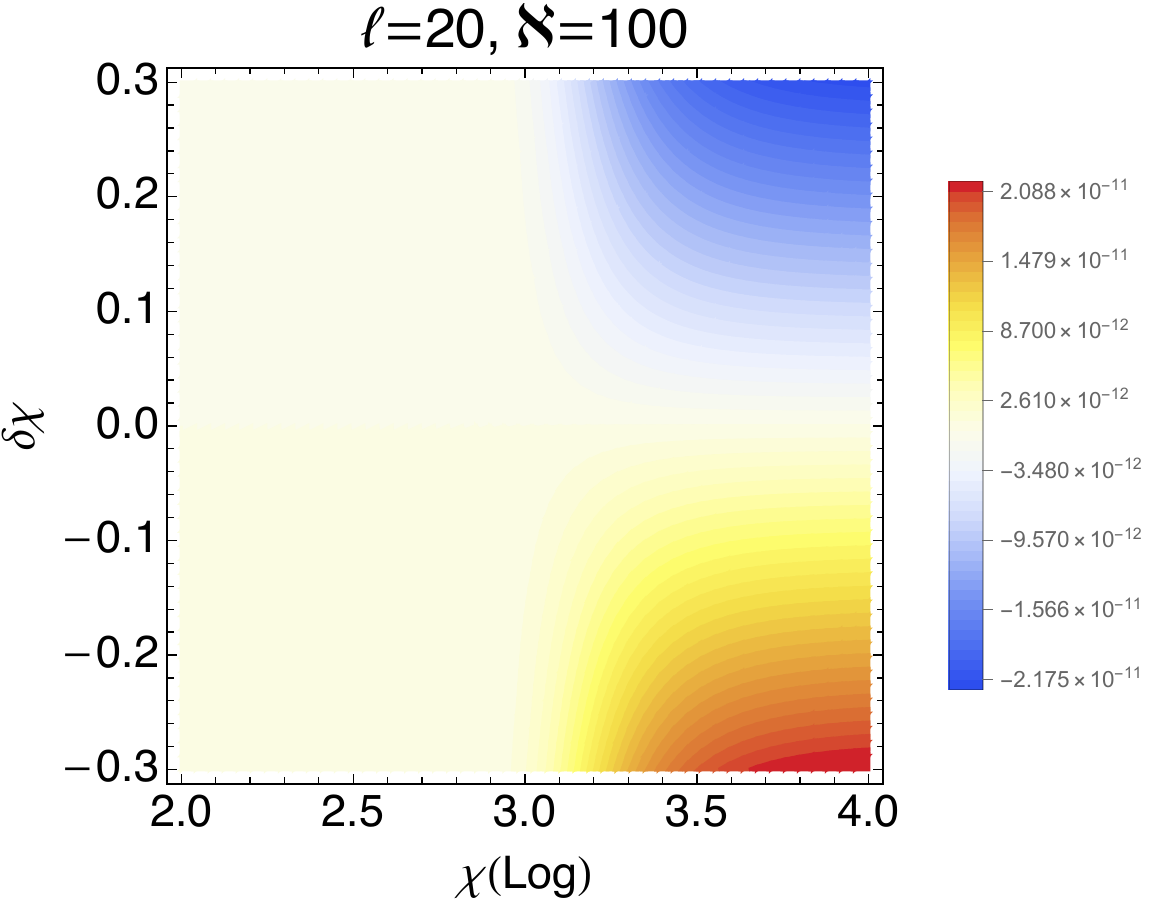}
\includegraphics[width=0.47\textwidth]{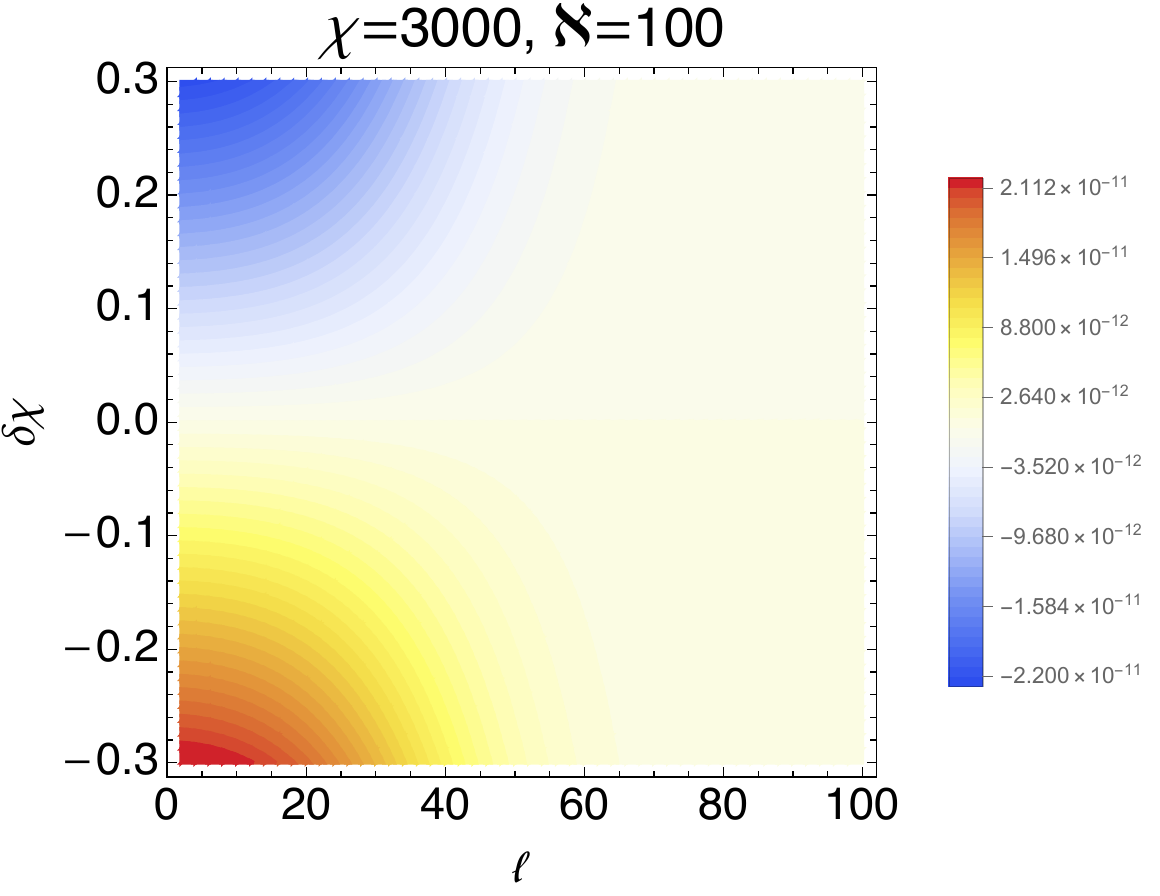}
\includegraphics[width=0.47\textwidth]{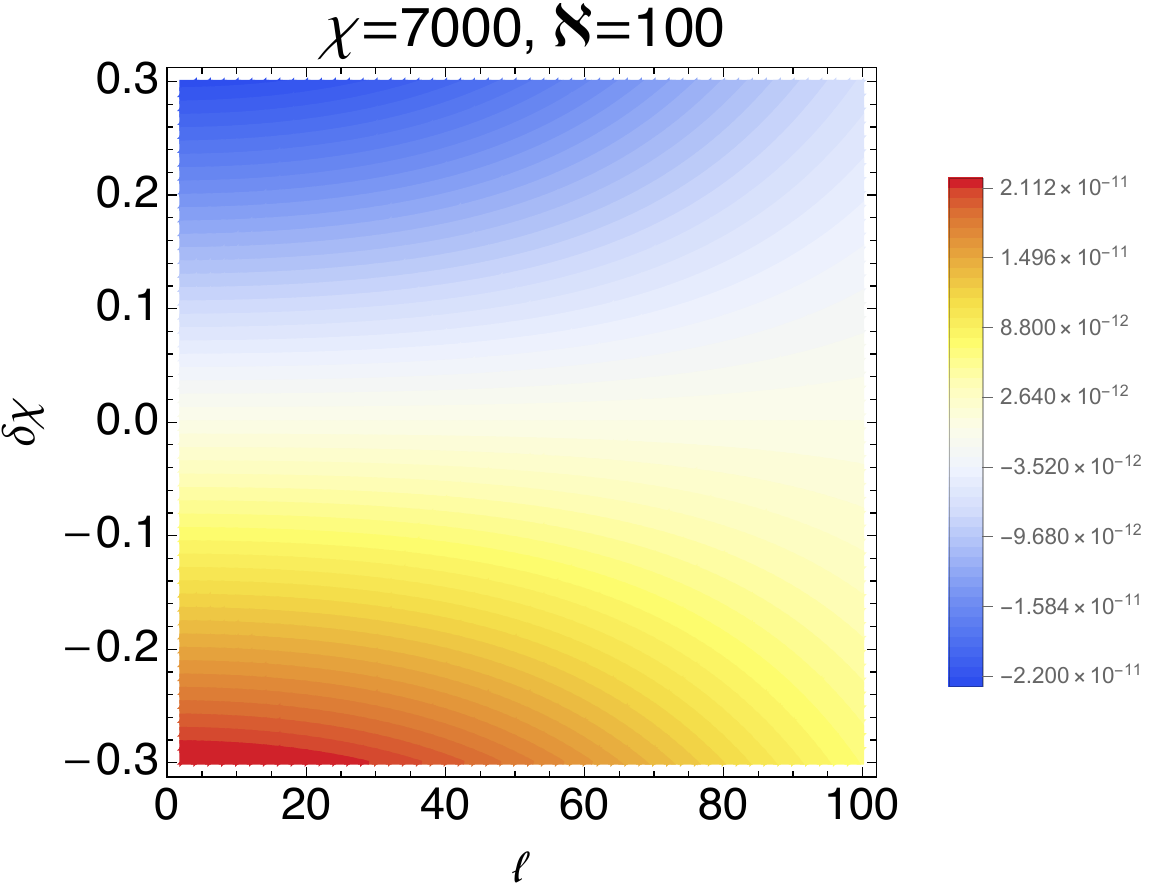}
\caption{Derivatives of Equation~\eqref{eq:cls38} with respect to $\delta\chi$.}
\label{fig:Cls38prime}
\end{figure*}

In Figures~\ref{fig:Cls38},~\ref{fig:Cls38prime} we show the spectrum from Equation~\eqref{eq:cls38} and its derivative with respect to $\delta\chi$, for different cases and dependencies. We do this to continue and deepen our investigation of the behavior of the unequal time spectra using the toy power spectrum.
The plots highlight where the derivative goes to zero and therefore we have maxima and minima of the spectrum. We can indeed see that the spectrum peaks for equal time correlations, and the amplitude of the derivative grows with larger $\delta\chi$ for larger and larger multipoles when going to higher redshift.

\begin{figure}[htb!]
\includegraphics[width=0.47\textwidth]{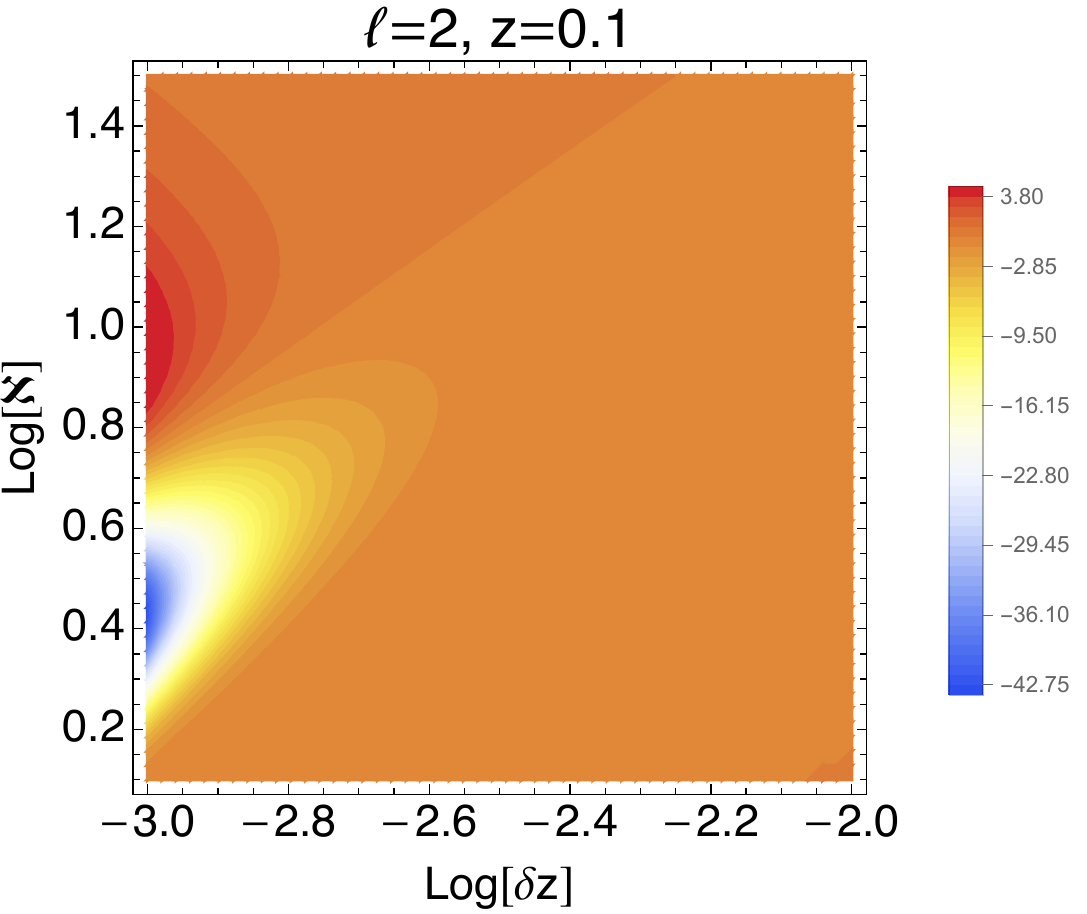}
\includegraphics[width=0.47\textwidth]{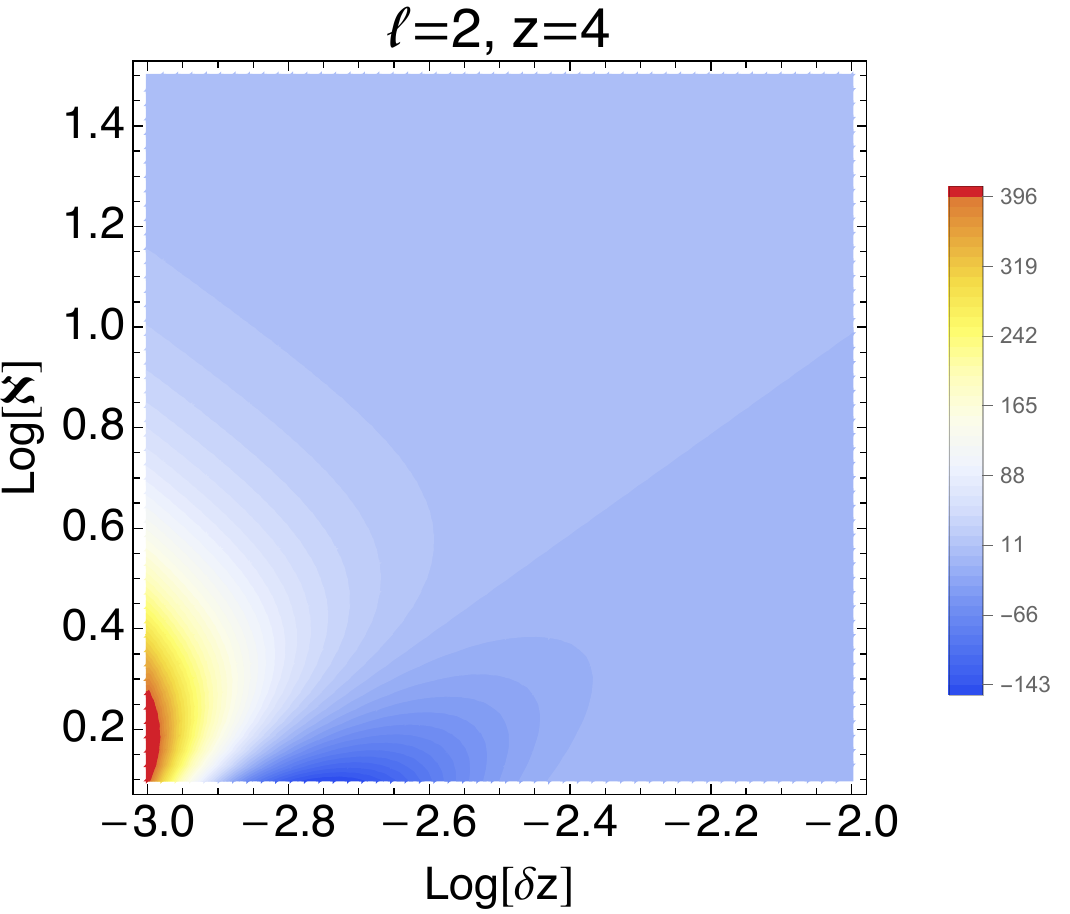}
\caption{Equation~\eqref{eq:CsLD} as a function of unequalness in time and power spectrum slope, for different values of $\ell$ and $z$.}
\label{fig:Csldep1}
\end{figure}

\begin{figure}[htb!]
\includegraphics[width=0.47\textwidth]{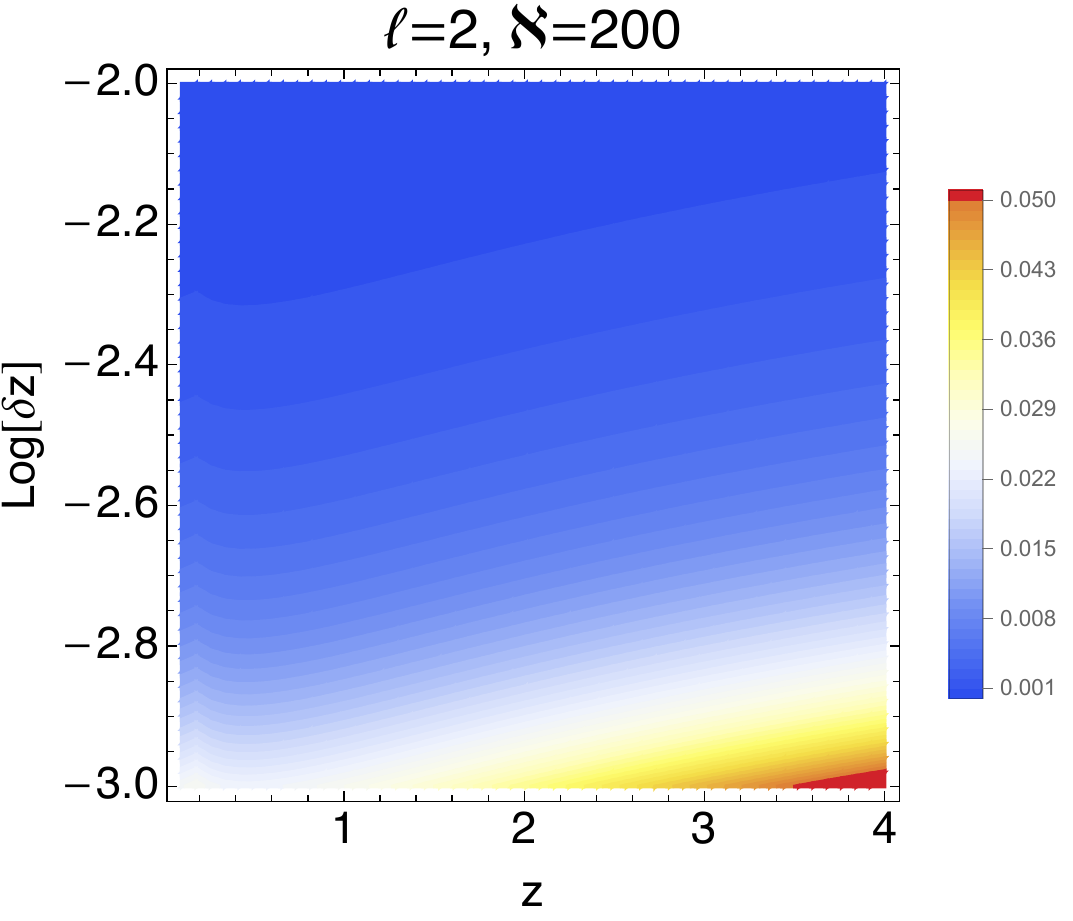}
\includegraphics[width=0.47\textwidth]{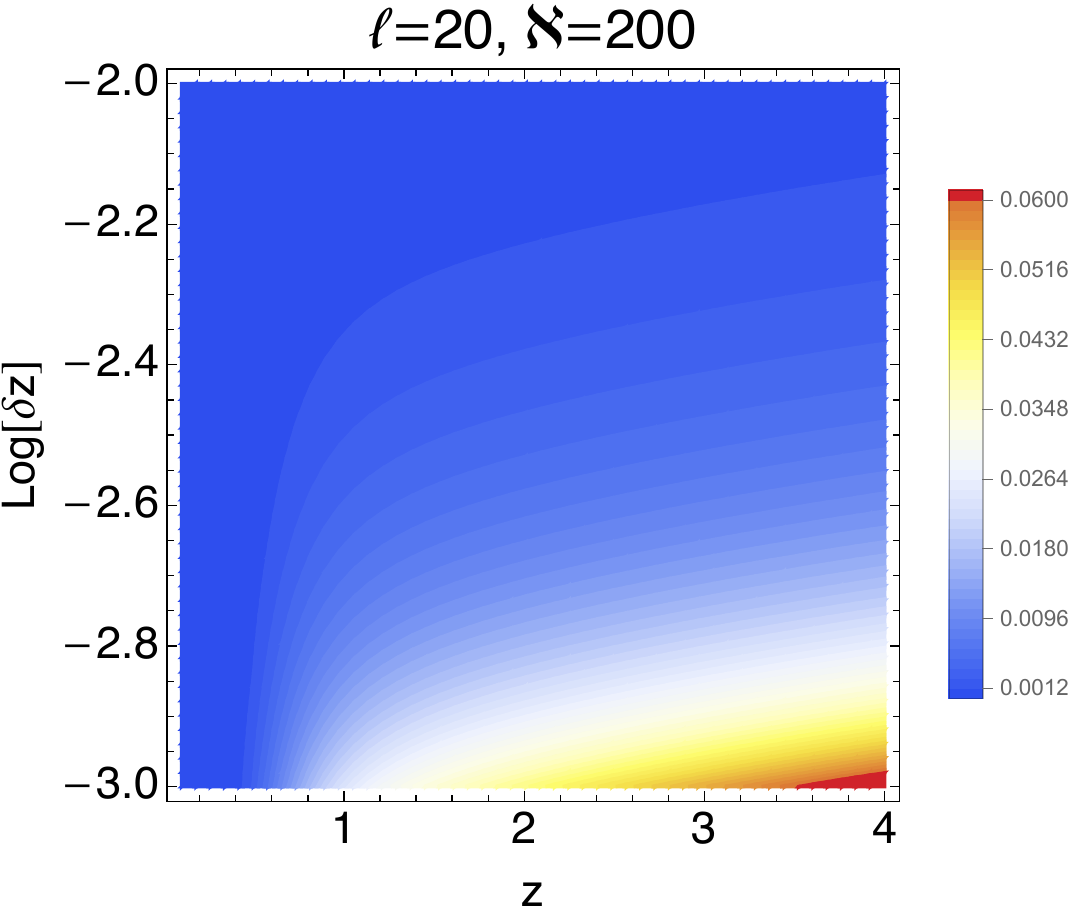}
\caption{Equation~\eqref{eq:CsLD} as a function of unequalness in time and redshift, for different values of $\ell$ and $\aleph$.}
\label{fig:Csldep2}
\end{figure}

\begin{figure}[htb!]
\includegraphics[width=0.47\textwidth]{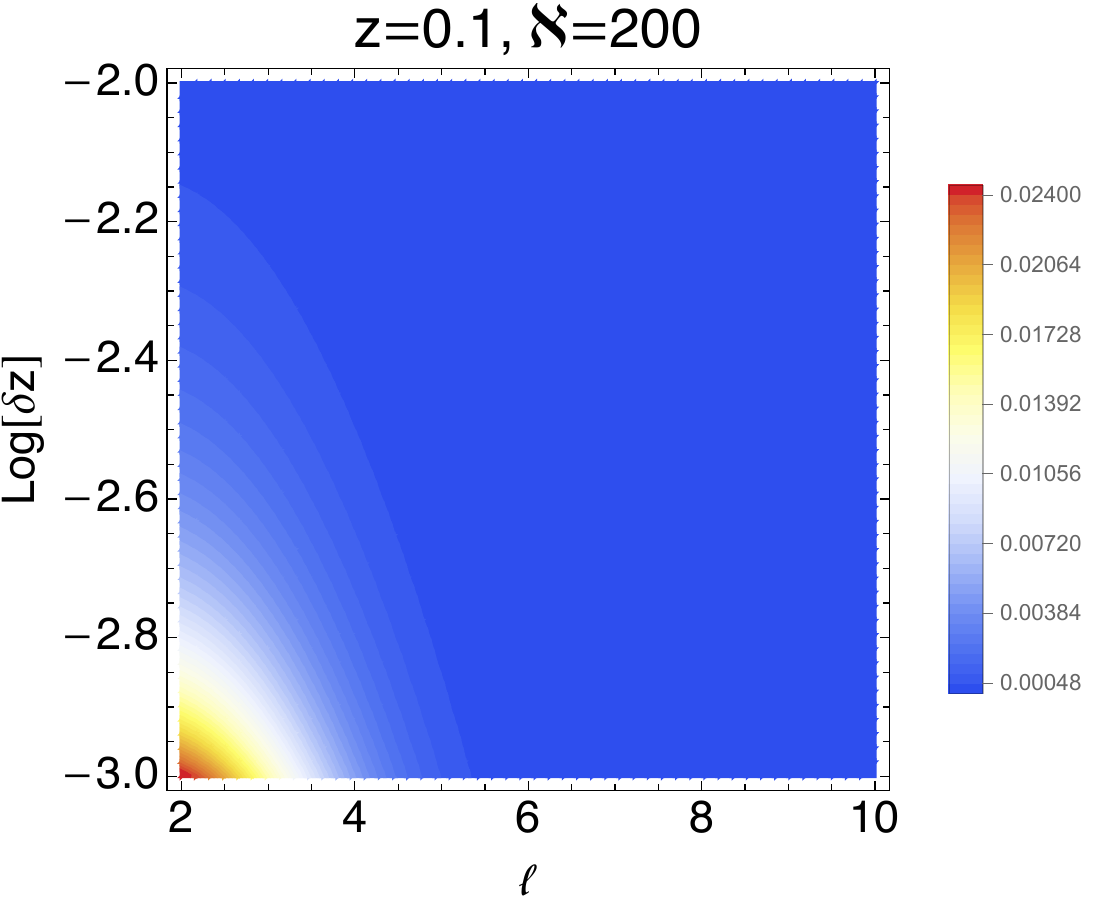}
\includegraphics[width=0.47\textwidth]{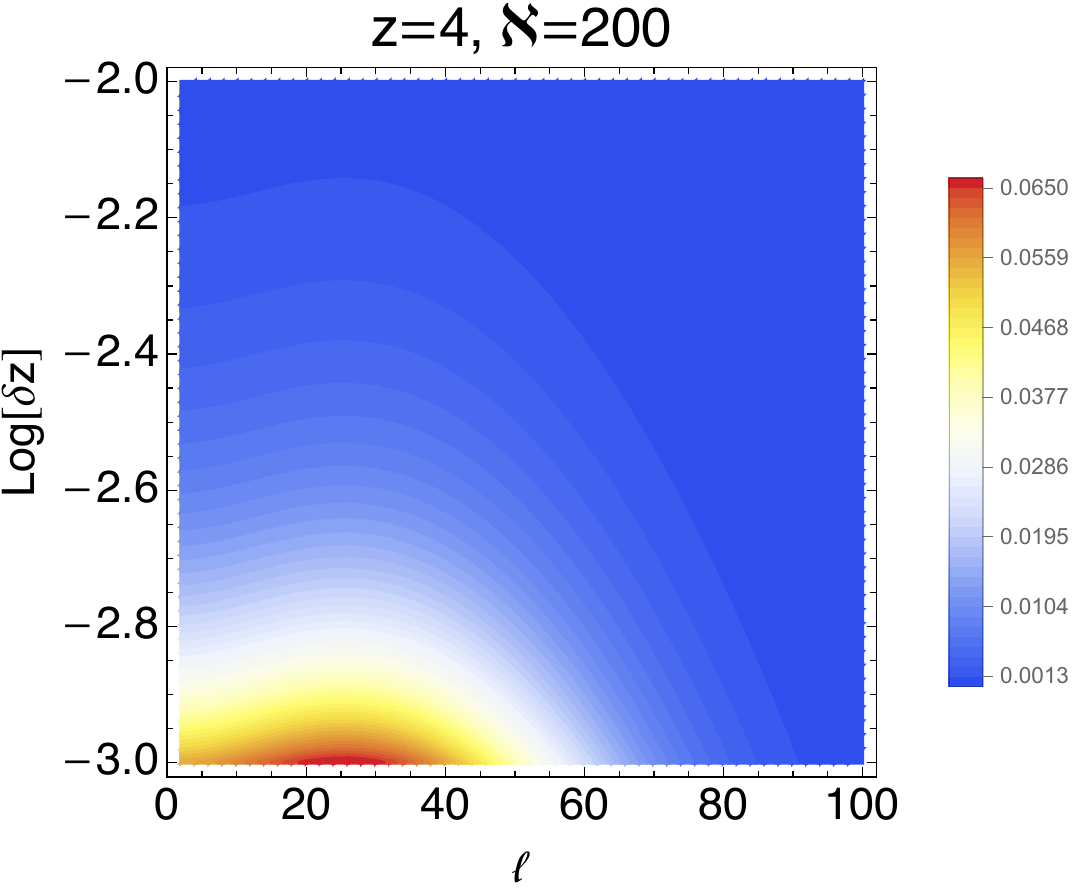}
\caption{Equation~\eqref{eq:CsLD} as a function of unequalness in time and $\ell$, for different values of $z$ and $z$.}
\label{fig:Csldep3}
\end{figure}

\begin{figure}[htb!]
\includegraphics[width=0.47\textwidth]{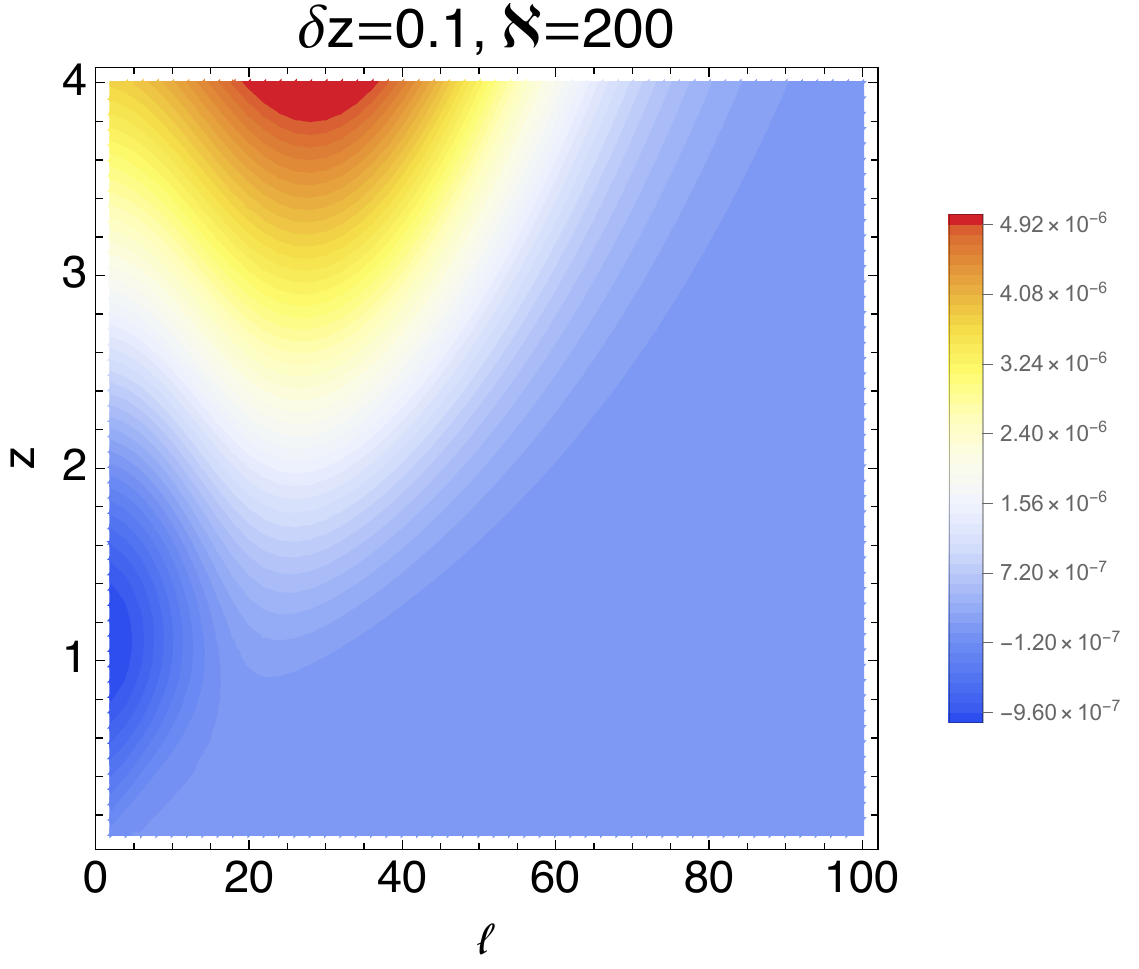}
\includegraphics[width=0.47\textwidth]{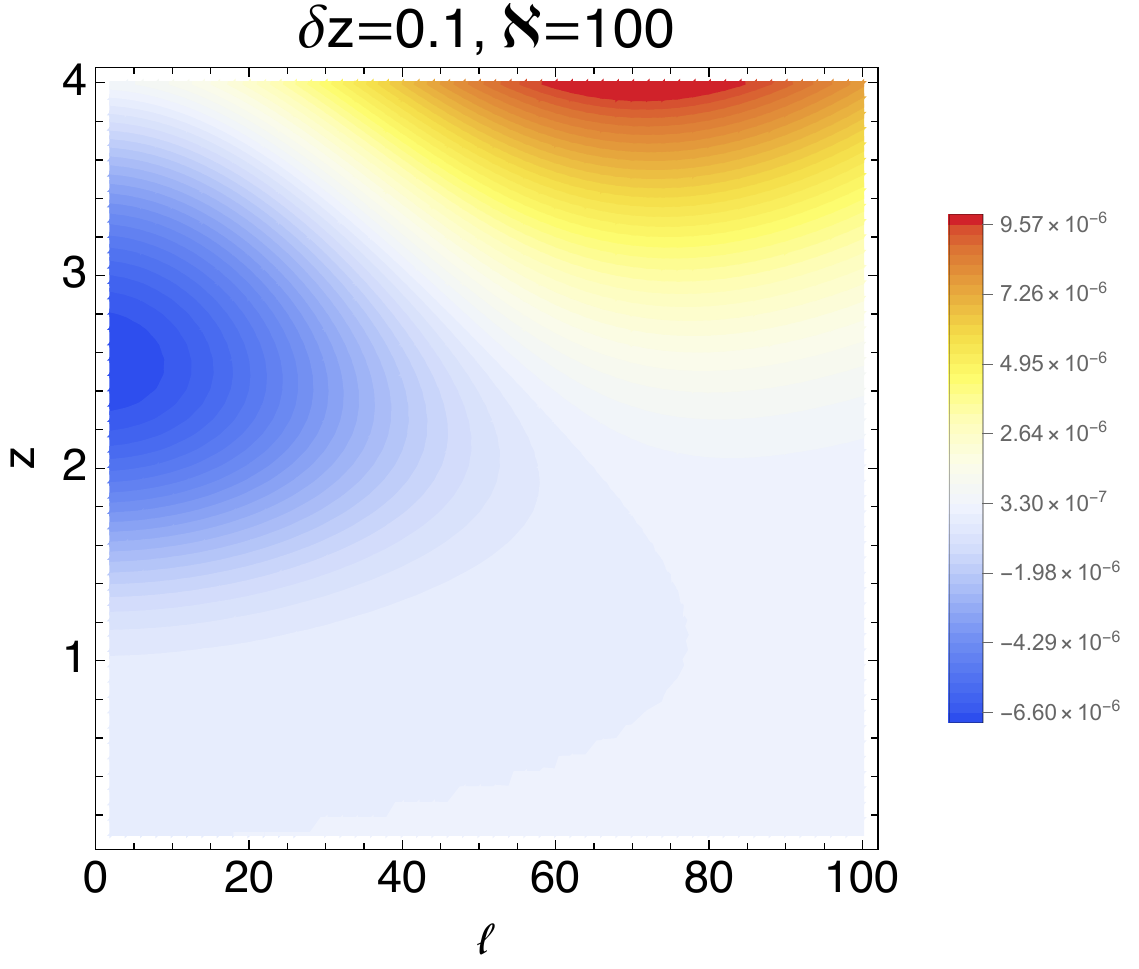}
\includegraphics[width=0.47\textwidth]{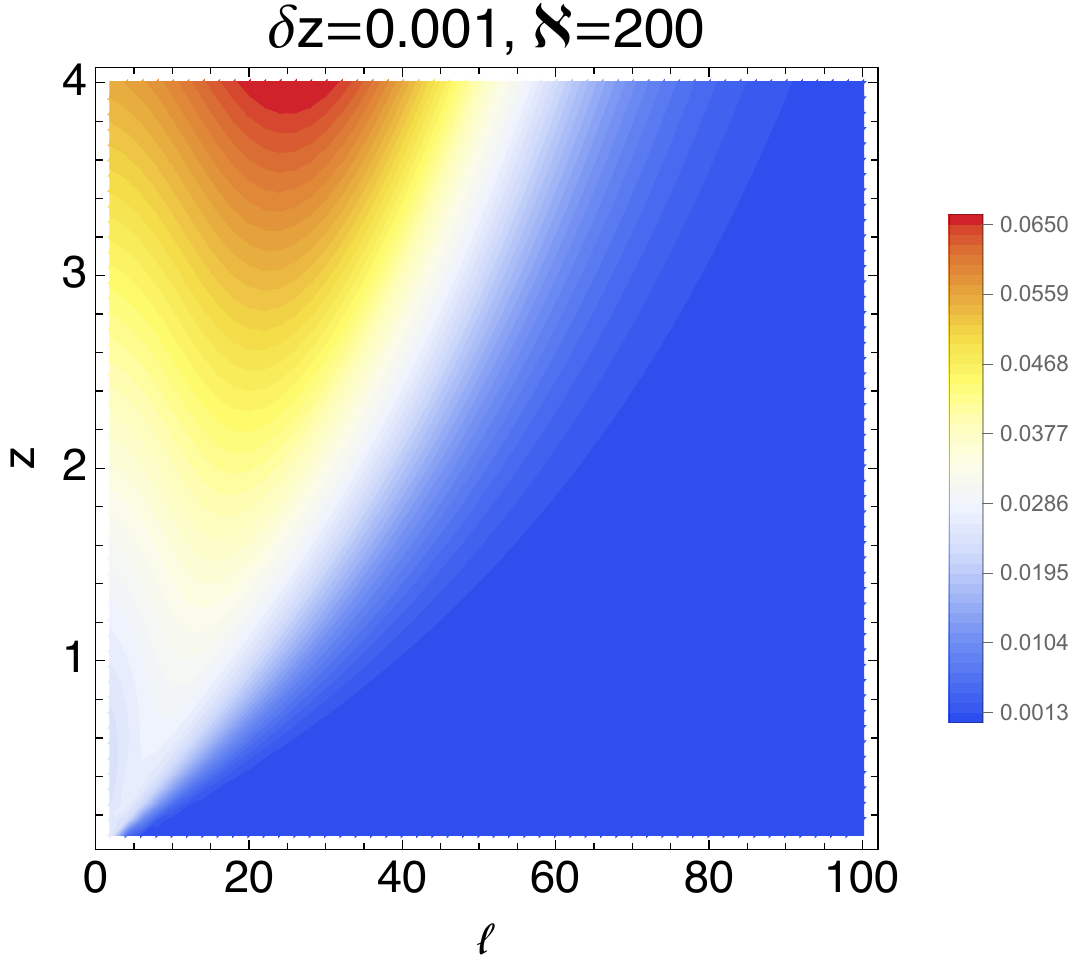}
\includegraphics[width=0.47\textwidth]{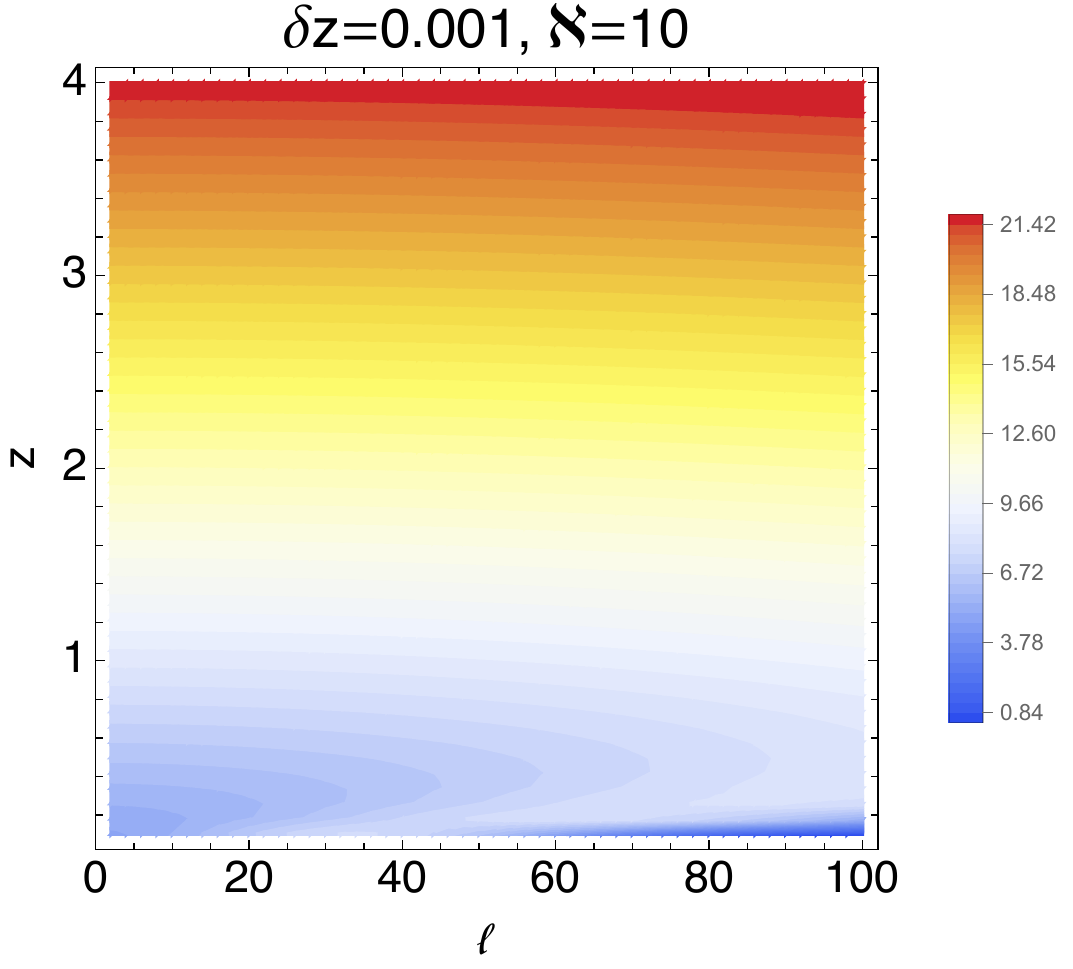}
\caption{Equation~\eqref{eq:CsLD} as a function of $z$ and $\ell$, for different values of $\delta z$ and $\aleph$.}
\label{fig:Csldep4}
\end{figure}

\begin{figure}[htb!]
\includegraphics[width=0.47\textwidth]{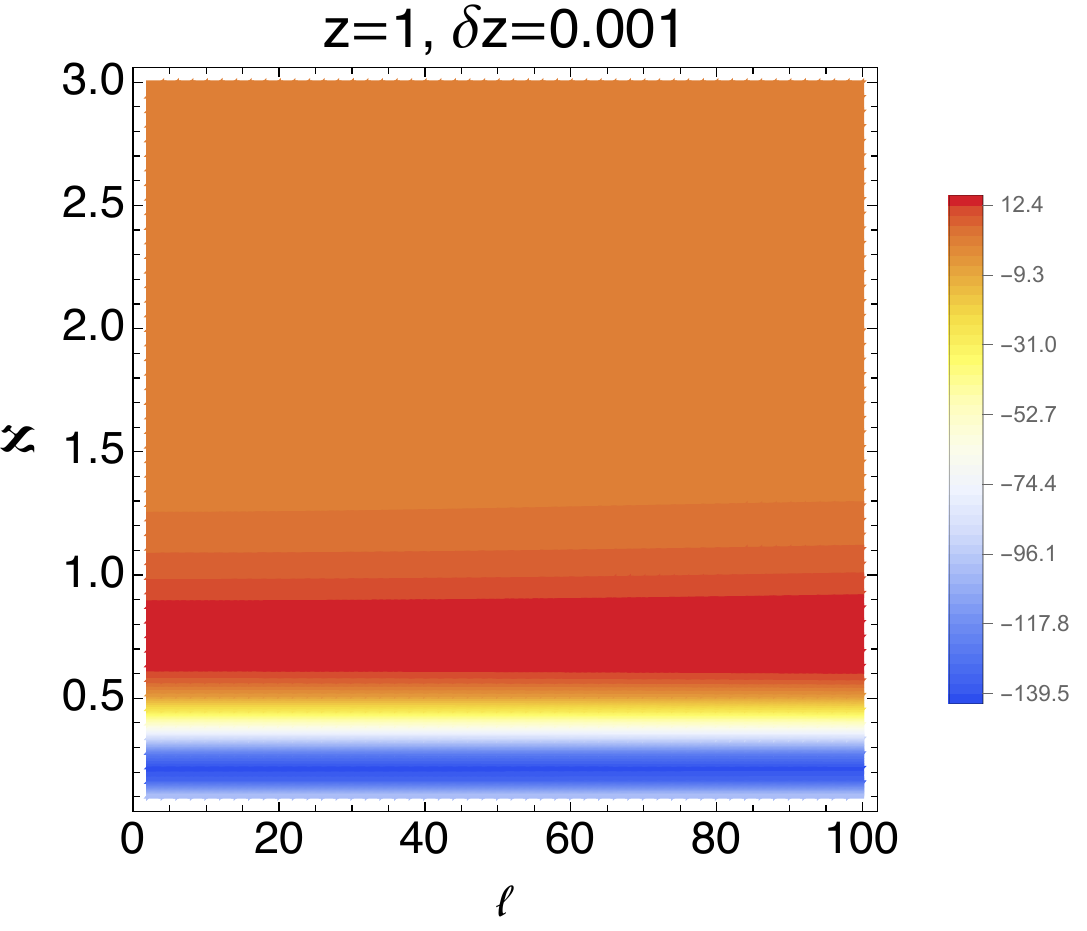}
\includegraphics[width=0.47\textwidth]{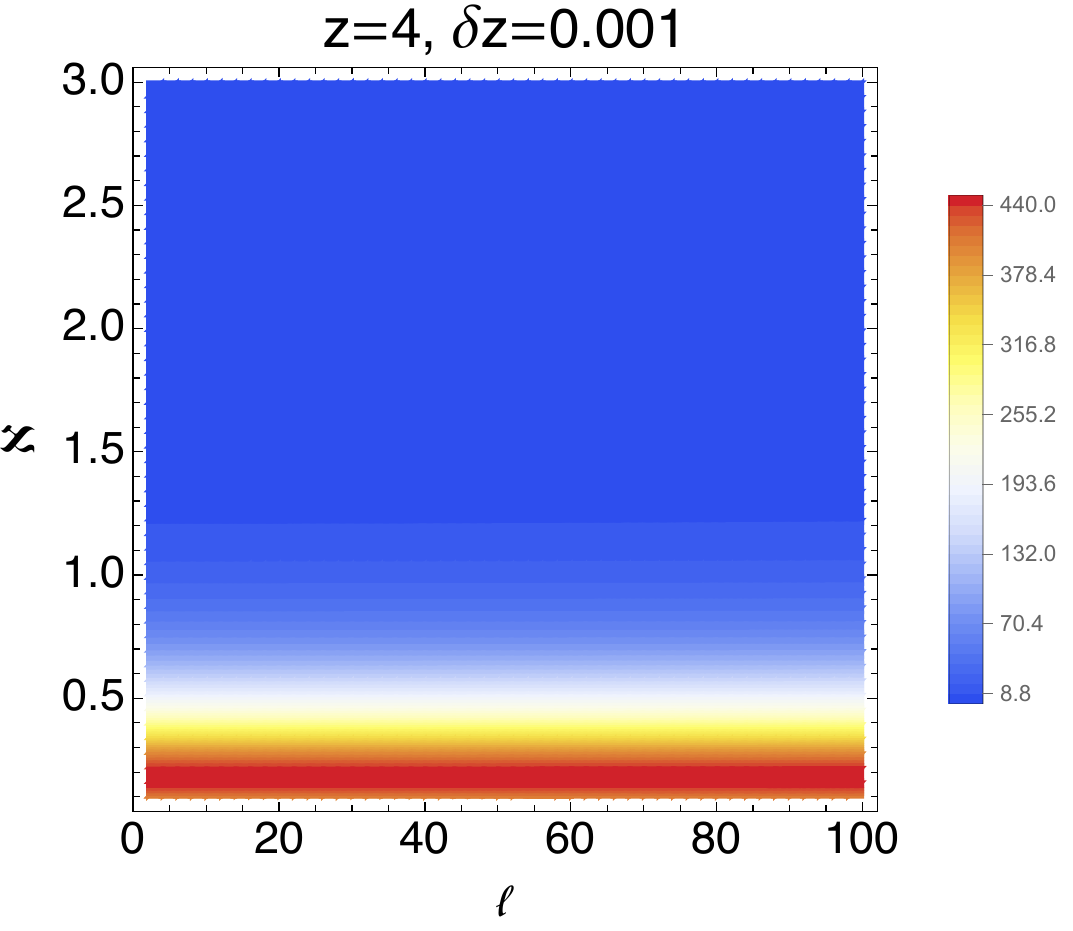}
\includegraphics[width=0.47\textwidth]{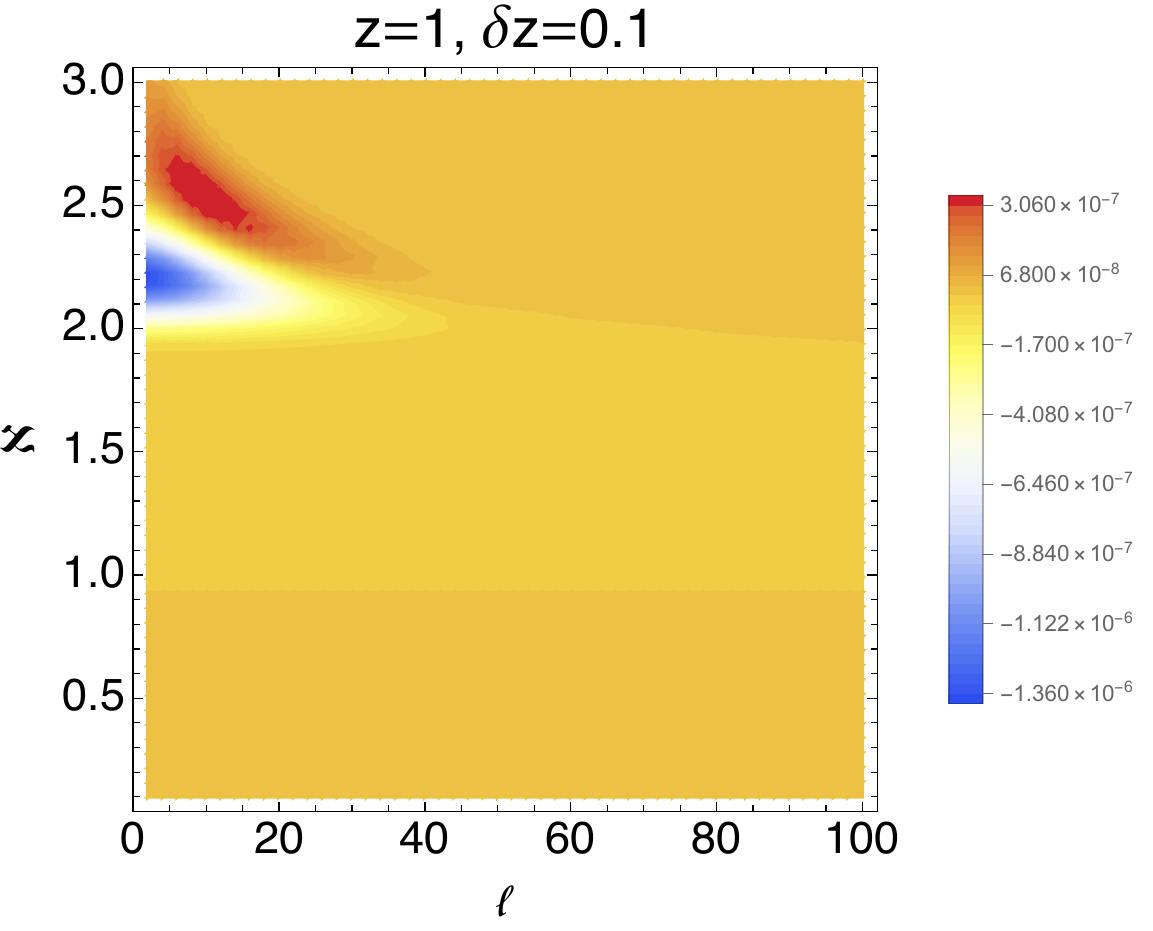}
\includegraphics[width=0.47\textwidth]{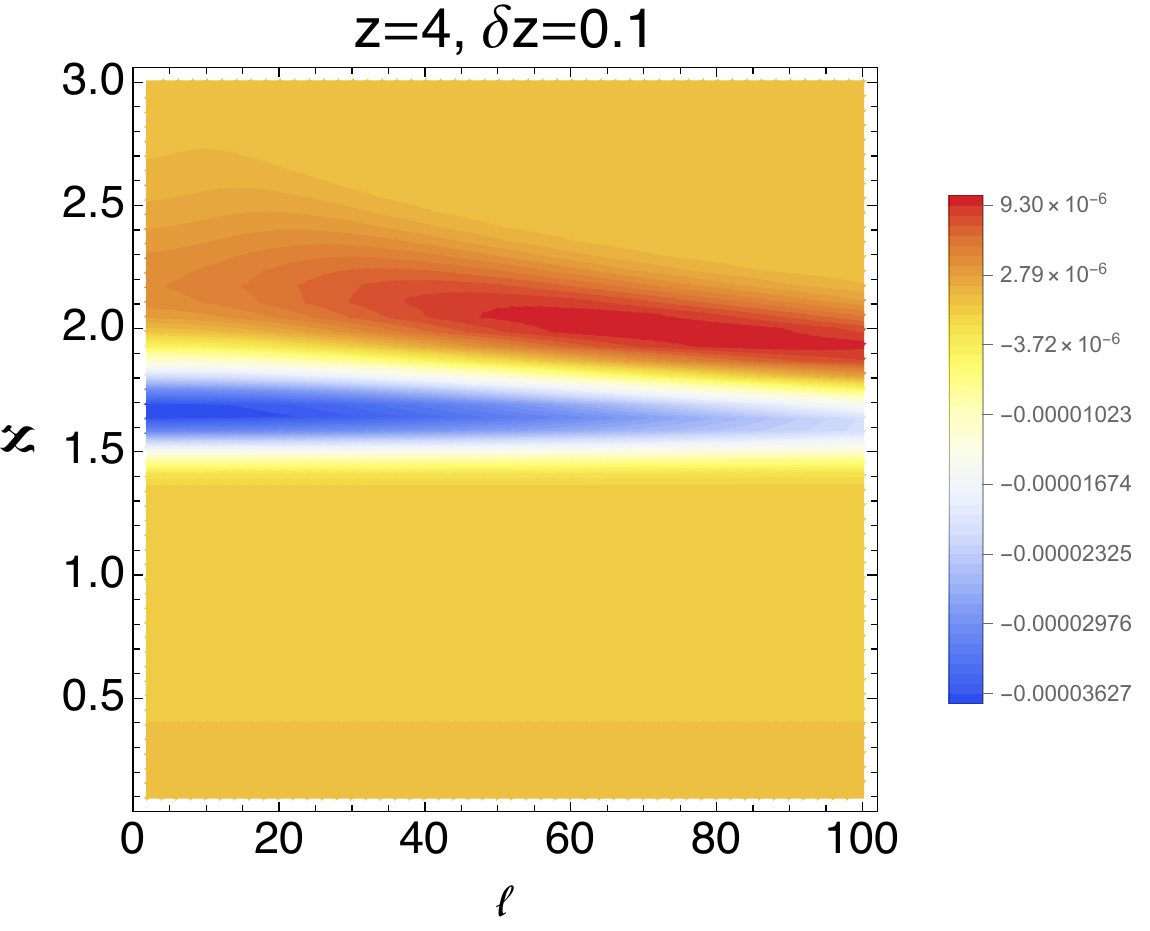}
\caption{Equation~\eqref{eq:CsLD} as a function of $\ell$ and power spectrum slope, for different values of $\delta z$ and $z$.}
\label{fig:Csldep6}
\end{figure}

Furthermore, in Figures~\ref{fig:Csldep1}-\ref{fig:Csldep6}, we show how the flat-sky power spectrum $\mathcal C_s(\ell, \chi, \delta\chi)$ of Equation~\eqref{eq:CsLD} depends on variations of the parameters, including redshift, scale, unequalness in time, equivalence scale.
In the plots the colored code goes from larger negative (blue), to around zero (white) to larger positive (red).
%The effects of different parameters on $\mathcal C_s(\ell, \chi, \delta\chi)$ can be translated into the $\Lambda$CDM case (which we will do in the follow-up papers).
Here the values assumed by specific parameters are not intended to be realistic for a particular scenario, but they provide a guidance to understand the behavior and regimes in which $\mathcal C_s$ is larger or smaller or changes sign. Those can then be remapped into $\Lambda$CDM or alternative models cases.

%==========================================================================
%=================%

%=================%

%==========================================================================

%==========================================================================
\section*{References}
%\bibliography{ms} 

\bibliography{Cave}

\end{document}